\documentclass{cernrep} 
\usepackage{texnames}
\usepackage[T1]{fontenc}

\newcommand{\beq}{\begin{equation}}
\newcommand{\eeq}{\end{equation}}
\newcommand{\bea}{\begin{eqnarray}}
\newcommand{\eea}{\end{eqnarray}}

\newcommand{\ra}{\rightarrow}

\newcommand{\gsim}{\lower.7ex\hbox{$
\;\stackrel{\textstyle>}{\sim}\;$}}
\newcommand{\lsim}{\lower.7ex\hbox{$
\;\stackrel{\textstyle<}{\sim}\;$}}

\newcommand{\GeV}{\,\mbox{GeV}}
\newcommand{\MeV}{\,\mbox{MeV}}
\newcommand{\matel}[3]{\langle #1|#2|#3\rangle}

\newcommand{\eod}{\end{document}}

\def\op{{\bf P}}
\def\oc{{\bf C}}
\def\ot{{\bf T}}
\def\cp{{\bf CP}}
\def\cpt{{\bf CPT}}

\begin{document}
\begin{flushright}
UND-HEP-07-BIG\hspace*{.08em}01\\
hep-ph/0701273\\

\end{flushright}
\vspace*{1.3mm}
\title{Flavour Dynamics \& \cp~Violation in the Standard Model: 
A Crucial Past -- and an Essential Future}
 
\author{Ikaros I. Bigi}

\institute{Physics Dept., University of Notre Dame du Lac, Notre Dame, IN 46556, U.S.A. \\
email address: ibigi@nd.edu}

\maketitle 

\begin{abstract}
Our knowledge of flavour dynamics has undergone a `quantum jump' since just before the 
turn of the millenium: direct \cp~violation has been firmly {\em established} in $K_L \to \pi \pi$ 
decays in 1999; the first \cp~asymmetry outside $K_L$ decays has been discovered in 2001 in 
$B_d \to \psi K_S$, followed by $B_d \to \pi^+\pi^-$, $\eta^{\prime}K_S$ and $B \to K^{\pm}\pi^{\mp}$, 
the last one establishing direct \cp~violation also in the beauty sector. Furthermore CKM dynamics 
allows a description of \cp~insensitive and sensitive $B$, $K$ and $D$ transitions that is 
impressively consistent also on the quantitative level. Theories of flavour dynamics that could 
serve as {\em alternatives} to CKM have been ruled out. Yet these novel successes of the Standard Model (SM) do not invalidate any of the theoretical arguments for the incompleteness of the SM. In 
addition we have also more direct evidence for New Physics, namely neutrino oscillations, the 
observed baryon number of the Universe, dark matter and dark energy. While the New Physics anticipated at the TeV scale is not likely to shed any light on the SM's mysteries of flavour, 
detailed and comprehensive studies of heavy flavour transitions will be essential in diagnosing salient 
features of that New Physics. Strategic principles for such studies will be outlined.   
\end{abstract}

\tableofcontents

In my lecture series I will sketch the past evolution of central concepts of the Standard 
Model (SM), which are of particular importance for its flavour dynamics. The reason is not primarily of 
a historical nature. I hope these sketches will illuminate the main message I want to convey, namely 
that we find ourselves in the midst of a great intellectual adventure: even with the recent novel successes of the SM the case for New Physics at the TeV (and at higher scales)   
is as strong as ever. While there is a crowd favourite for the TeV scale 
New Physics, namely some implementation of Supersymmetry (SUSY) -- an expectation I 
happen to share -- we better allow for many diverse scenarios. To deduce which one is realized in nature we will need all the experimental information we can get, including the impact of the New Physics 
on flavour dynamics. Yet based on the present successes of the SM, we cannot count on that impact being numerically massive. I will emphasize general principles for designing search strategies for New Physics over specific and detailed examples. For at a school like this we want to help you prepare yourself for a future leadership role;  
that requires that you do your own thinking rather than `out-source' it. 

The outline of my three lectures is as follows: 
\begin{itemize}
\item 
{\bf Lecture I:  "Flavour Dynamics in the Second Millenium ($\to$ 1999)"} -- Basics of flavour dynamics 
and \cp~violation, CKM theory, $K^0$ and $B^0$ oscillations, the SM `Paradigm of large 
\cp~violation in $B$ decays'. 
\item 
{\bf Lecture II: "Flavour Dynamics 2000 - 2006"} -- Verifying the SM `Paradigm of large 
\cp~violation in $B$ decays', praising EPR correlations \& hadronization, Heavy Quark Theory, 
extracting CKM parameters and CKM triangle fits.  
\item 
{\bf Lecture III: "Probing the Flavour Paradigm of the {\em Emerging New} Standard Model} -- Indirect 
searches for New Physics, `King Kong' scenarios (EDM's, charm, $\tau$ leptons) vs. precision 
probes (beauty), the case for a Super-Flavour Factory and a new generation of kaon experiments 
in HEP's future landscape. 

\end{itemize}
To a large degree I will follow the historical development, because it demonstrates best, why it is 
advantageous to listen to predictions from theory -- but also go against it at times! 

\vspace{0.3cm}

\section{Lecture I:  "Flavour Dynamics in the Second Millenium ($\to$ 1999)"}
\label{LECT1}

\begin{center}
Memento $\Delta S \neq 0$ dynamics:
\end{center}
\begin{itemize}
\item 
The `$\theta - \tau$ puzzle' -- the observation that two particles decaying into final states of 
opposite parity ($\theta \to 2 \pi$, $\tau \to 3 \pi$) exhibited the same mass and lifetime -- lead to the realization that parity was violated in weak interactions, and actually to a maximal degree in charged 
currents.  
\item 
The observation that the production rate of strange hadrons exceeded their decay rates by many 
orders of magnitude -- a feature that gave rise to the term `strangeness' -- was attributed to 
`associate production' meaning the strong and electromagnetic forces conserve this new 
quantum number `strangeness', while weak dynamics do not. Subsequently it gave rise to the notion of quark families. 
\item 
The great suppression of flavour changing neutral currents as evidenced by the tiny rates for 
$K_L \to \mu^+\mu^-$, $\gamma \gamma$ and the minute size for $\Delta M_K$, lead some daring 
spirits to postulate the existence of a new quantum number for quarks, namely charm. 
\item 
The observation of $K_L \to \pi^+\pi^-$ established that \cp~invariance was not fully implemented 
in nature and induced two other daring spirits to postulate the existence of yet another, the third, quark family, with the top quark, as we learnt later, being two hundred times heavier than kaons.

\end{itemize}
All these features, which are pillars of the SM {\em now}, represented `New Physics' at {\em that} time!

\subsection{On the Uniqueness of the SM}
\label{SMUNIQ}

A famous American Football coach once declared:"Winning is not the greatest thing -- 
it is the only thing!" This quote provides some useful criteria for sketching the status of the 
different components of the Standard Model (SM). It can be characterized by the carriers of its strong and electroweak forces that are described by {\em gauge} dynamics and the 
{\em mass matrices} for its quarks and leptons 
as follows: 
\beq 
{\rm SM}^* = SU(3)_C \times SU(2)_L \oplus {\rm `CKM'} (\oplus {\rm `PMNS'})  
\eeq
I have attached the asteriks to `SM' to emphasize the SM contains a very peculiar pattern of fermion 
mass parameters that is not illuminated at all by its gauge structure. 
Next I will address the status of these components. 

\subsubsection{QCD -- the `Only' Thing}
\label{QCD}

\subsubsubsection{`Derivation' of QCD}
\label{DERIVATION}

While it is important to subject QCD again and again to quantitative tests as the theory for the strong 
interactions, one should note that these serve more as tests of our computational control over 
QCD dynamics than of QCD itself. For its features can be inferred from a few general requirements 
and basic observations. A simplified list reads as follows:  
\begin{itemize}
\item 
Our understanding of chiral symmetry as a {\em spontaneously} realized one -- which allows treating 
pions as Goldstone bosons implying various soft pion theorems -- requires vector couplings for the gluons. 
\item 
The ratio  $R=\sigma (e^+ e^- \to {\rm had.})/\sigma (e^+e^- \to \mu ^+ \mu ^-)$ and the branching ratios 
for $\pi ^0 \to \gamma \gamma $, $\tau ^- \to e^- \bar \nu _e \nu _{\tau}$ and $B \to l \nu X_c$ point to the need for three colours. 
\item 
Colour has to be implemented as an {\em un}broken symmetry. Local gauge theories are the only 
known way to couple {\em massless spin-one} fields in a {\em Lorentz invariant} way. The basic challenge is easily stated: $4 \neq 2$; i.e., while Lorentz covariance requires four component to describe a spin-one field, the latter contains only two physical degrees of freedom for massless fields. 
(For massive vector fields one can go to their rest frame  to reduce and project out one 
component in a Lorentz invariant way to arrive at the  three physical degrees of freedom.)  
\item 
Combining confinement with asymptotic freedom requires a {\em non}-abelian gauge theory. 

\end{itemize}
In summary: for describing the strong interactions QCD is the {\em unique} choice among 
{\em local} quantum field theories. A true failure of QCD would thus create a genuine paradigm 
shift, for one had to adopt an {\em intrinsically non-}local description. It should be remembered that string theory was first put forward for describing the strong interactions. 

\subsubsubsection{`Fly-in-the-Ointment': the Strong \cp~Problem of QCD}
\label{FLY}

A theoretical problem arises for QCD from an unexpected quarter that is very relevant 
for our context here: QCD does {\em not automatically} conserve \op, \ot~ and \cp. To reflect the 
nontrivial topological structure of QCD's ground state one employs an 
{\em effective} Lagrangian containing an additional term to the usual QCD Lagrangian \cite{CPBOOK}: 
\beq 
{\cal L}_{eff} = {\cal L}_{QCD} + \theta \frac{g_S^2}{32\pi^2} G_{\mu \nu} 
\tilde G^{\mu \nu} \; ,  \; \tilde G_{\mu \nu} = \frac{i}{2} \epsilon _{\mu \nu \rho \sigma}G^{\rho \sigma} 
\label{LQCDEFF}
\eeq
Since $G_{\mu \nu} \tilde G^{\mu \nu}$ is a gauge invariant operator, its appearance in general cannot be forbidden, and what is not forbidden has to be considered allowed in a quantum field theory. It represents a total divergence, yet in QCD -- unlike in QED -- it cannot be ignored due 
to the topological structure of the ground state. 

Since under parity $\op$ and time reversal $\ot$ one has 
\beq 
G_{\mu \nu} \tilde G^{\mu \nu} \stackrel{\op, \ot}{\Longrightarrow} - G_{\mu \nu} \tilde G^{\mu \nu} 
\; ,   
\eeq 
the last term in Eq.(\ref{LQCDEFF}) violates $\op$ as well as $\ot$. Since 
$G_{\mu \nu} \tilde G^{\mu \nu}$ is flavour-{\em diagonal}, it generates an electric dipole moment 
(EDM) for the neutron.  From the upper bound on the latter 
\beq 
d_N < 0.63 \cdot 10^{-25} \; {\rm e \, cm} 
\label{NEUTEDM}
\eeq
one infers \cite{CPBOOK} 
\beq 
\theta < 10^{-9} \; . 
\label{THETABOUND} 
\eeq
Being the coefficient of a dimension-four operator $\theta$ can be renormalized to any value, even zero. Yet the modern view of renomalization is more demanding: requiring the renormalized value to be smaller than its `natural' one by {\em orders of magnitude} is frowned upon, since it requires 
{\em finetuning} between the loop corrections and the counterterms. This is what happens here. For  
purely within QCD the only intrinsically `natural' scale for $\theta$ is unity. If $\theta \sim 0.1$ 
or even $0.01$ were found, one would not be overly concerned.  Yet the bound of 
Eq.(\ref{THETABOUND}) is viewed with great alarm as very {\em unnatural} -- unless a symmetry 
can be called upon. If any quark were massless -- most likely the $u$ quark -- chiral rotations representing symmetry transformations in that case could be employed to remove 
$\theta$ contributions. Yet a considerable phenomenological body rules against such a scenario. 

A much more attractive solution would be provided by transforming $\theta$ from a fixed parameter into the manifestation of a {\em dynamical} field -- as is done 
for gauge and fermion masses through the Higgs-Kibble mechanism, see below -- and imposing a 
Peccei-Quinn symmetry would lead {\em naturally} to $\theta \ll {\cal O}(10^{-9})$. Alas -- this attractive solution does not come `for free': it requires the existence of axions. Those have not been observed despite great efforts to find them. 

This is a purely theoretical problem. Yet I consider the fact that it remains unresolved a significant chink 
in the SM$^*$'s armour. I still have not given up hope that `victory can be snatched from the jaws of defeat': establishing a Peccei-Quinn-type solution would be a major triumph for theory.

\subsubsubsection{Theoretical Technologies for QCD}
\label{TECH}

True theorists tend to think that by writing down, say, a Lagrangian one has defined a theory. 
Yet to make contact with experiment one needs theoretical technologies to infer observable quantities 
from the Lagrangian. That is the task that engineers and plumbers like me have set for themselves. 
Examples for such technologies are: 
\begin{itemize}
\item 
perturbation theory; 
\item 
chiral perturbation; 
\item 
QCD sum rules; 
\item 
heavy quark expansions (which will be described in some detail in Lecture II). 

\end{itemize}
Except for the first one they incorporate the treatment of nonperturbative effects. 

None of these can claim universal validity; i.e., they are all `protestant' in nature. There is only 
one `catholic' technology, namely lattice gauge theory 
\footnote{I hasten to add that lattice gauge theory -- while catholic in substance -- exhibits a different sociology: it has {\em not} developed an inquisition and deals with heretics in a rather gentle way.}: 
\begin{itemize}
\item 
It can be applied to nonperturbative dynamics in all domains (with the possible {\em practical} 
limitation concerning strong final state interactions). 
\item 
Its theoretical uncertainties can be reduced in a {\em systematic} way. 
\end{itemize}   
Chiral perturbation theory {\em is} QCD at low energies describing the dynamics of soft pions and kaons. The heavy quark expansions treating the nonperturbative effects in heavy flavour decays through an expansion in inverse powers of the heavy quark mass are tailor made  for describing 
$B$ decays; to which degree their application can be extended down to the charm scale is a more iffy 
question. Different formulations of lattice QCD can approach the 
nonperturbative dynamics at the charm scale from below as well as from above. The degree to which they yield the same results for charm provides an essential cross check on their numerical reliability. 
In that sense the study of charm decays serves as an important bridge between our understanding of nonperturbative effects in heavy and light flavours. 

\subsubsection{$SU(2)_L\times U(1)$ -- not even the Greatest Thing}
\label{SU(2)}

\subsubsubsection{Prehistory}
\label{PRE}

It was recognized from early on that the four-fermion-coupling of Fermi's theory for the weak forces  
yields an {\em effective} description only that cannot be extended to very high energies. The lowest order 
contribution violates unitarity around 250 GeV. Higher order contributions cannot be called upon to 
remedy the situation, since due to the theory being non-renormalizable those come with more 
and more untamable infinities. Introducing massive charged vector bosons softens the problem, yet does not solve it. Consider the propagator for a massive spin-one 
boson carrying momentum $k$: 
\beq 
\frac{- g_{\mu \nu} + \frac{k_{\mu}k_{\nu}}{M_W^2}}{k^2 - M_W^2}
\eeq
The second term in the numerator has great potential to cause trouble. For it can act like 
a coupling term with dimension $1/({\rm mass})^2$; this is quite analogous to the original ansatz of Fermi's theory and amounts to a non-renormalizable coupling. It is actually the {\em longitudinal} 
component of the vector boson that is at the bottom of this problem. 

This potential problem is neutralized, if these massive vector bosons couple to conserved currents. 
To guarantee the latter property, one needs a non-abelian gauge theory, which implies the 
existence of neutral weak currents.

\subsubsubsection{Strong Points}
\label{STRONG}

The requirements of unitarity, which is nonnegotiable, and of renormalazibility, which is to some degree, severely restrict  possible theories of the electroweak interactions. It makes the generation of mass a highly nontrivial one, as sketched below. There are other strong points as well: 

\noindent 
$\oplus$ Since there is a {\em single} $SU(2)_L$ group, there is a single set of gauge bosons. 
Their {\em self}-coupling controls also, how they couple to the fermion fields. As explain later in more detail, this implies the property of `weak universality'. 

\noindent $\oplus$ The SM truly {\em pre}dicted the existence of neutral currents characterized by one 
parameter, the weak angle $\theta_W$, and the masses of the $W$ and $Z$ bosons. 

\noindent 
$\oplus$: Most remarkably the $SU(2)_L\times U(1)$ gauge theory combines QED 
with a pure parity conserving vector coupling to a massless neutral force field with the weak interactions, 
where the charged currents exhibit {\em maximal} parity violation due to their $V-A$ coupling and a 
very short range due to $M_Z > M_W \gg m_{\pi}$.

\subsubsubsection{Generating Mass}
\label{MASSGEN}

A massive spin-one field with momentum $k_{\mu}$ and spin $s_{\mu}$ 
has four (Lorentz) components. Going into its rest frame one realizes that the Lorentz invariant constraint $k\cdot s = 0$ can be imposed, which leaves three independent components, as it has to be. 

A massless spin-one field is still described by four components, yet has only two physical degrees 
of freedom. It needs another physical degree of freedom to transmogrify itself into a massive field. 
This is achieved by having the gauge symmetry {\em realized spontaneously}. For the case at hand this 
is implemented through  an ansatz that should be -- although rarely is -- referred to as 
Higgs-Brout-Englert-Guralnik-Hagen-Kibble mechanism (HBEGHK). Suffice it to consider 
the simplest case of a complex 
scalar field $\phi$ with a potential invariant under $\phi (x) \to e^{i\alpha (x)}\phi (x)$, since this mechanism has been described in great detail in Pich's lectures \cite{PICH}: 
\beq 
V(\phi) = \lambda |\phi|^4  - \frac{m^2}{2} |\phi|^2 
\eeq
Its minimum is obviously not at $|\phi| = 0$, but at $\sqrt{m^2/4\lambda}$. Thus rather than having 
a {\em unique} ground state with $|\phi|=0$ one has an 
{\em infinity of different, yet equivalent} ground states with $|\phi| = \sqrt{m^2/4\lambda}$.  To understand the physical content of such a scenario, one 
considers oscillations of the field around the minimum of the potential: oscillations in the radial direction 
of the field $\phi$ represent a scalar particle with mass; in the polar direction (i.e. the phase of 
$\phi$) the potential is at its minimum, i.e. flat, and 
the corresponding field component constitutes a {\em massless} field. 

It turns out that this massless scalar field can be combined with the two transverse components 
of a $M=0$ spin-one gauge field to take on the role of the latter's longitudinal component leading to the 
emergence of a {\em massive} spin-one field. Its mass is thus controlled by the 
nonperturbative quantity $\langle 0|\phi|0\rangle$. 

Applying this generic construction to the SM one finds that a priori both $SU(2)_L$ doublet and triplet 
Higgs fields could generate masses for the weak vector bosons. The ratio {\em observed} for the 
$W$ and $Z$ masses is fully consistent with only doublets contributing. Intriguingly enough such 
doublet fields can eo ipso generate fermion masses as well. 

In the SM one adds a single complex scalar doublet field to the mix of vector boson and fermion 
fields. Three of its four components slip into the role of the longitudinal components of 
$W^{\pm}$ and $Z^0$; the fourth one emerges as an independent physical field -- `the' Higgs 
field. Fermion masses are then given by the product of the single vacuum expectation value 
(VEV) $\langle 0|\phi|0\rangle$ and their Yukawa couplings -- a point we will return to.

\subsubsubsection{Triangle or ABJ Anomaly}
\label{ABJ}

The diagram with an internal loop of only fermion lines, to which three external axial vector 
(or one axial vector and two vector) lines are attached, generates a `quantum anomaly' 
\footnote{It is referred to as `triangle' anomaly due to the form of the underlying diagram or 
A(dler)B(ell)J(ackiw) anomaly due to the authors that identified it \cite{ABJ}.}: it 
removes a {\em classical} symmetry as expressed through the existence of a conserved current. 
In this specific case it affects the conservation of the axialvector current $J_{\mu}^5$. Classically 
we have $\partial ^{\mu}J^5_{\mu} = 0$ for {\em massless} fermions; yet the triangle anomaly leads to 
\beq 
\partial ^{\mu}J^5_{\mu} =  \frac{g_S^2}{16\pi ^2} G \cdot \tilde G  \neq 0
\label{TRIANOM}
\eeq
even for massless fermions; $G$ and $\tilde G$ denote the gluonic field strength tensor and its 
dual, respectively, as introduced in Eq.(\ref{LQCDEFF}). 

While by itself it yields a finite result on the right hand side of Eq.(\ref{TRIANOM}), it destroys the renormalizability of the theory. It cannot be 
`renormalized away' (since in four dimensions it cannot be regularized in a gauge invariant way). 
Instead it has to be neutralized by requiring that adding up this contribution from all types of fermions in the theory yields a vanishing result. 

For the SM this requirement can be expressed very concisely that all electric charges of the fermions 
of a given family have to add up to zero.  This imposes a connection between the charges of quarks 
and leptons, yet does not explain it.

\subsubsubsection{Theoretical Deficiencies}
\label{DEFEC}

With all the impressive, even  amazing successes of the SM, it is natural to ask why is the community 
not happy with it. There are several drawbacks: 

\noindent 
$\ominus$ Since the gauge group is $SU(2)_L \times U(1)$, only partial unification has been 
achieved. 

\noindent $\ominus$ 
The HBEGHK mechanism is viewed as providing merely an `engineering' solution, in particular since the physical Higgs field has not been observed yet. Even if or when it is, theorists in particular will 
not feel relieved, since scalar dynamics induce {\em quadratic} mass renormalization and are viewed 
as highly `unnatural', as exemplified through the gauge hierarchy problem. This concern has 
lead to the conjecture of New Physics entering around the TeV scale, which has 
provided the justification for the LHC and the motivation for the ILC. 

\noindent $\ominus$
{\em maximal} violation of parity is implemented for the charged weak currents `par ordre du mufti'
\footnote{A French saying describing a situation, where a decision is imposed on someone with no 
explanation and no right of appeal.}, i.e. 
based on the data with no deeper understanding. 

\noindent $\ominus$
Likewise neutrino masses had been set to zero  `par ordre du mufti'.

\noindent $\ominus$
The observed quantization of electric charge is easily implemented and is instrumental in neutralizing 
the triangle anomaly  -- yet there is no understanding of it. 

One might say these deficiencies are merely `warts' that hardly detract from the beauty of the SM.  
Alas -- there is the whole issue of family replication!

\subsubsection{The Family Mystery}
\label{FAMILY}

The twelve known quarks and leptons are arranged into three families. Those families possess identical gauge couplings and are distinguished only by their mass terms, i.e. their Yukawa couplings. 
We do not understand this family replication or why there are three families. It is not even clear 
whether the number of families represents a fundamental quantity  or is due to the more or less accidental interplay of complex forces as one encounters when analyzing the structure of nuclei. 
The only hope for a theoretical understanding we can spot on the horizon is superstring or 
M theory -- which is merely a euphemistic way of saying we have no clue. 

Yet the circumstantial evidence that we miss completely a central element of Nature's `Grand Design' is even stronger in view of the strongly hierarchical pattern in the masses for up- and down-type quarks, charged leptons and neutrinos and the CKM parameters as discussed later.

\subsection{Basics of \op, \oc, \ot, \cp~ and \cpt}
\label{BASICSDISC}

\subsubsection{Definitions}
\label{DEFDISC}

{\em Parity transformations} flip the sign of position vectors $\vec r$ while leaving the time coordinate $t$ 
unchanged: 
\beq 
(\vec r, t) \stackrel{\op}{\longrightarrow} (-\vec r, t) 
\eeq
Momenta change their signs as well, yet orbital and other angular momenta do not:
\beq 
\vec p  \stackrel{\op}{\longrightarrow} -\vec p   \; \; vs. \; \; \vec l \equiv \vec r \times \vec p  
\stackrel{\op}{\longrightarrow} \vec l
\eeq
Parity odd vectors -- $\vec r$, $\vec p$ -- and parity even ones -- $\vec l$ -- are referred to as 
polar and axial vectors, respectively. 
Likewise one talks about scalars $S$ and pseudoscalars $P$ with 
$S \stackrel{\op}{\longrightarrow} S$ and $P\stackrel{\op}{\longrightarrow} -P$. Examples are 
$S = \vec p_1 \cdot \vec p_2$, $\vec l_1 \cdot \vec l_2$ and $P = \vec l_1 \cdot \vec p_2$. 
Parity transformations are equivalent to mirror transformations followed by a rotation. They are described by a linear operator \op. 

{\em Charge conjugation} exchanges particles and antiparticles and thus flips the sign of all charges 
like electric charge, hyper-charge etc. It is also described by a linear operator \oc. 

{\em Time reversal} is operationally defined as a reversal of motion 
\beq 
(\vec p, \vec l) \stackrel{\ot}{\longrightarrow} - (\vec p, \vec l)  \; , 
\eeq
which follows from $(\vec r, t) \stackrel{\ot}{\longrightarrow} (\vec r,-t)$. While the Euclidean scalar 
$\vec l_1 \cdot \vec p_2$ is invariant under the time reversal operator \ot, the triple correlations 
of (angular) momenta are not: 
\beq 
\vec v_1 \cdot (\vec v_2 \times \vec v_3)  \stackrel{\ot}{\longrightarrow} 
- \vec v_1 \cdot (\vec v_2 \times \vec v_3) \; \; {\rm with} \; \; \vec v = \vec p, \vec l \; . 
\eeq
The expectation value of such triple correlations accordingly are referred to as \ot~{\em odd moments}. 

In contrast to \op~ or \oc~ the \ot~operator is {\em anti}linear: 
\beq 
\ot (\alpha |a\rangle + \beta |b\rangle ) = \alpha ^* \ot |a \rangle + \beta ^* \ot |b \rangle
\eeq
This property of \ot~is enforced by the commutation relation $[X,P] = i \hbar$, since 
\bea 
\ot ^{-1} [X,P]\ot &=& - [X,P] \\
\ot ^{-1} i\hbar \ot  &=& - i \hbar 
\eea
The anti-linearity of \ot~implies three important properties: 
\begin{itemize}
\item 
\ot~violation manifests itself through complex phases. \cpt~invariance then implies that also 
\cp~violation enters through complex phases in the relevant couplings. For \ot~or 
\cp~violation to become observable in a decay transition one thus needs the contribution 
from two different, yet coherent amplitudes.  
\item 
While a non-vanishing \op~odd moment establishes unequivocally \op~violation, this is {\em not} 
necessarily 
so for \ot~odd moments; i.e., even \ot~invariant dynamics can generate a non-vanishing 
\ot~odd moment. \ot~being antilinear comes into play when the transition amplitude is described 
{\em through second} 
(or even higher) order in the effective interaction, i.e. when final state interactions are 
included denoted symbolically by 
\beq 
\ot ^{-1} ({\cal L}_{eff}\Delta t + \frac{i}{2}({\cal L}_{eff}\Delta t)^2 + ...)\ot = 
{\cal L}_{eff}\Delta t -  \frac{i}{2}({\cal L}_{eff}\Delta t)^2 + ... \neq 
{\cal L}_{eff}\Delta t +  \frac{i}{2}({\cal L}_{eff}\Delta t)^2 + ... 
\eeq  
even for $[\ot, {\cal L}_{eff}]=0$. 
\item 
`Kramer's degeneracy' \cite{KRAMER}: 
With \ot~being anti-unitary,  the Hilbert space -- for \ot~invariance -- 
can be decomposed into two disjoint sectors, one with \ot$^2 = 1$ and the other with 
\ot$^2 = -1$, and the latter one is at least doubly degenerate in energy. 

It turns out that 
for bosonic states one has \ot$^2 = 1$ and for fermionic ones \ot$^2 = -1$. The amazing thing is that 
the necessary anti-unitarity of the \ot~operator already anticipates the existence of fermions and 
bosons -- without any reference to spin. Maybe a better way of expressing it is as follows. While 
nature seems to be fond of realizing mathematical structures, it does so in a very efficient way: 
it can have bosons -- states symmetric under permutation of identical particles -- and fermions, which are 
antisymmetric; it can contain states with half integer and integer spin, and finally it allows for 
states with \ot$^2 = \pm 1$. It implements all these structures and does so in the most efficient way, 
namely by bosons [fermions] carrying 
[half] integer spin and \ot$^2 = +[-]1$.  

Kramer's degeneracy has practical applications as well, for example in solid state physics: 
consider electrons inside an external electrostatic field. Such a field breaks 
rotational invariance; thus angular momentum is no longer conserved. yet no matter how complicated this field is, for an {\em odd} number of electrons there always has to be at least two-fold degeneracy.

\end{itemize}

\subsubsection{{\em Macroscopic} \ot~Violation or `Arrow of Time'}
\label{ARROWW}

Let us consider a simple example from classical mechanics: the motion of billiard ball(s) across a 
billiard table in three different scenarios.  

\noindent (i) Watching  a movie showing a {\em single} ball role around and bounce off the walls of the table one could not decide whether one was seeing the events in the actual time sequence or in the reverse order, i.e. whether one was seeing the movie running backwards. For both sequences are possible and equally likely. 

\noindent (ii)  Seeing one ball move in and hit another ball at rest leading to both balls moving off in different directions is a possible and ordinary sequence. The reverse -- two balls moving in from different 
directions, hitting each other with one ball coming to a complete rest and the other one moving off in a different direction -- is still a possible sequence yet a rather unlikely one since it requires fine tuning between the momenta of the two incoming billiard balls. 

\noindent (iii) One billiard ball hitting a triangle of ten billiard balls at rest and scattering them in all 
directions is a most ordinary sequence for anybody but the most inept billiard player.  The reverse sequence -- eleven billiard balls coming in from all different directions, hitting each other in such a way that ten come to rest in a neatly arranged triangle while the eleventh one moves off -- is a practically impossible one, since it requires a most delicate fine tuning of the initial conditions. 

There are countless other examples of one time sequence being ordinary while the reversed one is 
(practically) impossible -- take $\beta$ decay $n \to p e^-\bar \nu$, the scattering of a plane wave off an object leading to an outgoing spherical wave in addition to the continuing plane wave or the challenge of parking a car in a tight spot compared with the relative ease to get out of it. These 
daily experiences do not tell us anything about \ot~violation in the underlying dynamics; they reflect 
asymmetries in the {\em macroscopic} initial conditions, which are of a statistical nature. 

Yet a central message of my lectures is that {\em microscopic} \ot~violation has been observed, 
i.e. \ot~violation that resides in the basic dynamics of the SM. It is conceivable though that in a more 
complete theory it reflects an asymmetry in the initial conditions in some higher sense.

\subsection{The Very Special Role of \cp~Invariance and its Violation}
\label{SPECIALCP}

While the discovery of \op~violation in the weak dynamics in 1957 caused a well documented shock 
in the community, even the theorists quickly recovered. Why then was the 
discovery of \cp~violation in 1964 not viewed as a `deja vue all over again' in the language of Yogi Berra? There are several reasons for that as illustrated by the following statements: 

\begin{itemize}
\item 
Let me start with an analogy from politics.  In my days as a student -- at a time long ago and a place far away -- politics was hotly debated. One of the subjects drawing out the greatest 
passions was the questions of what distinguished the `left' from the `right'. If you listened to it, you quickly found out that people almost universally defined `left' and `right' in terms of `positive' and 
`negative'. The only problem was they could not quite agree  who the good guys and the bad guys are. 

There arises a similar  conundrum  when considering decays like $\pi \to e \nu$. When saying 
that a pion decay produces a {\em left} handed charged lepton one had $\pi ^- \to e_L^- \bar \nu$ in 
mind. However $\pi^+ \to e^+_R \nu$ yields a {\em right} handed charged lepton.  
`Left' is thus defined in terms of `negative'. No matter how much \op~is violated, \cp~invariance 
imposes equal rates for these $\pi^{\pm}$ modes, and it is untrue to claim that nature makes an absolute 
distinction between `left' and `right'. The situation is analogous to the saying that `the thumb is 
left on the right hand' -- a correct, yet useless statement, since circular. 

\cp~violation is required to define `matter' vs. `antimatter', `left' vs. `right', `positive' vs. `negative' 
in a convention independent way. 

\item 
Due to the almost unavoidable \cpt~symmetry violation of \cp~implies one of \ot. 

\item 
It is the smallest {\em observed} violation of a symmetry as expressed through 
\beq 
{\rm Im} M^K_{12} \simeq 1.1 \cdot 10^{-8} \; {\rm eV}  \; \leftrightarrow \;  
\frac{{\rm Im} M^K_{12}}{M_K} \simeq 2.2 \cdot 10^{-17} 
\eeq
\item 
It is one of the key ingredients in the Sakharov conditions for baryogenesis 
\cite{DOLGOV}: to obtain the 
observed baryon number of our Universe as a {\em dynamically generated} quantity rather than 
an arbitrary initial condition one needs baryon number violating transitions with \cp~violation to occur in a period, where our Universe had been out of thermal equilibrium. 

\end{itemize}

\subsection{Flavour Dynamics and the CKM Ansatz}
\label{FLAVDYN}

\subsubsection{The GIM Mechanism}
\label{GIMMECH}

A striking feature of (semi)leptonic kaon decays are the huge suppression of strangeness changing 
neutral current modes: 
\beq 
\frac{\Gamma (K^+ \to \pi^+e^+e^-)}{\Gamma (K^+ \to \pi^0 e^+\nu)} \sim 6 \cdot 10^{-6}\; , \; 
\frac{\Gamma (K_L \to \mu^+ \mu^-)}{\Gamma (K^+ \to \mu^+ \nu)} \sim 3\cdot 10^{-9}
\eeq
Embedding weak charged currents with their Cabibbo couplings 
\bea 
\nonumber 
J^{(+)}_{\mu} &=& {\rm cos}\theta _C \bar d_L \gamma _{\mu}u_L + 
{\rm sin}\theta _C \bar s_L \gamma _{\mu}u_L \\
J^{(-)}_{\mu} &=& {\rm cos}\theta _C \bar u_L \gamma _{\mu}d_L + 
{\rm sin}\theta _C \bar u_L \gamma _{\mu}s_L
\label{CABCUR}
\eea   
into an $SU(2)$ gauge theory to arrive at a renormalizable theory requires neutral currents of a structure as obtained from the commutator of $J^{(+)}_{\mu}$ and $J^{(-)}_{\mu}$. Using for 
the latter the expressions of Eq.(\ref{CABCUR}) one arrives unequivocally at 
\beq 
J_{\mu}^{(0)} = ... + {\rm cos}\theta_C {\rm sin}\theta _C (\bar s_L \gamma _{\mu} d_L + 
\bar d_L \gamma _{\mu} s_L) \; , 
\label{SChNC}
\eeq 
i.e., strangeness changing neutral currents. Yet their Cabibbo suppression is not remotely sufficient 
to make them compatible with these observed super-tiny branching ratios  
\footnote{The observed huge suppression of strangeness changing neutral currents actually 
led to some speculation that also flavour {\em conserving} neutral currents are greatly suppressed.}.  
The huge discrepancy between observed and expected branching ratios lead 
some daring spirits \cite{GIM} to postulate a fourth quark 
\footnote{A fourth quark had been originally introduced by Glashow and Bjorken to regain quark-lepton correspondence by completing the second quark family.} with quite specific properties to complete 
a second quark family in such a way that no strangeness changing neutral currents arise at 
{\em tree} level. The name `charm' derives from this feature of warding off the evil of 
strangeness changing neutral currents rather than an anticipated relation to beauty.

Yet I remember there was great skepticism felt in the community maybe best expressed 
by the quote: "Nature is smarter than Shelley (Glashow) -- she can do without charm quarks." 
\footnote{The fact that nature needed charm after all does not prove the inverse of this quote, 
of course.} These remarks can indicate how profound a shift in paradigm were begun by the 
observation of scaling in deep inelastic lepton-nucleon scattering and completed by the discovery of 
the $J/\psi$ in 1974 and its immediate aftermatch.  

\subsubsection{Quark Masses and \cp~Violation}
\label{QMASSES}

Let us consider the mass terms for the up- and down-type quarks as expressed through 
matrices ${\cal M}_{U/D}$ and vectors of quark fields $U^F= (u,c,t)^F$ and $D^F=(d,s,b)^F$ 
in terms of the {\em flavour} eigenstates denoted 
by the superscript $F$: 
\beq 
{\cal L}_M \propto \bar U_L^F{\cal M}_U U_R^F + \bar D_L^F{\cal M}_D D_R^F  \; . 
\label{MASSLAG}
\eeq
 A priori there is no reason 
why the matrices ${\cal M}_{U/D}$ should be diagonal. Yet applying bi-unitary rotations 
${\cal J}_{U/D,L}$ will 
allow to diagonalize them
\beq 
{\cal M}_{U/D}^{\rm diag} =     {\cal J}_{U/D,L} {\cal M}_{U,D}{\cal J}_{U/D,R}^{\dagger}
\eeq
and obtain the {\em mass} eigenstates of the quark fields: 
\beq 
U^m_{L/R} = {\cal J}_{U,L/R} U_{L/R}^F \; , \; \; D^m_{L/R} = {\cal J}_{D,L/R} D_{L/R}^F
\eeq
The eigenvalues of ${\cal M}_{U/D}$ represent the masses of the quark fields. The measured 
values exhibit a very peculiar pattern that seems unlikely to be accidental being so hierarchical 
for up- and down-type quarks, charged and neutral leptons. 

Yet again, there is much more to it. Consider the neutral current coupling
\beq 
{\cal L}_{NC}^{U[D]} \propto \bar g_Z \bar U^F[\bar D^F] \gamma _{\mu} U^F[D^F] Z^{\mu} 
\label{LAGNC}
\eeq
It keeps its form when expressed in terms of the mass eigenstates
\beq 
{\cal L}_{NC}^{U[D]} \propto \bar g_Z \bar U^m[\bar D^m] \gamma _{\mu} U^m[D^m] Z^{\mu} \; ; 
\label{LAGNC2} 
\eeq 
i.e., there are {\em no} flavour changing neutral currents. This important property is referred to 
as the `generalized' GIM mechanism \cite{GIM}. 

However for the charged currents the situation is quite different: 
\beq 
{\cal L}_{CC} \propto \bar g_W \bar U^F_L \gamma _{\mu} D^F W^{\mu} = 
\bar g_W \bar U^m_L \gamma _{\mu} V_{CKM}D^mW^{\mu}
\label{LAGCC}
\eeq
with 
\beq 
V_{CKM} = {\cal J}_{U,L}{\cal J}_{D,L}^{\dagger} 
\eeq
There is no reason, why the matrix $V_{CKM}$ should be the identity matrix or even 
diagonal \footnote{Even if some speculative dynamics were to enforce an alignment 
between the $U$ and $D$ quark fields at some high scales causing their mass matrices 
to get diagonalized by the same bi-unitary transformation, this alignment would probably 
get upset by renormalization down to the electroweak scales.}. It means the charged 
current couplings of the mass eigenstates will be modified in an observable way. In which way and by how much this happens requires further analysis since the phases of fermion fields are not necessarily 
observables. Such an analysis was first given by Kobayashi and Maskawa \cite{KM}. 

Consider $N$ families. $V_{CKM}$ then represents an $N \times N$ matrix that has to be unitary 
based on two facts: 
\begin{itemize}
\item 
The transformations ${\cal J}_{U/D,L/R}$ are unitary by construction. 
\item 
As long as the carriers of the weak force are described by a {\em single} local gauge group -- 
$SU(2)_L$ in this case -- they have to couple to all other fields in a way fixed by their 
{\em self}coupling. This was already implied by Eq.(\ref{LAGCC}), when writing the weak coupling 
$\bar g_W$ as an overall factor.  

\end{itemize} 
The unitarity of $V_{CKM}$ implies {\em weak universality}, as addressed later in more detail. 
There are actually $N$ such relations characterized by 
\beq 
\sum _j |V(ij)|^2 = 1\; , \; \; i = 1, ..., N 
\label{WU}
\eeq
These relations are important, yet insensitive to weak phases; thus they provide no {\em direct} 
information on \cp~violation.  

{\em Violations} of weak universality can be implemented by adding dynamical layers to the SM. 
So-called horizontal gauge interactions, which differentiate between families and induce 
flavour-changing neutral currents, will do it. Another admittedly ad-hoc possibility is to introduce 
a separate $SU(2)_L$ group for each quark family while allowing the gauge bosons from the 
different $SU(2)_L$ groups to mix with each other. This mixing can be set up in such a way that the lightest 
mass eigenstates couple to all fermions with approximately universal strength. Weak universality 
thus emerges as an approximate symmetry. Flavour changing neutral currents are again induced, and they can generate electric dipole moments. 

After this aside on weak universality let us return to  $V_{CKM}$. There are $N^2 - N$ 
{\em orthogonality} relations:
\beq 
\sum _j V^*(ij)V(jk) = 0 \; , \; \; i\neq k 
\eeq
Those are very sensitive to complex phases and tell us {\em directly} about \cp~violation. 

An $N \times N$ complex matrix contains $2N^2$ real parameters; the unitarity constraints 
reduce it to $N^2$ independent real parameters. Since the phases of quark fields like other 
{\em fermion} fields can be rotated freely, $2N-1$ phases can be removed from 
${\cal L}_{CC}$ (a {\em global} phase rotation of all quark fields has no impact on 
${\cal L}_{CC}$). Thus we have $(N-1)^2$ {\em independent physical} parameters. 
Since an $N\times N$ {\em orthogonal} matrix has $N(N-1)/2$ angles, we conclude that 
an $N\times N$ {\em unitary} matrix contains also $(N-1)(N-2)/2$ {\em physical phases}. This 
was the general argument given by Kobayashi and Maskawa. Accordingly: 
\begin{itemize}
\item 
For $N=2$ families we have one angle -- the Cabibbo angle -- and zero phases. 
\item 
For $N=3$ families we obtain three angles and one irreducible phase; i.e. a three family 
ansatz can support \cp~violation with a single source -- the `CKM phase'. 
PDG suggests a "canonical" parametrization for the $3\times 3$ CKM 
matrix: 
$$ 
{\bf V}_{CKM} = 
\left(  
\begin{array}{ccc} 
V(ud) & V(us) & V(ub) \\
V(cd) & V(cs) & V(cb) \\
V(td) & V(ts) & V(tb) 
\end{array} 
\right) 
$$
\beq 
= \left( 
\begin{array}{ccc} 
c_{12}c_{13} & s_{12}c_{13} & s_{13}e^{-i \delta _{13}}  \\
- s_{12}c_{23} - c_{12}s_{23}s_{13}e^{i \delta _{13}} &
c_{12}c_{23} - s_{12}s_{23}s_{13}e^{i \delta _{13}} & 
c_{13}s_{23} \\
s_{12}s_{23} - c_{12}c_{23}s_{13}e^{i \delta _{13}} &
- c_{12}s_{23} - s_{12}c_{23}s_{13}e^{i \delta _{13}} &
c_{13}c_{23} 
\end{array}
\right) 
\label{PDGKM} 
\eeq 
where 
\beq 
c_{ij} \equiv {\rm cos} \theta _{ij} \; \; , \; \;  
s_{ij} \equiv {\rm sin} \theta _{ij}
\eeq  
with $i,j = 1,2,3$ being generation labels. 

This is a completely general, yet not unique parametrisation: a 
different set of 
Euler angles could be chosen; the phases can be shifted around 
among the matrix elements 
by using a different phase convention. 
\item 
For even more families we encounter a proliferation of angles and phases, namely six angles 
and three phases for $N=4$. 

\end{itemize}
These results obtain by algebraic means can be visualized graphically: 
\begin{itemize}
\item 
For $N=2$ we have two weak universality conditions and two orthogonality relations: 
\bea 
\nonumber 
V^*(ud)V(us) + V^*(cd)V(cs) &=& 0 \\
V^*(us)V(ud) + V^*(cs)V(cd) &=& 0
\eea
While the CKM angles can be complex, there can be no nontrivial phase 
($\neq 0,\pi$) between their observable combinations; i.e., there can be no 
\cp~violation for two families in the SM. 
\item 
For three families  the orthogonality relations read 
\beq 
\sum _{j=1}^{j=3} V^*(ij)V(jk) = 0 \; , \; \; i\neq k
\eeq
There are six such relations, and they represent triangles in the complex plane with in general 
nontrivial relative angles. 
\item 
While these six triangles can and will have quite different shapes, as we will describe later 
in detail, they all have to possess the {\em same area}, namely \cite{JARL} 
\bea
\nonumber  
{\rm area(every\; triangle)} &=& \frac{1}{2}J \\
J =  {\rm Im}[V(ud)V(cs)V^*(us)V^*(cd)]   
\label{JARL}
\eea
{\em If} $J=0$, one has obviously no nontrivial angles, and there is no \cp~violation. The 
fact that all triangles have to possess the same area reflects the fact that for three families 
there is but a {\em single} CKM phase. 
\item 
Only the angles, i.e. the relative phases matter, but not the overall orientation of the triangles 
in the complex plane. That orientation merely reflects the phase convention for the quark fields. 

\end{itemize}
{\em If} any pair of up-type or down-type quarks were mass {\em degenerate}, then 
{\em any} linear combination of those two would be a mass eigenstate as well. Forming 
different linear combinations thus represents symmetry transformations, and with this 
{\em additional} symmetry one can further reduce the number of physical parameters. 
For $N=3$ it means \cp~violation could still not occur. 

The CKM implementation of \cp~violation depends on the form of the quark mass matrices 
${\cal M}_{U,D}$, not so much on how those are generated. Nevertheless something can be inferred 
about the latter: within the SM all fermion masses are driven by a {\em single} VEV; to obtain an 
irreducible relative phase between different quark couplings thus requires such a phase in 
quark Yukawa couplings; this means that in the SM \cp~violation arises in dimension-{\em four} 
couplings, i.e., is `hard'. 

\subsubsection{`Maximal' \cp~Violation?}
\label{MAXCPV}

As already mentioned charged current couplings with their $V-A$ structure break parity and 
charge conjugation {\em maximally}. Since due to \cpt~invariance \cp~violation is expressed 
through couplings with complex phases, one might say that maximal \cp~violation 
is characterized by complex phases of $90^o$. However this would be fallacious: for by changing 
the phase {\em convention} for the quark fields one can change the phase of a given 
CKM matrix element and even rotate it away; it will of course re-appear in other matrix elements. 
For example $|s\rangle \to e^{i\delta _s}|s\rangle$ leads to $V_{qs} \to  e^{i\delta _s}V_{qs}$ 
with $q=u,c,t$. In that sense the CKM phase is like the `Scarlet Pimpernel': "Sometimes here, 
sometimes there, sometimes everywhere." 

One can actually illustrate with a general argument, why there can be no straightforward definition for maximal \cp~violation. Consider neutrinos: Maximal \cp~violation means there are 
$\nu _L$ and $\bar \nu _R$, yet no $\nu _R$ or $\bar \nu _L$  
\footnote{To be more precise: $\nu_L$ and $\bar \nu _R$ couple to weak gauge bosons, 
$\nu _R$ or $\bar \nu _L$ do not.}. Likewise there are $\nu_L$ and $\bar \nu_R$, but 
not $\bar \nu _L$ or $\nu _R$. One might then suggest that maximal \cp~violation means 
that $\nu_L$ exists, but $\bar \nu _R$ does not. Alas -- \cpt~invariance already enforces the existence 
of both. 

Similarly -- and maybe more obviously -- it is not clear what maximal \ot~violation would mean although 
some formulations have entered daily language like the `no future generation', the `woman 
without a past' or the `man without a future'. 

\subsubsection{Some Historical Remarks}
\label{HISTREM}

\cp~violation was discovered in 1964 through the observation of $K_L \to \pi^+\pi^-$, yet it was not realized for a number of years that dynamics known at {\em that} time could {\em not} generate it. 
We should not be too harsh on our predecessors for that oversight: as long as one did not have a 
renormalizable theory for the weak interactions and thus had to worry about {\em infinities} 
in the calculated rates, one can be excused for ignoring a seemingly marginal rate with a branching 
ratio of $2\cdot 10^{-3}$. Yet even after the emergence of the renormalizable 
Glashow-Salam-Weinberg model its {\em phenomenological} incompleteness was not recognized right 
away. There is a short remark by Mohapatra in a 1972 paper invoking the need for right handed 
currents to induce \cp~violation. 

It was the 1973 paper by Kobayashi and Maskawa \cite{KM} that fully stated the inability of even a 
two-family SM to produce \cp~violation and that explained what had to be added to it: right-handed 
charged currents, extra Higgs doublets -- or (at least) a third quark family. Of the three options 
Kobayashi and Maskawa listed, their name has been attached only to the last one as the CKM 
description. They were helped by the `genius loci' of Nagoya University: 
\begin{itemize}
\item 
Since it was the home of the Sakata school and the Sakata model of elementary particles quarks 
were viewed as physical degrees of freedom from the start. 
\item 
It was also the home of Prof. Niu who in 1971 had observed 
\cite{NIU} a candidate for a charm decay in emulsion 
exposed to cosmic rays and actually recognized it as such. The existence of charm, its association 
with strangeness and thus of two complete quark families were thus taken for granted at Nagoya. 

\end{itemize}

\subsection{Meson-antimeson Oscillations -- on the Power of Quantum Mysteries}
\label{MYST}

After the conceptual exposition of the SM I return to the historical development. 
With respect to meson-antimeson oscillations nature has treated us like a patient teacher does with  
somewhat dense students: she has provided us not with one, but 
with three meson systems that exhibit oscillations, 
namely the $K^0 - \bar K^0$, $B_d - \bar B_d$ and $B_s - \bar B_s$ complexes; as we will discuss in some detail, those three systems present complementary perspectives on oscillations. 
I would like to add that these phenomena by and large followed theoretical predictions -- yet the 
most revolutionary feature, namely the first manifestation of \cp~violation in particle decays, was 
outside the `theoretical' horizon at that time. 

As already mentioned, `strange' hadrons obtained their name from the observation that their production rate exceeds their decay rate by many orders of magnitude. This feature was explained by assigning them an internal quantum number strangeness $S=\pm 1$ and postulating that only the weak interactions can produce $\Delta S \neq 0$ transitions. One has two different neutral kaons: 
$K^0$ and $\bar K^0$ with $S=1$ and $S=-1$, respectively. Then the question arises: How does 
one verify it experimentally? 

\begin{figure}[t]
\vspace{7.0cm}
\includegraphics{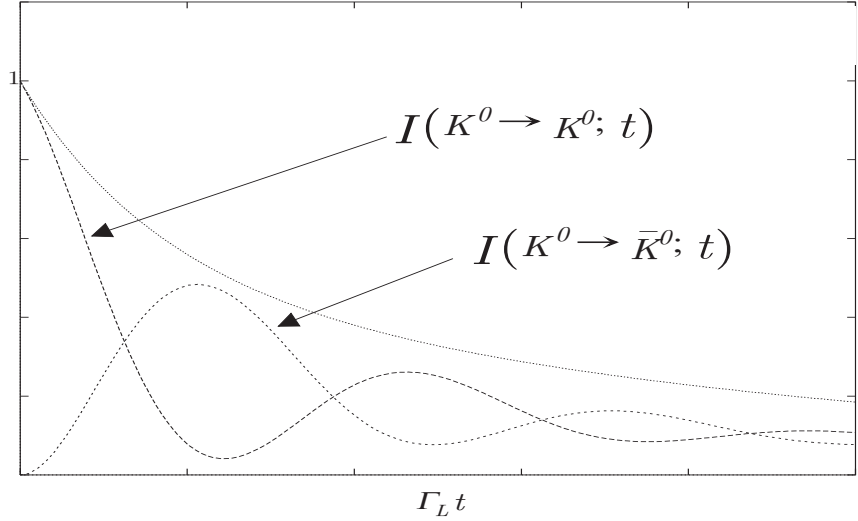}
 \caption{
      The probabilities of finding a $K^0$ and a $\bar K^0$ in an {\em initial} $K^0$ 
      beam as a function of time.  \label{Oscillfig} }
\end{figure}

The answer to this challenge came in the form of oscillations and represents one of the glory pages of particle physics. Symmetry considerations allows to derive many essential features of 
oscillations without solving any equations explicitly. 
With{\em out} weak interactions $K^0$ and $\bar K^0$ 
have, due to \cpt~invariance, equal masses and lifetimes (the latter being infinite at this point). 
With the weak $\Delta S\neq 0$ forces  `switched on' the two neutral kaon mass eigenstates 
will be linear combinations of $K^0$ and $\bar K^0$ and thus carry no definite strangeness. 
\cp~invariance implies the mass eigenstates to be \cp~eigenstates as well. With the 
{\em definition} 
\beq 
\cp~|K^0\rangle = |\bar K^0\rangle 
\label{CPCONV}
\eeq
one has for the \cp~even and odd states 
\beq 
|K_{\pm}\rangle = \frac{1}{\sqrt{2}} [|K^0 \rangle \pm |\bar K^0\rangle 
\; \; {\rm with} \; \; \Delta M_K \equiv M(K_-) - M(K_+) \neq 0 \neq 
\Delta \Gamma_K = \Gamma (K_+) - \Gamma (K_-) 
\label{KPLUSMIN}
\eeq
\cp~symmetry also constrains the decay modes  
\beq 
|K_+\rangle \to 2 \pi \; , \; \; 2\pi \not \leftarrow |K_-\rangle \to 3 \pi \; , 
\eeq
since $\pi^+\pi^-$ and $2\pi^0$ are \cp~even, whereas $\pi^+\pi^-\pi^0$ can be \cp~odd 
and $3\pi^0$ has to be. (With $M_K< 4 m_{\pi}$ $K \to 4\pi$ cannot occur.)   
Such difference leads to $\tau (K_+) \neq \tau (K_-)$. A kinematical `accident' 
intervenes at this point: Since the kaon mass is barely above the three pion threshold and thus 
$K_- \to 3 \pi$ greatly suppressed by phase space its lifetime is much longer than for 
$K_+$. Their lifetime ratio is actually as large as 570; 
accordingly one refers to them as $K_L$ and $K_S$ with the subscripts $L$ and $S$ referring to 
`long'- and `short'-lived. Thus one predicts the following nontrivial scenario: if one starts 
with a pure beam of, say, $K^0$, one finds different components in the decay rate evolution 
depending on the nature of the final state: 
\begin{itemize}
\item 
In $ K^{\rm neut} \to $ pions two distinct components will emerge, namely $K^{\rm neut} \to 2 \pi$ and 
$K^{\rm neut} \to 3 \pi$ following two separate exponential functions in (proper) time
controlled by the lifetimes $\tau (K_S)$ and $\tau (K_L)$, respectively.  
\item 
Tracking the flavour-{\em specific} (semi)leptonic modes instead, one encounters a considerably more 
complex situation {\em not} described by simple exponential functions in time. The mathematics 
involved is rather straightforward though. Using Eq.(\ref{KPLUSMIN}) and the fact that 
$K_{\pm}$ are mass eigenstates (in the limit of \cp~invariance) we obtain for the time evolution of 
the amplitude of 
an 
initially pure $K^0$ beam
\bea 
|K^0(t)\rangle &=& \frac{1}{\sqrt{2}} [|K_+(t)\rangle + |K_-(t)\rangle] \\
\nonumber
&=& 
 \frac{1}{\sqrt{2}}e^{(iM(K_+) - \frac{1}{2}\Gamma_+)t} [|K_+\rangle + 
 e^{(i\Delta M_K + \frac{1}{2}\Delta \Gamma_K)t}|K_-\rangle] \\
 \nonumber
 &=& \frac{1}{2}e^{(iM(K_+) - \frac{1}{2}\Gamma_+)t}
 \left[ \left(1+ e^{(i\Delta M_K + \frac{1}{2}\Delta \Gamma_K)t}\right) |K^0 \rangle
 +   \left(1- e^{(i\Delta M_K + \frac{1}{2}\Delta \Gamma_K)t}\right) |\bar K^0 \rangle
 \right]  
\eea
The probability for the initial $K^0$ to decay as a $K^0$ or a $\bar K^0$ is then given by
\bea 
{\rm Prob}(K^0 \to K^0; t) &=& \frac{1}{4} e^{-\Gamma_+t}\left( 1 + e^{\Delta \Gamma_Kt} 
+ 2e^{\frac{1}{2}\Delta \Gamma_Kt} {\rm cos}\Delta M_Kt  
\right)\\
{\rm Prob}(K^0 \to \bar K^0; t) &=& \frac{1}{4} e^{-\Gamma_+t}\left( 1 + e^{\Delta \Gamma_Kt} 
- 2e^{\frac{1}{2}\Delta \Gamma_Kt} {\rm cos}\Delta M_Kt  
\right)
\label{REGENEXP}
\eea
The phenomenon that a state that is initially absent in a beam traveling through vacuum re-emerges, 
Eq.(\ref{REGENEXP}),  
is often called `spontaneous regeneration'. 

These expressions are shown in Fig.\ref{Oscillfig}: the decay rate for the `right-sign' leptons $K^0 \to l^+\nu \pi^+$ at first drops off 
faster than follows from $e^{-\Gamma_St}$, an exponential dependence on the time of decay , then bounces back up etc., i.e. `oscillates' -- hence the name. The rate for the 
`wrong-sign' transitions $K^0 \to l^-\nu \pi^+$, which has to start out at zero for $t=0$ rises quickly, yet turns around dropping down, before bouncing back up again etc. It provides the complement for $K^0 \to l^+\nu \pi^-$, i.e. the rate for the sum of both modes should exhibit a simple exponential behaviour.   
\end{itemize}   

\noindent 
These predictions given by Gell-Mann and Pais first assuming \oc~conservation and relaxing it later to 
\cp~symmetry were verified experimentally with impressive numerical sensitivity \cite{PDG06}: 
\beq 
\Delta M_K \equiv M_{K_L} - M_{K_S} = (3.483 \pm 0.006) \cdot 10^{-12} \; {\rm MeV}
\label{DELTMKEXP}
\eeq
(English speakers can rely on a simple mnemonic to remember which state is heavier: 
`L' stands for {\em larger} mass and {\em longer} lifetime, whereas `S' denotes 
{\em smaller} and {\em shorter}.)  
This number is a striking demonstration for the sensitivity reached when quantum mechanical interference can be tracked over macroscopic distances, i.e. flight paths of meters or even hundreds of meters.
Using the kaon mass as yardstick one can re-express Eq.(\ref{DELTMKEXP}) 
\beq 
\frac{\Delta M_K}{M_K} = \frac{M_{K_L} - M_{K_S}}{M_K} = 7.7 \cdot 10^{-15} \; , 
\label{DMKRATIO}
\eeq
which is obviously a most striking number. 
A hard-nosed reader can point out that Eq.(\ref{DMKRATIO}) vastly overstates the point since the kaon 
mass generated largely by the strong interactions has no intrinsic connection with 
$\Delta M_K$ generated by the weak interactions and that calibrating 
$\Delta M_K$ by, say, the mass of an elephant is not truly more absurd.

The more relevant yardstick for the oscillation rates is indeed provided by the weak decay rate 
\cite{PDG06}
\bea 
x_K &=& \frac{\rm oscillation\; rate}{\rm decay\; rate} = \frac{\Delta M_K}{\bar \Gamma _K} 
\simeq 0.945 \pm 0.003 \\
y_K &=& \frac{\Delta \Gamma _K}{2\bar \Gamma _K}\simeq 0.996 \; \; \; \; {\rm with} \; \; 
\bar \Gamma_K = \frac{1}{2}(\Gamma _{K_S} + \Gamma _{K_L})
\eea 

\subsubsection{The Shock of 1964 -- \cp~Violation Surfaces}
\label{SHOCK}

1964 was an excellent year for high energy physics: (i) The Higgs mechanism for the `spontaneous 
realization' of a symmetry was first developed. (ii) The quark model (and the first elements of current 
algebra) were first suggested. (iii) The charm quark was first introduced to implement quark-lepton 
symmetry. (iv) The nonrelativistic $SU(6)$ symmetry combining 
$SU(3)_{Fl}$ with the spin $SU(2)$ was proposed for hadron spectroscopy. (v) The first 
$e^+e^-$ storage ring was inaugurated in Frascati. (vi) The $\Omega^-$ baryon was found at 
Brookhaven National Lab, which was viewed as essential validation for the `Eightful Way' of 
$SU(3)_{Fl}$ symmetry. The modern perspective on it has changed: Being composed of three strange quarks it exhibits rather directly the need for colour as a new internal degree of freedom -- together 
with other observables like $R$ and $\Gamma (\pi^0 \to 2\gamma)$ as already mentioned in 
Sect.\ref{DERIVATION}. 
(vii) \cp~violation was discovered at the same lab through the observation that $K_L$ mesons 
can decay both into three and two pion final states, albeit the latter with the tiny branching ratio of 
$0.23\%$ only. 

The theoretical concepts listed under items (i) - (iii) and the experimental tool of item (v) turn out to be 
crucial for the subject matter of these lectures. 

It is a fact of life that if one wants to see what moves physicists, one should {\em not} focus on what they 
say (rarely a good indicator for scientists in general), but on what they do. Point in case: 
How much this discovery shook the HEP community is best gauged by noting the efforts made to 
reconcile the observation of $K_L \to \pi^+\pi^-$ with \cp~invariance: 
\begin{itemize}
\item 
To infer that $K_L \to \pi \pi$ implies \cp~violation one has to invoke the superposition principle of 
quantum mechanics. One can introduce \cite{ROOS} {\em non}linear terms into the Schr\" odinger 
equation in such a way as to allow $K_L \to \pi^+\pi^-$ with \cp~invariant dynamics. While completely ad hoc, it is possible in principle. Such efforts were ruled out by further data, most decisively by 
$\Gamma (K^0(t) \to \pi^+\pi^-) \neq \Gamma (\bar K^0(t) \to \pi^+\pi^-)$. 
\item 
One can try to emulate the success of Pauli's neutrino hypothesis. An apparent violation of 
energy-momentum conservation had been observed in $\beta$ decay $n \to p e^-$, since 
the electron exhibited a {\em continuous} momentum spectrum. Pauli postulated that the reaction actually was 
\beq 
n \to p e^- \bar \nu 
\eeq  
with $\bar \nu$ a neutral and light particle that had escaped direct observation, yet led to a continuous 
spectrum for the electron. I.e., Pauli postulated a new particle -- and a most whimsical one at that -- to save a symmetry, namely the one under  translations in space and time responsible for the 
conservation of energy and momentum. Likewise it was suggested that the real reaction was 
\beq 
K_L \to \pi^+\pi^- U 
\label{UPART}
\eeq 
with $U$ a neutral and light particle with {\em odd} intrinsic \cp~parity. 
I.e., a hitherto unseen particle was introduced to save a symmetry.  
This attempt at evasion was also soon rejected 
experimentally (see Homework \# 1). This represents an example of the ancient Roman saying: 

\begin{center} 
"Quod licet Jovi, non licet bovi." \\
"What is allowed Jupiter, is not allowed a bull."
\end{center} 
I.e., we mere mortals cannot get away with speculations like `Jupiter' Pauli. 

\end{itemize} 
\begin{center}
{\bf Homework \# 1} 
\end{center} 
\noindent 
What was the conclusive argument to rule out the reaction of Eq.(\ref{UPART}) taking place even for 
a very tiny  $U$ mass? 
\begin{center}
{\bf End of Homework \# 1} 
\end{center} 

Notwithstanding these attempts at evasion, the finding of the Fitch-Cronin experiment -- 
namely that $K_L \to \pi^+\pi^-$ does occur -- were soon widely accepted, since, in the words 
of Pram Pais the `perpetrators' were considered `real pros'.  Yet they induced a feeling of a certain frustration. Parity emerged as violated `maximally' in the charged weak currents that involve 
{\em left}-handed, but {\em no right}-handed neutrinos; thus it followed Luther's dictum "Peccate Fortiter!",  i.e. "Sin boldly!". In contrast \cp~violation, while having an even more profound impact 
on nature's basic design as indicated above, appeared as a `near-miss' as suggested by 
the rarity of the observed transition: BR$(K_L \to \pi^+\pi^-) \simeq 0.0023$. Actually we do not know 
how to give a an unambiguous definition of `maximal' \cp~violation, as  explained in 
Sect. \ref{MAXCPV}.  

From the discovery in 1964 till the 1973 Kobayashi-Maskawa paper there was no theory of 
\cp~violation. Worse still, it was not even recognized -- apart from a short remark in a paper by 
Mohapatra -- that the dynamics known at that time were 
insufficient to implement \cp~violation. It should be noted that Wolfenstein's `Superweak Model', which 
will be sketched below, is {\em not} a theory, not even a model -- it is a classification scheme, not more and not less. 

Yet despite the lack of a true theoretical underpinning, the relevant phenomenology was quickly developed.   

\subsubsection{Phenomenology of \cp~Violation, Part I}
\label{PHENOMPI}

The discussion here will be given in terms of strangeness $S$, yet can be generalized to any 
other flavour quantum number $F$ like beauty, charm, etc. 

Weak dynamics can drive $\Delta S = 1 \& 2$ transitions, i.e. decays and oscillations. 
While the underlying theory has to account  for both, it is useful to differentiate between them on the 
phenomenological level. The interplay between $\Delta S = 1\& 2$ affects also \cp~violation and how it 
can manifest itself. Consider $K_L \to \pi \pi $: while $\Delta S =2$ dynamics transform the the flavour 
eigenstates $K^0$ and $\bar K^0$ into mass eigenstates $K_L$ and $K_S$, $\Delta S=1$ forces produce the decays into pions. 
\beq 
[K^0 \stackrel{\Delta S =2}{\longleftrightarrow} \bar K^0] \Rightarrow 
K_L \stackrel{\Delta S=1}{\longrightarrow} \pi \pi 
\eeq
Both of these reactions can exhibit \cp~violation, which is usually 
expressed as follows: 
\bea 
\nonumber 
\eta_{+-[00]} &\equiv& \frac{T(K_L \to \pi^+\pi^- [\pi^0\pi^0])}{T(K_S \to \pi^+\pi^- [\pi^0\pi^0])} \\
\eta_{+-} &\equiv& \epsilon_K + \epsilon^{\prime}\; , \; \; \; 
\eta_{00} \equiv \epsilon_K - 2 \epsilon^{\prime} 
\eea
Both $\eta_{+-}, \eta_{00} \neq 0$ signal \cp~violation; $\epsilon_K$ is common to both 
observables and reflects the \cp~properties of the state mixing, i.e. in $\Delta S =2$ dynamics; 
$\epsilon^{\prime}$ on the other hand differentiates between the two final states and parametrizes 
\cp~violation in $\Delta S =1$ dynamics. With an obvious lack in Shakespearean flourish 
$\epsilon_K \neq 0$ is referred to as `indirect' or `superweak' \cp~violation and 
$\epsilon^{\prime}\neq 0$ as `direct' \cp~violation. As long as \cp~violation is seen only through a 
single mode of a neutral meson -- in this case {\em either} $K_L \to \pi^+\pi^-$ {\em or} 
$K_L \to \pi^0\pi^0$ -- the  distinction between direct and indirect \cp~violation is somewhat 
arbitrary, as explained later for $B_d$ decays. 

Five types of \cp~violating observables have emerged through $K^0 - \bar K^0$ oscillations: 
\begin{enumerate}
\item 
{\em Existence} of a transition: $K_L \to \pi^+\pi^-,\, \pi^0\pi^0$; 
\item 
An {\em asymmetry} due to the {\em initial} state: $K^0 \to \pi^+\pi^-$ vs. $\bar K^0 \to \pi^+\pi^-$; 
\item 
An {\em asymmetry} due to the {\em final} state: $K_L \to l^+\nu \pi^-$ vs. $K_L \to l^- \bar \nu \pi^+$, 
$K_L \to \pi^+\pi^-$ vs. $K_L \to \pi^0 \pi^0$; 
\item 
A {\em microscopic} \ot~asymmetry: rate$(K^0 \to \bar K^0)$ $\neq$ rate$(\bar K^0 \to K^0)$; 
\item 
A \ot~{\em odd correlation} in the final state: $K_L \to \pi^+\pi^- e^+e^-$. 

\end{enumerate}
We know now that all these observables except $|\eta_{+-}| \neq |\eta_{00}|$ are predominantly (or even exclusively) given by $\epsilon_K$, i.e. indirect \cp~violation. The asymmetry in semileptonic $K_L$ 
decays has been measured to be 
\beq 
\delta _l\equiv 
\frac{\Gamma (K_L \ra l^+ \nu \pi ^-) - 
\Gamma (K_L \ra l^- \bar \nu \pi ^+)}
{\Gamma (K_L \ra l^+ \nu \pi ^-) + 
\Gamma (K_L \ra l^- \bar \nu \pi ^+)}  = 
(3.27 \pm 0.12)\cdot 10^{-3} \; , 
\label{SLDIFFDATA} 
\eeq
averaged over electrons and muons. This measurement provides a {\em convention independent} 
definition of "+" vs. "-", hence of "matter" -- $l^-$ -- vs. "antimatter" -- $l^+$ -- and of 
"left" -- $l^-_L$ -- vs. "right" -- $l^+_R$. 
\footnote{This definition can be communicated to a far-away civilization using 
{\em un}polarized radio signals. Such a communication is of profound academic as well as 
practical value: when meeting such a civilization in outer space, one better finds out, whether they are made of matter or antimatter; otherwise the first handshake might be the last one as well.} 

To describe oscillations in the presence of \cp~violation one turns to solving a nonrelativistic 
Schr\" odinger equation, which I formulate for the general case of a pair of neutral 
mesons $P^0$ and $\bar P^0$ with flavour quantum number $F$; it can denote 
a $K^0$, $D^0$ or $B^0$ \cite{WIGNERAPPROX}:  
\beq 
i\frac{d}{dt} \left( 
\begin{array}{ll}
P^0 \\
\bar P^0
\end{array}  
\right)  = \left( 
\begin{array}{ll}
M_{11} - \frac{i}{2} \Gamma _{11} & 
M_{12} - \frac{i}{2} \Gamma _{12} \\ 
M^*_{12} - \frac{i}{2} \Gamma ^*_{12} & 
M_{22} - \frac{i}{2} \Gamma _{22} 
\end{array}
\right) 
\left( 
\begin{array}{ll}
P^0 \\
\bar P^0
\end{array}  
\right) 
\label{SCHROED} 
\eeq
\cpt~ invariance imposes 
\beq 
M_{11}= M_{22} \; \; , \; \; \Gamma _{11} = \Gamma _{22} \; . 
\label{CPTMASS}
\eeq  
\begin{center} 
{\bf Homework \# 2}: 
\end{center}
Which physical situation is 
described by an equation analogous to Eq.(\ref{SCHROED}) 
where however the two diagonal matrix elements differ 
{\em without} violating \cpt? 
\begin{center} 
{\bf End of Homework \# 2}
\end{center} 
The subsequent discussion might strike the reader as overly 
technical, yet I hope she or he will bear with me since 
these remarks will lay important groundwork for a proper 
understanding of \cp~asymmetries in $B$ decays as well. 

The mass eigenstates obtained through diagonalising this matrix 
are given by (for details see \cite{LEE,BOOK}) 
\bea 
 |P_A\rangle &=& 
\frac{1}{\sqrt{|p|^2 + |q|^2}} \left( p |P^0 \rangle + 
q |\bar P^0\rangle \right) \\  
|P_B\rangle &=& 
\frac{1}{\sqrt{|p|^2 + |q|^2}} \left( p |P^0 \rangle -  
q |\bar P^0\rangle \right) 
\label{P1P2}
\eea 
with eigenvalues 
\bea 
M_A - \frac{i}{2}\Gamma _A &=& M_{11} - \frac{i}{2} \Gamma _{11} 
+\frac{q}{p}\left( M_{12} - \frac{i}{2} \Gamma _{12}\right) \\    
M_B - \frac{i}{2}\Gamma _B &=& M_{11} - \frac{i}{2} \Gamma _{11} 
- \frac{q}{p}\left( M_{12} - \frac{i}{2} \Gamma _{12}\right) 
\label{EV} 
\eea 
as long as 
\beq 
\left( \frac{q}{p}\right) ^2 = 
\frac{M_{12}^* - 
\frac{i}{2} \Gamma ^*_{12}}
{M_{12} - 
\frac{i}{2} \Gamma _{12}} 
\label{Q/PSQ} 
\eeq 
holds. I am using letter subscripts 
$A$ and $B$ for labeling the 
mass eigenstates rather than numbers $1$ and $2$ 
as it is usually done. For I want to 
avoid confusing them with the matrix indices 
$1,2$ in $M_{ij} - \frac{i}{2}\Gamma _{ij}$. 
 
Eqs.(\ref{EV}) yield for the differences in mass and width 
\bea 
\Delta M &\equiv& M_B - M_A = 
-2 {\rm Re} \left[ \frac{q}{p}(M_{12} - 
\frac{i}{2}\Gamma _{12})\right]  \\
\Delta \Gamma  &\equiv& \Gamma _A - 
\Gamma _B = 
-2 {\rm Im}\left[ \frac{q}{p}(M_{12} - 
\frac{i}{2}\Gamma _{12})\right] 
\label{DELTAEV} 
\eea
Note that the subscripts $A$, $B$ have been swapped in 
going from $\Delta M$ to $\Delta \Gamma$! This is 
done to have both quantities {\em positive} 
for kaons. 

In expressing the mass eigenstates $P_A$ and $P_B$ 
explicitely in terms of the flavour eigenstates -- 
Eqs.(\ref{P1P2}) -- one needs $\frac{q}{p}$. There 
are two solutions to Eq.(\ref{Q/PSQ}):  
\beq 
\frac{q}{p} = \pm 
\sqrt{\frac{M_{12}^* - \frac{i}{2} \Gamma _{12}^*}
{M_{12} - \frac{i}{2} \Gamma _{12}}}
\label{Q/P} 
\eeq
There is actually a more general ambiguity than this 
binary one. For antiparticles
are defined up to a phase only: 
\beq 
{\bf CP} |P^0 \rangle = 
\eta | \bar P^0 \rangle  \; \; \; {\rm with} 
\; \; |\eta| =1
\eeq 
Adopting a different phase convention will change 
the phase for $M_{12} - \frac{i}{2} \Gamma _{12}$ 
as well as 
for $q/p$: 
\beq 
|\bar P^0 \rangle \ra e^{i\xi}|\bar P^0 \rangle \; 
\Longrightarrow \; 
(M_{12}, \Gamma _{12}) \ra e^{i\xi} 
(M_{12}, \Gamma _{12}) \; 
\& \; 
\frac{q}{p} \ra e^{-i\xi} \frac{q}{p} \; , 
\eeq
yet leave $(q/p) (M_{12} - \frac{i}{2} \Gamma _{12})$ 
invariant -- as it has to be since the eigenvalues, 
which are observables, depend on this combination, see 
Eq.(\ref{EV}). Also $\left| \frac{q}{p}\right|$ is an 
observable; its {\em deviation} from unity is one 
measure of \cp~violation in $\Delta F =2$ dynamics.

By {\em convention} most authors pick the 
{\em positive} sign in Eq.(\ref{Q/P}) 
\beq 
\frac{q}{p} = +  
\sqrt{\frac{M_{12}^* - \frac{i}{2} \Gamma _{12}^*}
{M_{12} - \frac{i}{2} \Gamma _{12}}} \; . 
\label{QPPOS} 
\eeq 
Up to this point the two states 
$|P_{A,B}\rangle$ are merely 
{\em labelled} by their subscripts. 
Indeed $|P_A\rangle$ and $|P_B\rangle$ switch places 
when selecting the minus rather than the plus sign in 
Eq.(\ref{Q/P}). 

One can define the labels $A$ and $B$ such that 
\beq 
\Delta M \equiv M_B - M_A > 0
\label{DMPOS}
\eeq 
is satisfied. Once this {\em convention} 
has been adopted, it becomes a sensible question 
whether 
\beq 
\Gamma _B > \Gamma _A  \; \; \; {\rm or} \; \; \; 
\Gamma _B < \Gamma _A 
\eeq 
holds, i.e. whether the heavier state is shorter or 
longer lived. 

One can write the general 
mass eigenstates in terms of the 
\cp~eigenstates as well: 
\bea 
|P_A \rangle &=& \frac{1}{\sqrt{1+|\bar \epsilon |^2}} 
\left( |P_+ \rangle + \bar \epsilon |P_-\rangle \rangle 
\right)  
\; \; \; , \; CP|P_{\pm}\rangle = \pm |P_{\pm}\rangle \\  
|P_B \rangle &=& \frac{1}{\sqrt{1+|\bar \epsilon |^2}} 
\left( |P_- \rangle + \bar \epsilon |P_+\rangle \rangle 
\right) 
\; ; 
\eea
$\bar \epsilon = 0$ means that the mass and 
\cp~eigenstates coincide, i.e. \cp~is conserved in 
$\Delta F=2$ dynamics driving $P - \bar P$ 
oscillations. With the phase between the 
orthogonal states $|P_+\rangle$ and 
$|P_-\rangle$ arbitrary, the phase of 
$\bar \epsilon$ can be changed at will and is not an 
observable; $\bar \epsilon$ can be expressed in terms of 
$\frac{q}{p}$, yet in a way that depends on the 
convention for the phase of antiparticles. For 
${\bf CP}|P\rangle = \pm |\bar P\rangle$ one has 
\bea 
|P_+\rangle &=& \frac{1}{\sqrt{2}} 
\left( |P^0 \rangle \pm |\bar P^0 \rangle \right) \\
|P_-\rangle &=& \frac{1}{\sqrt{2}} 
\left( |P^0 \rangle \mp |\bar P^0 \rangle \right) \\
\bar \epsilon &=& 
\frac{1 \mp \frac{q}{p}}{1 \pm  \frac{q}{p}} 
\eea  
\begin{center} 
{\bf Homework \# 3}: 
\end{center}
While $\langle \bar P^0|P^0 \rangle = 0$ holds -- e.g., $\langle \bar K^0|K^0 \rangle = 0$ --, 
one has $\langle K_L|K_S \rangle \neq 0$ and in general $\langle P_B | P_A\rangle \neq 0$. 
Calculate it and interprete your result. 
\begin{center} 
{\bf End of Homework \# 3}
\end{center} 
Later we will discuss how to evaluate $M_{12}$ and thus 
also $\Delta M$ within a given 
theory for the $P-\bar P$ complex. The examples just listed 
illustrate that some care has to be applied in interpreting 
such results. For expressing mass eigenstates explicitely 
in terms of flavour eigenstates involves some conventions. 
Once adopted we have to stick with a convention; yet our 
original choice cannot influence observables.

Let me recapitulate the relevant points: 
\begin{itemize}
\item 
The labels of the two mass eigenstates $P_A$ and $P_B$ can 
be chosen such that 
\beq 
M_{P_B} > M_{P_A} 
\eeq 
holds. 
\item 
Then it becomes an {\em empirical} question whether 
$P_A$ or $P_B$ are longer lived: 
\beq 
\Gamma _{P_A} > \Gamma _{P_B}  \; \; \; {\rm or} 
\; \; \; 
\Gamma _{P_A} < \Gamma _{P_B} \; \; ? 
\eeq 
\item 
In the limit of \cp~invariance one can also raise the 
question whether it is the \cp~even or the odd state 
that is heavier. 
\item 
We will see later that within a {\em given theory} 
for $\Delta F =2$ dynamics  
one can calculate 
$M_{12}$, including its sign, if phase conventions 
are treated consistently. To be more 
specific: adopting a phase convention for 
$\frac{q}{p}$ and having ${\cal L}(\Delta F =2)$ one 
can calculate $\frac{q}{p} \left( M_{12} - 
\frac{i}{2}\Gamma _{12}\right) = 
\frac{q}{p}\matel{P^0}{{\cal L}(\Delta F =2)}{\bar P^0}$. 
Then one assigns the labels $B$ and $A$ such that 
$\Delta M = M_B - M_A = 
-2{\rm Re}\frac{q}{p} \left( M_{12} - 
\frac{i}{2}\Gamma _{12}\right)$ turns out to be 
{\em positive}! 
\end{itemize}

\subsection{CKM -- From a General Ansatz to a Specific Theory}
\label{ANSAT}

Electroweak forces can be dealt with perturbatively. Consider the $\Delta S =1$ 
four-fermion  transition operator: $(\bar u_L\gamma ^{\mu}s_L)(\bar d_L \gamma _{\mu}u_L)$. 
It constitutes a dimension-{\em six} operator. Yet placing such an operator -- or any other operator 
with dimension larger than four -- into the Lagrangian creates {\em non}renormalizable 
interactions. What happened is that we have started out from a renormalizable Lagrangian 
\beq 
{\cal L}_{CC} = g_W \bar q^{(i)}_L\gamma_{\mu}q_L^{(j)}W^{\mu} \; , 
\label{LCCEX}
\eeq
iterated it to second order in $g_W$ with $(q^{(i)},q^{(j)}) = (u,s) \& (u,d)$ 
and then `integrated out' the heavy field, namely in this case the vector boson field $W^{\mu }$. 
That way one arrives at an effective Lagrangian containing only light quarks as `active' fields. 

Such effective field theories have experienced a veritable renaissance in the last ten years. 
Constructing them in a self-consistent way is greatly helped by adopting a Wilsonian prescription: 
\begin{itemize}
\item 
First one defines a field theory ${\cal L}(\Lambda _{UV})$ at a high ultraviolet scale 
$\Lambda _{UV} \gg$ germane scales of theory like $M_W$, $m_Q$ etc. 
\item 
For applications characterized by physical scales $\Lambda_{phys}$ one renormalizes the theory from 
the cutoff $\Lambda _{UV}$ down to $\Lambda_{phys}$. In doing so one integrates out the 
{\em heavy} degrees of freedom, i.e. with masses exceeding $\Lambda_{phys}$ -- like 
$M_W$ -- to arrive at an 
{\em effective low energy} field theory using the operator product expansion (OPE) as a tool: 
\beq 
{\cal L}(\Lambda _{UV})  \Rightarrow {\cal L}(\Lambda_{phys} ) = 
\sum _i c_i(\Lambda_{phys} , \Lambda _{UV}, M_W, ...) 
{\cal O}_i (\Lambda_{phys} )
\eeq
\begin{itemize}
\item 
The {\em local} operators ${\cal O}_i (\Lambda_{phys} )$ contain the {\em active} dynamical 
fields, i.e. those with frequencies below ${\cal O}_i (\Lambda_{phys} )$.

\item 
Their c number coefficients $c_i(\Lambda_{phys} , \Lambda _{UV}, M_W, ...)$ provide the gateway 
for heavy degrees of freedom with frequencies exceeding ${\cal O}_i (\Lambda_{phys} )$ to enter. 
They are shaped by short-distance dynamics and therefore usually computed perturbatively.

\end{itemize}
\item 
Lowering the value of ${\cal O}_i (\Lambda_{phys} )$ in general changes the form of the Lagrangian: 
${\cal L}(\Lambda ^{(1)}_{phys}) \neq {\cal L}(\Lambda ^{(2)}_{phys}) $ for 
$\Lambda ^{(1)}_{phys} \neq \Lambda ^{(2)}_{phys}$. In particular integrating out heavy degrees of 
freedom will induce higher-dimensional operators to emerge in the Lagrangian.  In the example 
above integrating the $W$ field from the dimension-four term in 
Eq.(\ref{LCCEX}) produces dimension six four-quark operators. 
\item 
As a matter of principle observables cannot depend on the choice of $\Lambda_{phys}$; the latter
primarily provides just a demarkation line: 
\beq 
{\rm short \; distances} < 1/\Lambda_{phys} < {\rm long \; distances} 
\eeq
\end{itemize}
In practice, however, its value must be chosen judiciously due to limitations of our 
(present) computational abilities: on one hand we want to be able to calculate radiative 
corrections perturbatively and thus require $\alpha_S (\Lambda_{phys}) < 1$. Taken by itself it would 
suggest to choose $ \Lambda_{phys}$ as large as possible. Yet on the other hand 
we have to evaluate hadronic matrix elements; there $\Lambda_{phys}$ can provide an UV cutoff on 
the momenta of the hadronic constituents. Since the tails of hadronic wave functions cannot be 
obtained from, say, quark models in a reliable way, one wants to pick 
$\Lambda_{phys}$ as low as possible. More specifically for heavy flavour hadrons one can 
expand their matrix elements in powers of $\Lambda_{phys}/m_Q$. Thus one encounters 
a Scylla \& Charybdis situation. A reasonable middle course can be steered by picking 
$\Lambda_{phys} \sim 1$ GeV, and hence I will denote this quantity and this value by 
$\mu$.

Some concrete examples might illuminate these remarks. 
\begin{itemize} 
\item 
Iterating the coupling of Eq.(\ref{LCCEX}) leads to an effective current-current coupling 
$(\bar u_L\gamma ^{\mu}s_L)(\bar d_L \gamma _{\mu}u_L)$ at low energies, i.e. scales below 
$M_W$.  QCD radiative corrections have to be included: they affect the strength of these effective 
weak transition operators significantly, since they represent an expansion in $\alpha_S$ 
multiplied by a numerically large logarithm log$(M_W/\mu)$ rather than merely $\alpha_S$;    
they also create different types of such operators. On the tree graph 
level there is one $\Delta S=1$ operator, namely 
$(\bar u_L \gamma _{\mu}s_L)(\bar d_L \gamma ^{\mu}u_L)$.  
Including one-loop diagrams where a gluon is exchanged between quark lines one obtains 
${\cal O}(\alpha_S)$ contributions to the original 
$(\bar u_L \gamma _{\mu}s_L)(\bar d_L \gamma ^{\mu}u_L)$ operator -- and to the new coupling 
$(\bar u_L \gamma _{\mu}t^is_L)(\bar d_L \gamma ^{\mu}t^iu_L)$, where the 
$t^i$ denote the generators of colour $SU(3)$. I.e., the two operators 
$O^{1\times 1}= (\bar u_L \gamma _{\mu}s_L)(\bar d_L \gamma ^{\mu}u_L)$ and 
$O^{8\times 8}= (\bar u_L \gamma _{\mu}t^is_L)(\bar d_L \gamma ^{\mu}t^iu_L)$, 
where the former [latter] represents the product of two colour-singlet[octet] currents, mix under 
QCD renormalization already on the one-loop level: 
\beq 
(\bar u_L \gamma _{\mu}s_L)(\bar d_L \gamma ^{\mu}u_L) \; \; \; 
\stackrel{{\rm QCD\, 1-loop\, renormalization}}\Longrightarrow \; \; \; 
c_{1\times 1}O^{1\times 1} + c_{8\times 8}O^{8\times 8}
\eeq
with $c_{1\times 1} = 1 +{\cal O}(\alpha_S)$, whereas $c_{8\times 8} = {\cal O}(\alpha_S)$. Since 
some of these $\alpha_S$ corrections are actually enhanced by numerically sizeable 
log$(M_W/\mu)$ factors, they are quite significant. Therefore one wants to identify the 
{\em multiplicatively} renormalized transition operators with 
\beq 
\tilde O \; \; \; \; \; 
\stackrel{{\rm QCD\, 1-loop\, renormalization}}\Longrightarrow 
\; \; \; \; \; \tilde c \; \tilde O
\eeq
This can be done even without brute-force computations by relying on isospin arguments: consider 
the weak scattering process between quarks 
\beq 
s_L + u_L \to u_L + d_L
\eeq 
proceeding in an S wave. 
It can be driven by two $\Delta S=1$ operators, namely 
\beq 
O_{\pm} = \frac{1}{2} \left[ (\bar u_L\gamma_{\mu} s_L)(\bar d_L \gamma^{\mu}u_L) \pm 
(\bar d_L\gamma_{\mu} s_L)(\bar u_L \gamma^{\mu}u_L) \right] 
\eeq
The operator $O_+ [O_-]$ produces an $ud$ pair in the final state that is [anti]symmetric in isospin and thus 
carries $I=1[I=0]$; since the initial $su$ pair carries $I=1/2$, $O_+[O_-]$ generates 
$\Delta I = 1/2\&3/2$ [only $\Delta I =1/2]$ transitions. 

With QCD conserving isospin, its radiative corrections cannot mix the operators $O_{\pm}$, 
which therefore are {\em multiplicatively} renormalized:  
\beq 
O_+\; [O_-] \; \; \; \; \; 
\stackrel{{\rm QCD\, 1-loop\, renormalization}}\Longrightarrow 
\; \; \; \; \;  c_+  \, O_+\; [c_- \, O_-]  \; . 
\eeq 
and therefore 
\beq 
{\cal L}_{eff}^{(0)}(\Delta S =1)= O_+ + O_- 
\; \; \;  
\stackrel{{\rm QCD\, 1-loop\, ren.}}\Longrightarrow 
\; \; \;   {\cal L}_{eff}(\Delta S =1)=c_+  \, O_+ +c_- \, O_-
\eeq
with $c_{\pm} = 1 + {\cal O}(\alpha _S)$. 

Integrating out those loops containing a $W$ line in addition to the gluon line and two 
quark lines yields terms $\propto \alpha_S {\rm log}(M_W^2/\mu ^2)$, which are not necessarily 
small. Using the renormalization group equation to sum those terms terms one finds on the leading 
log level 
\beq 
c_{\pm} = \left[\frac{ \alpha_S(M_W^2)}{\alpha_S(\mu ^2)}
\right]^{\gamma_{\pm}} \; , \; \; \gamma _+= \frac{6}{33-2N_F}=-\frac{1}{2}\gamma_- \; . 
\label{LEADLOG1}
\eeq
I.e., 
\footnote{The expressions of Eq.(\ref{LEADLOG1}) hold in the `leading log approximation'; 
including terms $\sim \alpha_S^{n+1}{\rm log}^n(M_W^2/\mu ^2)$ modifies them, yet 
$c_- > 1 > c_+$ and  $c_- c_+^2 \simeq 1$ still hold.} 
\beq 
c_- > 1 > c_+ \; , \; \; c_- c_+^2 = 1
\label{LEADLOG2} 
\eeq
That means that QCD radiative corrections provide a quite sizeable $\Delta I=1/2$ 
enhancement. Corresponding effects arise for ${\cal L}_{eff}(\Delta C/B=1)$. 

QCD radiative corrections create yet another effect, namely they lead to the emergence 
of `Penguin' operators. Without the gluon line their diagram would decompose into two 
{\em disconnected} parts and thus not contribute to a transition operator. These Penguin 
diagrams can drive only $\Delta I = 1/2$ modes. Furthermore in the loop all three quark families 
contribute; the diagram thus contains the irreducible CKM phase -- i.e. it generates 
{\em direct} \cp~violation in strange decays. Similar effects arise in beauty, but not necessarily in 
charm decays. 
\item 
Consider 
$K^0 - \bar K^0$ oscillations, which represent $\Delta S=2$ transitions. As explained in Sect.\ref{MYST}  
those are driven by the off-diagonal elements of a `generalized mass matrix': 
\beq 
{\cal M}_{12} = M_{12} + \frac{i}{2}{ \Gamma}_{12} = 
\matel{K^0}{{\cal L}_{eff}(\Delta S =2)}{\bar K^0}
\eeq
The observables $\Delta M_K$ and $\epsilon_K$ are given in terms of Re$M_{12}$ and 
Im$M_{12}$, respectively. 
In the SM ${\cal L}_{eff} (\Delta S=2)$, which generates $M_{12}$, is produced by iterating 
two $\Delta S=1$ operators: 
\beq 
{\cal L}_{eff} (\Delta S=2) = {\cal L}(\Delta S=1) \otimes 
{\cal L}(\Delta S=1) 
\eeq  
This leads to the well known 
quark box diagrams, which generate a {\em local} $\Delta S=2$ operator. 
The contributions that do 
{\em not} depend on the mass of the internal quarks cancel against 
each other due to the GIM mechanism, which leads to highly convergent diagrams. 
Integrating over the internal 
fields, namely the $W$ bosons and the top and charm quarks 
\footnote{The up quarks act merely as a subtraction term here.} 
then yields a convergent result: 
$$  
{\cal L}_{eff}^{box}(\Delta S=2, \mu ) = 
\left( \frac{G_F}{4\pi }\right) ^2 \cdot 
$$ 
\beq  
\left[  \xi _c^2 E(x_c) \eta _{cc} + 
\xi _t^2 E(x_t) \eta _{tt} + 
2\xi _c \xi _t E(x_c, x_t) \eta _{ct}
 \right]  [\alpha _S(\mu ^2)]^{- \frac{6}{27}} 
\left( \bar d \gamma _{\mu}(1- \gamma _5) s\right) ^2 
+ h.c. 
\label{LAGDELTAS2}
\eeq  
with $\xi _i$ denoting combinations of KM parameters 
\beq 
\xi _i = V(is)V^*(id) \; , \; \; i=c,t \; ; 
\eeq  
$E(x_i)$ and $E(x_c,x_t)$ reflect the box loops with equal and 
different internal quarks, respectively \cite{INAMI}:  
\beq 
E(x_i) = x_i 
\left(   
\frac{1}{4} + \frac{9}{4(1- x_i)} - \frac{3}{2(1- x_i)^2} 
\right) 
- \frac{3}{2} \left( \frac{x_i}{1-x_i}\right) ^3 
{\rm log} x_i 
\label{TOPBOX}
\eeq  
$$  
E(x_c,x_t) = x_c x_t 
\left[ \left( 
\frac{1}{4} + \frac{3}{2(1- x_t)} - \frac{3}{4(1- x_t)^2} \right) 
\frac{{\rm log} x_t}{x_t - x_c} + (x_c \leftrightarrow x_t) - 
\right. 
$$ 
\beq 
\left. - \frac{3}{4} \frac{1}{(1-x_c)(1- x_t)} \right]  
\eeq 
\beq 
x_i = \frac{m_i^2}{M_W^2}  \; . 
\eeq  
The $\eta _{ij}$ represent the QCD radiative corrections from 
evolving the effective Lagrangian from $M_W$ down to 
the internal quark mass. 
The factor $[\alpha _S(\mu ^2)]^{-6/27}$ 
reflects the fact that a scale 
$\mu$ must be introduced at which the four-quark operator 
$\left( \bar s \gamma _{\mu}(1- \gamma _5) d\right) ^2 $ is 
defined. This dependance on the auxiliary variable 
$\mu$ drops out when one takes the matrix element of this 
operator (at least when one does it correctly).  
Including next-to-leading log 
corrections one finds (for $m_t \simeq 180$ GeV) \cite{BURAS}: 
\beq 
\eta _{cc} \simeq 1.38 \pm 0.20 \; , \; \; 
\eta _{tt} \simeq 0.57 \pm 0.01 \; , \; \; 
\eta _{cc} \simeq 0.47 \pm 0.04 
\eeq                       
\footnote{There is also a {\em non}-local $\Delta S=2$ 
operator generated from the iteration of ${\cal L}(\Delta S=1)$. 
While it  
presumably provides a major contribution to $\Delta m_K$, 
it is not sizeable for $\epsilon _K$ within the KM ansatz,  
as be inferred from the observation that 
$|\epsilon ^{\prime}/\epsilon _K|\ll 0.05$.} 
The dominant contributions  
for $\Delta M(K)$ and $\epsilon_K$ are produced, when (in addition to the $W^{\pm}$ pair) the {\em internal} quarks are charm and top, respectively. In either case the {\em internal} quarks are 
heavier than the {\em external} ones: $m_d, m_s \ll m_c, m_t$, and 
evaluating the Feynman diagrams indeed corresponds to integrating out the heavy fields. 

The situation is qualitatively very similar for $\Delta M(B^0)$, and in some sense even simpler: 
for within the SM by far the leading contribution is due to internal top quarks. Evaluating 
the quark box diagram with internal $W$ and top quark lines corresponds to integrating those 
heavy degrees of freedom out in a straightforward way leading to: 
\beq   
{\cal L}_{eff}^{box}(\Delta B=2, \mu ) \simeq 
\left( \frac{G_F}{4\pi }\right) ^2 M_W^2\cdot     
\xi _t^2 E(x_t) \eta _{tt} 
\left( \bar q \gamma _{\mu}(1- \gamma _5) b\right) ^2 
+ h.c. 
\label{LAGDELTAS2}
\eeq  
with $q = d,s$. 
\begin{center}
{\bf Homework \# 4} 
\end{center} 
\noindent 
When one calculates $\Delta M(B)$ as a function of the top mass employing the quark box 
diagram, one finds, see Eq.(\ref{TOPBOX}) 
\beq 
\Delta M(B) \propto \left(   \frac{m_t}{M_W}    \right) ^2  \; \; {\rm for} \; \; m_t \gg M_W
\label{HW4}
\eeq
The factor on the right hand side reflects the familiar GIM suppression for $m_t \ll M_W$; 
yet for $m_t \gg M_W$ it constitutes a (huge) enhancement! It means that a low 
energy observable, namely $\Delta M(B)$, is controlled more and more by a state or field at 
asymptotically high scales. Does this not violate decoupling theorems and even common sense? 
Does it violate decoupling -- and if so, why is it allowed to do so -- or not? 
\begin{center}
{\bf End of Homework \# 4} 
\end{center} 

While quark box diagrams contribute also to $\Gamma_{12}(\Delta S=2)$, it would be absurd to assume they are significant. For $\Gamma_K$ is dominated by the impact of hadronic phase 
space causing $\Gamma (K_{neut} \to 2 \pi) \gg \Gamma (K_{neut} \to 3 \pi)$. Yet even beyond that it is 
unlikely that such a computation would make much sense: to contribute to $\Delta \Gamma_K$ 
the internal quark lines in the quark box diagram have to be $u$ and $\bar u$ quarks, i.e. 
{\em lighter} than the external quarks $s$ and $\bar s$. That means calculating this 
Feynman diagram  does not correspond to integrating out the heavy degrees of freedom. 
For the same reason (and others as explained later in more detail) computing quark box diagrams 
tells us little of value concerning $D^0 - \bar D^0$ oscillations, since the internal quarks on the leading 
CKM level  -- $s$ and $\bar s$ -- are lighter than the external charm quarks. 

A new and more intriguing twist concerning quark box diagrams occurs when addressing $\Delta \Gamma$ for $B^0$ mesons. Those diagrams again do not generate a {\em local} operator, since the internal charm quarks carry less than half the mass of the external $b$ quarks. 
Nevertheless it can be conjectured that the on-shell $\Delta B=2$ transition operator generating 
$\Delta \Gamma _B$ is largely shaped by short distance dynamics.

\end{itemize}

The main message of these more technical considerations was to show that while QCD conserves 
flavour, it has a highly nontrivial impact on flavour transitions by not only affecting the strength of 
the bare weak operator, but also inducing new types of weak transition operators already on the 
perturbative level. In particular, QCD creates a source of {\em direct} \cp~violation in strange decays  naturally, albeit with a significantly reduced strength.

\subsection{The SM Paradigm of Large \cp~Violation in $B$ Decays}
\label{SMPARAD}

\subsubsection{Basics}
\label{CPBASICS}

As pointed out in Sect.\ref{DEFDISC} for an observable \cp~asymmetry to emerge in a decay one 
needs two different, yet coherent amplitudes to contribute. In 1979 it was pointed out that 
$B^0 - \bar B^0$ oscillations are well suited to satisfy this requirement for final states $f$  
that can be fed both by $B^0$ and $\bar B^0$ decays, in particular since those oscillation 
rates were expected to be sizable \cite{CARTER}: 
\beq  
B^0 \Rightarrow \bar B^0 \to f \leftarrow B^0  \hspace{1cm} vs. \hspace{1cm} 
\bar B^0 \Rightarrow B^0 \to \bar f \leftarrow B^0
\label{BASICIDEA}
\eeq
In 1980 it was predicted \cite{BS80}that in particular $B_d \to \psi K_S$ should exhibit such a 
\cp~asymmetry larger by two orders of magnitude than the corresponding one in  
$K^0 \to 2\pi$ vs. $\bar K^0 \to 2 \pi$, if CKM theory provides the main driver of $K_L \to \pi^+\pi^-$;  
even values close to 100 \% were suggested as conceivable. The analogous mode 
$B_s \to \psi \phi$ should however show an asymmetry not exceeding the few percent level. 

It was also suggested that in rare modes like $\bar B_d \to K^- \pi^+$ sizable {\em direct} \cp~violation 
could emerge due to intervention of `Penguin' operators \cite{SONI}. 

We now know that these predictions were rather prescient. It should be noted that at the time 
of these predictions very little was known about $B$ mesons. 
While their existence had been inferred from the 
discovery of the $\Upsilon (1S - 4S)$ family at FNAL in 1977ff,  none of their exclusive decays had been 
identified, and their lifetime were unknown as were a forteriori their oscillation rates. Yet the 
relevant formalism for \cp~asymmetries involving $B^0 - \bar B^0$ oscillations was already fully given. 

Decay rates for \cp~conjugate channels can be expressed as follows: 
\beq  
\begin{array} {l} 
{\rm rate} (B(t) \ra f) = e^{-\Gamma _Bt}G_f(t) \\  
{\rm rate} (\bar B(t) \ra \bar f) = 
e^{-\Gamma _Bt}\bar G_{\bar f}(t)  
\end{array} 
\label{DECGEN}
\eeq 
where \cpt~invariance has been invoked to assign the same lifetime 
$\Gamma _B^{-1}$ to $B$ and $\bar B$ hadrons. Obviously if 
\beq
\frac{G_f(t)}{\bar G_{\bar f}(t)} \neq 1 
\eeq 
is observed, \cp~violation has been found. Yet one should 
keep in mind that this can manifest itself in two (or three) 
qualitatively different ways: 
\begin{enumerate} 
\item 
\beq 
\frac{G_f(t)}{\bar G_{\bar f}(t)} \neq 1 
\; \; {\rm with} \; \; 
\frac{d}{dt}\frac{G_f(t)}{\bar G_{\bar f}(t)} =0 \; ; 
\label{DIRECTCP1}
\eeq   
i.e., the {\em asymmetry} is the same for all times of decay. This 
is true for {\em direct} \cp~violation; yet, as explained later, it also 
holds for \cp~violation {\em in} the oscillations.  
\item 
\beq 
\frac{G_f(t)}{\bar G_{\bar f}(t)} \neq 1 
\; \; {\rm with} \; \; 
\frac{d}{dt}\frac{G_f(t)}{\bar G_{\bar f}(t)} \neq 0 \; ; 
\label{DIRECTCP2}
\eeq   
here the asymmetry varies as a function of the time of decay. 
This can be referred to as \cp~violation {\em involving} oscillations. 
\end{enumerate} 

A straightforward application of quantum mechanics with its linear superposition principle yields   
\cite{CPBOOK} for $\Delta \Gamma = 0$, which holds for $B^{\pm}$ and 
$\Lambda_b$ exactly and for $B_d$ to a good approximation 
\footnote{Later I will address the scenario with $B_s$, where $\Delta \Gamma$ presumably reaches 
a measurable level.}:
\beq 
\begin{array}{l}
G_f(t) = |T_f|^2 
\left[ 
\left( 1 + \left| \frac{q}{p}\right| ^2|\bar \rho _f|^2 \right) + 
\left( 1 - \left| \frac{q}{p}\right| ^2|\bar \rho _f|^2 \right) 
{\rm cos}\Delta m_Bt 
- 2 ({\rm sin}\Delta m_Bt) {\rm Im}\frac{q}{p} \bar \rho _f 
\right]  \\ 
\bar G_{\bar f}(t) = |\bar T_{\bar f}|^2 
\left[ 
\left( 1 + \left| \frac{p}{q}\right| ^2|\rho _{\bar f}|^2 \right) + 
\left( 1 - \left| \frac{p}{q}\right| ^2|\rho _{\bar f}|^2 \right) 
{\rm cos}\Delta m_Bt 
- 2 ({\rm sin}\Delta m_Bt) {\rm Im}\frac{p}{q} \rho _{\bar f}  
\right]  
\end{array}
\eeq 
The amplitudes for the instantaneous $\Delta B=1$ 
transition into a 
final state $f$ are denoted by 
$T_f = T(B \ra f)$ and $\bar T_f = T(\bar B \ra f)$ and  
\beq 
\bar \rho _f = \frac{\bar T_f}{T_f} \; \; , 
\rho _{\bar f} = \frac{T_{\bar f}}{\bar T_{\bar f}} \; \; , 
\frac{q}{p} = \sqrt{\frac{M_{12}^* - \frac{i}{2} \Gamma _{12}^*}
{M_{12} - \frac{i}{2} \Gamma _{12}}}
\eeq

Staring at the general expression is not always very illuminating; 
let us therefore consider three limiting cases: 
\begin{itemize}
\item
$\Delta m_B = 0$, i.e. {\em no} $B^0- \bar B^0$ oscillations: 
\beq 
G_f(t) = 2|T_f|^2 \; \; , \; \; 
\bar G_{\bar f}(t) = 2|\bar T_{\bar f}|^2 
\leadsto \frac{\bar G_{\bar f}(t)}{G_{ f}(t)} = 
\left|
\frac{\bar T_{\bar f}}{T_{ f}}
\right|^2 \; \; , \frac{d}{dt}G_f (t) \equiv 0 \equiv 
\frac{d}{dt}\bar G_{\bar f} (t) 
\eeq 
This is explicitely what was referred to above as {\em direct} 
\cp~violation. 
\item 
$\Delta m_B \neq  0$   
and $f$ a flavour-{\em specific} final state with {\em no} 
direct \cp~violation; i.e., 
$T_{f} = 0 = \bar T_{\bar f}$ and $\bar T_f = T_{\bar f}$   
\footnote{For a flavour-specific mode one has in general 
$T_f \cdot T_{\bar f} =0$; the more intriguing case arises  
when one considers a transition that requires oscillations 
to take place.}: 
\beq 
\begin{array} {c} 
G_f (t) = \left| \frac{q}{p}\right| ^2 |\bar T_f|^2 
(1 - {\rm cos}\Delta m_Bt )\; \; , \; \; 
\bar G_{\bar f} (t) = \left| \frac{p}{q}\right| ^2 |T_{\bar f}|^2 
(1 - {\rm cos}\Delta m_Bt) \\ 
\leadsto 
\frac{\bar G_{\bar f}(t)}{G_{ f}(t)} = 
\left| \frac{q}{p}\right| ^4 
\; \; , \; \; \frac{d}{dt} \frac{\bar G_{\bar f}(t)}{G_{ f}(t)} 
\equiv 0  
\; \; , \; \; \frac{d}{dt} \bar G_{\bar f}(t) \neq 0 \neq 
\frac{d}{dt} G_ f(t)
\end{array} 
\eeq 
This constitutes \cp~violation {\em in the 
oscillations}. For the \cp~conserving decay into the 
flavour-specific 
final state is used merely to track the flavour identity of the 
decaying meson. This situation can therefore be denoted also 
in the following way: 
\beq 
\frac{{\rm Prob}(B^0 \Rightarrow  \bar B^0; t) - 
{\rm Prob}(\bar B^0 \Rightarrow  B^0; t)}
{{\rm Prob}(B^0 \Rightarrow \bar B^0; t) + 
{\rm Prob}(\bar B^0 \Rightarrow  B^0; t)} = 
\frac{|q/p|^2 - |p/q|^2}{|q/p|^2 + |p/q|^2} = 
\frac{1- |p/q|^4}{1+ |p/q|^4} 
\eeq 

\item 
$\Delta m_B \neq  0$ with $f$ now being a  
flavour-{\em non}specific final state -- a final state 
{\em common} 
to $B^0$ and $\bar B^0$ decays -- of a special nature, namely 
a \cp~eigenstate -- $|\bar f\rangle = {\bf CP}|f\rangle = 
\pm |f\rangle $ -- {\em without} direct \cp~violation --  
$|\bar \rho _f| = 1 = |\rho _{\bar f}| $: 
\beq 
\begin{array} {c} 
G_f(t) = 2 |T_f|^2 
\left[ 1 - ({\rm sin}\Delta m_Bt) \cdot 
{\rm Im} \frac{q}{p} \bar \rho _f 
\right] \\  
\bar G_f(t) = 2 |T_f|^2 
\left[ 1 + ({\rm sin}\Delta m_Bt )\cdot 
{\rm Im} \frac{q}{p} \bar \rho _f 
\right] \\ 
\leadsto 
\frac{d}{dt} \frac{\bar G_f(t) }{G_f(t)} \neq 0\\
\frac{\bar G_f(t) - G_f(t)}{\bar G_f(t) + G_f(t)} = 
({\rm sin}\Delta m_Bt )\cdot {\rm Im} \frac{q}{p} \bar \rho _f 
\end{array} 
\label{GGBAR}
\eeq 
is the concrete realization of what was called \cp~violation 
{\em involving oscillations}. 

\noindent 
For $f$ still denoting a \cp~eigenstate, yet with $|\bar \rho_f| \neq 1$ one has the more complex 
asymmetry expression
\beq 
\frac{\bar G_f(t) - G_f(t)}{\bar G_f(t) + G_f(t)} = S_f\cdot  ({\rm sin}\Delta m_Bt ) - 
C_f \cdot ({\rm cos}\Delta m_Bt )
\label{CSASYM}
\eeq
with 
\beq 
S_f = \frac{2{\rm Im}\frac{q}{p}\bar \rho_{\pi^+\pi^-}}{1+\left|\frac{q}{p}\bar \rho_{\pi^+\pi^-}\right| ^2} \; , \; \; 
C_f = \frac{1 - \left|\frac{q}{p}\bar \rho_{\pi^+\pi^-}\right| ^2}
{1+\left|\frac{q}{p}\bar \rho_{\pi^+\pi^-}\right| ^2} 
\eeq
\end{itemize} 
For the decays of neutral mesons the following general statement is relevant, at least conceptually. 
\begin{center}
{\bf Theorem}: 
\end{center}

\noindent 
Consider a beam with an arbitrary combination of neutral mesons $P^0$  and $\bar P^0$ decaying into a final state $f$ that is a \cp~eigenstate. {\em If} the decay rate evolution in (proper) time $t$ is 
{\em not} described by a {\em single exponential}, i.e. 
\beq 
{\rm rate}(P^0/\bar P^0(t) \to f) \neq K  e^{-\Gamma t}  \; \; \; {\rm with} \; \; \; 
\frac{d}{dt} K \equiv 0
\eeq 
for any real $\Gamma$, then \cp~invariance is violated.  
\begin{center}
{\bf Homework \# 5} 
\end{center} 
\noindent 
Prove this theorem. 
\begin{center}
{\bf End of Homework \# 5} 
\end{center} 

An obvious, yet still useful criterion for \cp~observables is that they must be 
`re-phasing' invariant under $|\bar B^0\rangle \to e^{-i\xi}|\bar B^0\rangle$. The expressions 
above show there are three 
classes of such observables:
\begin{itemize}
\item 
An asymmetry in the {\em instantaneous} transition amplitudes for \cp~conjugate modes:
\beq 
|T(B \to f )| \neq |T(\bar B \to \bar f)| \hspace{1cm} 
\Longleftrightarrow \hspace{1cm} \Delta B =1 \; .  
\label{class1}
\eeq
It reflects pure $\Delta B = 1$ dynamics and thus amounts 
to {\em direct} \cp~violation. 
Those modes are most likely to be nonleptonic; in the SM they practically have to be. 

\item 
\cp~violation {\em in} $B^0 - \bar B^0$ oscillations: 
\beq 
|q| \neq |p| \hspace{1cm} 
\Longleftrightarrow \hspace{1cm} \Delta B =2 \; .  
\label{class2}
\eeq 
It requires \cp~violation {\em in} $\Delta B =2$ dynamics. 
The theoretically cleanest modes here are semileptonic ones due to the SM 
$\Delta Q = \Delta B$ selection rule.  

\item 
\cp~asymmetries {\em involving} oscillations
\footnote{This condition is formulated for the simplest case of $f$ being a \cp~eigenstate.}:  
\beq 
{\rm Im}\frac{q}{p}\bar \rho (f) \neq 0 \; ,\; \;  \bar \rho (f) = \frac{T(\bar B \to f)}{T(B\to f)}\hspace{1cm} 
\Longleftrightarrow \hspace{1cm} \Delta B =1 \& 2 \; . 
\label{class3}
\eeq
Such an effect requires the interplay of $\Delta B=1 \& 2$ forces. 

While $C_f \neq 0$ unequivocally signals {\em direct} \cp~violation in Eq.(\ref{CSASYM}), 
the interpretation of 
$S_f \neq 0$ is more complex. (i) As long as one has measured $S_f$ only in a single mode, the distinction 
between {\em direct} and {\em indirect} \cp~violation -- i.e. \cp~violation in $\Delta B =1$ and 
$\Delta B =2$ dynamics -- is convention dependent, since a change in phase for $\bar B^0$ -- 
$|\bar B^0 \rangle \to e^{-i\xi} |\bar B^0 \rangle$ -- leads to $\bar \rho_f \to e^{-i\xi}\bar \rho_f$ and 
$(q/p) \to e^{i\xi} (q/p)$, i.e. can shift any phase from $(q/p)$ to $\bar \rho_f$ and back while 
leaving $(q/p)\bar \rho_f$ invariant. However once $S_f$ has  been measured for two different 
final states $f$, then the distinction becomes truly meaningful independent of theory: 
$S_{f_1} \neq S_{f_2}$ implies $(q/p)\bar \rho_{f_1} \neq (q/p)\bar \rho_{f_2}$ and thus 
$\bar \rho_{f_1} \neq \bar \rho_{f_2}$, i.e. \cp~violation in the $\Delta B =1$ sector. One should note 
that this {\em direct} \cp~violation might not generate a $C_f$ term. For 
$\bar \rho_{f_1} = e^{i\phi_1}$ and $\bar \rho_{f_2} = e^{i\phi_2}$  causing $S_{f_1} \neq S_{f_2}$ 
would both lead to $C_{f_1} = 0 = C_{f_2}$.

\end{itemize}
Once the final state consists of more than two pseudoscalar or one pseudoscalar and one vector meson, 
it contains more dynamical information than expressed through the decay width into it, as can be 
described through a Dalitz plot. 
\begin{itemize}
\item 
Accordingly one can have a \cp~asymmetry in {\em final state distributions} of $B$ mesons, as discussed later. 
There is a precedent for such an effect, namely a \ot~odd correlation that has been observed between 
the $\pi^+ -\pi^-$ and $e^+ -e^-$ planes in the rare mode $K_L \to \pi^+\pi^- e^+e^-$, the size of which 
can be inferred from $K_L \to \pi^+\pi^-$. 
\end{itemize}

\subsubsection{The First Central Pillar of the Paradigm: Long Lifetimes}
\label{BLIFE}

Beauty, the existence of which had been telegraphed by the discovery of the $\tau$ as the third 
charged lepton 
was indeed observed exhibiting a surprising feature: starting in the early 1980's its lifetime was 
found to be about $10^{-12}$ sec.  This was considered `long'. For one can get an estimate 
for $\tau (B)$ by relating it to the muon lifetime: 
\beq 
\tau (B) \simeq \tau _b \sim \tau (\mu ) \left( \frac{m(\mu )}{m(b)}  \right) ^5 \frac{1}{9} 
\frac{1}{|V(cb)|^2} \simeq 3 \cdot 10^{-14} \left| \frac{{\rm sin}\theta_C}{V(cb)}  \right|^2 \; {\rm sec}
\eeq
One had expected $|V(cb)|$ to be suppressed, since it represents an out-of-family coupling. 
Yet  one had assumed without deeper reflection that $|V(cb)| \sim {\rm sin}\theta_C$ -- what 
else could it be? The measured value for $\tau (B)$ however pointed to  
$|V(cb)| \sim |{\rm sin}\theta_C|^2$.  By the end of the millenium one had obtained a rather accurate value: 
$\tau (B_d) = (1.55\pm 0.04) \cdot 10^{-12}$ s. Now the data have become even more precise: 
\beq 
\tau (B_d) = (1.530 \pm 0.009) \cdot 10^{-12} \; {\rm s}\; , \; \; 
\tau (B^{\pm}) /\tau (B_d) = 1.071 \pm 0.009 
\eeq
The lifetime {\em ratio}, which reflects the impact of hadronization, had been predicted 
\cite{MIRAGE} successfully 
well before data of the required accuracy had been available.

\subsubsection{Oscillations of $B_d$ \& $B_s$ Mesons -- Exactly like for Kaons, only Different}
\label{BOSCILL}

The general phenomenology of $B^0 - \bar B^0$ oscillations posed no mystery since the 
beginning of thinking about it, since it follows a close qualitative -- though not quantitative -- analogy 
with kaon oscillations described above. One obvious difference arises in the lifetime ratios of the two mass eigenstates: the huge disparity in the $K_L$ and $K_S$ lifetimes -- 
$\tau (K_L) \sim 600 \tau (K_S)$ -- is due to the kinematical `accident' that the kaon is barely above the 
three pion threshold; this does {\em not} have an analogue for the heavier mesons, where one expects 
on general grounds $\Delta \Gamma \ll 1$, to be quantified below. 

The most general observable signature of oscillations is the apparent violation of some selection rule. 
In the SM one has 
\beq 
l^-\bar \nu X_c^+ \not \leftarrow B^0 \to l^+ \nu X_c^-  \not \leftarrow \bar B^0 \to l^-\bar \nu X_c^+
\label{SELRULE} 
\eeq
Yet oscillations can circumvent it in the following way: 
\beq 
B^0 \Longrightarrow \bar B^0 \to l^- \nu +X_c^+ \; , \;  \bar B^0 \Longrightarrow B^0 \to l^+ \nu X_c^-  
\eeq
where "$\Longrightarrow$" and "$\to$" denote the $\Delta B =2$ oscillation and $\Delta B=1$ 
direct transitions, respectively. This apparent violation of the selection rule exhibits a characteristic 
dependence on the time of decay analogue to that of Eq.(\ref{REGENEXP}) where 
$\Delta \Gamma =0$ has been set for simplicity: 
\bea 
{\rm rate} (B^0 \to l^- X_c^+; t) &\propto& \frac{1}{2} e^{-\Gamma_B t} (1- {\rm cos}\Delta M_Bt) \\
{\rm rate} (B^0 \to l^+ X_c^+; t) &\propto& \frac{1}{2} e^{-\Gamma_B t} (1+ {\rm cos}\Delta M_Bt)
\eea 
Integrating over all times of decay one finds for the ratio of wrong- to right-sign leptons and for the  
probability of wrong-sign leptons
\bea
\nonumber
r_B &=& \frac{\Gamma (\bar B^0 \to l^+\nu X_c^-)}{\Gamma (\bar B^0 \to l^-\nu X_c^+)} = 
\frac{x_B^2}{2 + x_B^2}\; \; , \; \; \; 
x_B = \frac{\Delta M(B_d)}{\Gamma (B_d)} \\
\chi_B &=&  \frac{\Gamma (\bar B^0 \to l^+\nu X_c^-)}{\Gamma (\bar B^0 \to l^{\pm}\nu X_c^{\mp})} = 
\frac{r_B}{1+ r_B}
\label{CHI}
\eea
The quantities $r_B$ and $\chi_B$ thus represent the violation of the selection rule of 
Eq.(\ref{SELRULE}) `on average'.  
{\em Maximal} oscillations can be defined as $ x \gg 1$ and thus $r \to 1$ and $\chi \to 1/2$. 

\subsubsubsection{$B_d - \bar B_d$ Oscillations}
\label{BDCASE}

Present data yield for $B_d$ mesons: 
\beq 
x_d \equiv x_{B_d} = 0.776 \pm 0.008 \; , \; \; \chi_d = 0.188 \pm 0.003
\label{XDEXP}
\eeq

Huge samples of beauty mesons can be obtained in $p\bar p$ or $pp$ collisions at high energies, 
which yield {\em in}coherent pairs of $B$ mesons. Two cases have to be distinguished: 
\begin{itemize}
\item 
\beq 
p \bar p \to B^+\bar B_d + X /B^- B_d + X
\eeq
leading to a single beam of neutral $B$ mesons, for which Eq.(\ref{CHI}) applies. 
\item 
\beq 
p \bar p \to B_d \bar B_d + X \; , 
\eeq
when both $B$ mesons can oscillate -- actually into each other -- leading to {\em like}-sign 
di-leptons
\beq 
p \bar p \to B_d \bar B_d + X \Longrightarrow B_dB_d/\bar B_d \bar B_d + X \to 
l^{\pm}l^{\pm} + X^{\prime} \; . 
\eeq
Its relative probability can be expressed as follows 
\beq 
\frac{{\rm rate}(p \bar p \to B_d \bar B_d + X  \to l^{\pm}l^{\pm} + X^{\prime})}
{{\rm rate}(p \bar p \to B_d \bar B_d + X  \to ll + X^{\prime})} = 2 \chi _d (1- \chi _d) 
\label{INCOH}
\eeq
meaning that {\em like}-sign di-leptons require one $B$ meson to have oscillated into its antiparticle at 
its time of decay, while the other one has not. 
\item 
In 
\beq 
e^+e^- \to B_d \bar B_d 
\eeq
one encounters the {\em coherent} production of two neutral beauty mesons. As discussed in 
detail in Sect.\ref{EPRIMPORT} EPR correlations combine with the requirement of Bose-Einstein statistics to make the 
pair act as a {\em single} oscillating system leading to \cite{BS80} 
\beq 
\frac{{\rm rate}(e^+e^- \to B_d \bar B_d \to l^{\pm}l^{\pm}+X)}
{{\rm rate}(e^+e^- \to B_d \bar B_d \to ll+X^{\prime})} = \chi _d
\label{COH}
\eeq
\end{itemize}
For the measured value of $x_d$ the two expressions in Eqs.(\ref{INCOH}) and (\ref{COH}) yield 
\beq 
\chi_d = 0.188 \pm 0.003 \; \; vs. \; \; 2\chi_d (1- \chi_d) \simeq 0.305 \; ; 
\eeq
i.e., the two ratios of like-sign dileptons to all dileptons emerging from the decays of a {\em coherently} 
and {\em incoherently} produced  
$B_d \bar B_d$ pair differ by a factor of almost two due to EPR correlations as explained below.

One predicts on rather general grounds that  
${\cal L}(\Delta B =2)$ is dominated by short distance dynamics and more specifically by the quark box diagram to a higher degree than ${\cal L}(\Delta S =2)$. 
It is often stated that $B_d - \bar B_d$ oscillations were found to proceed much faster than predicted. 
Factually this is correct -- yet one should note the main reason for it. The prediction 
for $x_B$ depends very much on the value of the top quark mass $m_t$, see Eq.(\ref{HW4}) for a rough 
scaling law. In the early 1980's there had been the experimental claim by the UA1 collaboration that top quarks had been discovered in $p\bar p$ collisions with a mass $m_t = 40 \pm 10$ GeV. With 
$x_B \propto m_t^2$ and $r_B \propto x_B^2 \propto m_t^4$ for moderate values of $x_B$, one finds 
$r_B$ increases by more than one order of magnitude when going from $m_t =40$ GeV to $170$ GeV!  
Once the ARGUS collaboration discovered $B_d - \bar B_d$ oscillations with $x_d \sim 0.7$ theorists 
quickly concluded that top quarks had to be much heavier than previously considered, namely 
$m_t > 100$ GeV. This was the first indirect evidence for top quarks being `super-heavy'. A second 
and more accurate indirect evidence came later from studying electroweak radiative corrections 
at LEP.

Since $x = \Delta M/\Gamma$ denotes the ratio between the oscillation and decay rates, 
$x=1$ represents the optimal realization of the scenario sketched in Eq.(\ref{BASICIDEA}) 
for obtaining a \cp~asymmetry, 
namely to rely on oscillations to provide a second coherent amplitude of a comparable effective 
strength.  This statement can be made more quantitative by integrating the asymmetry of 
Eq.(\ref{GGBAR}) over all times of decay $t$: 
\bea
 {\rm rate}\left( B^0 (t) \to f \right) &=& 
 Ke^{-\Gamma _B t} \left( 1 - A\cdot {\rm sin}(x\Gamma _Bt) \right) \; , \; \; 
 x = \frac{\Delta M_B}{\Gamma _B} \\
\int _0^{\infty} dt \, {\rm rate}\left( B^0 (t) \to f\right) &=& 
\frac{K}{\Gamma_B} \left( 1 - A\cdot  \frac{x}{1+x^2}
\right)  
\eea
The oscillation induced factor $x/(1+x^2)$ is {\em maximal} for $x=1$; i.e., with 
Eq.(\ref{XDEXP}) nature has given us an almost optimal stage for observing \cp~violation in 
$B_d$ decays.

\subsubsubsection{The `hot' news: $B_s - \bar B_s$ oscillations}
\label{HOT}
For a moment I will deviate considerably from the historical sequence by presenting the `hot' news 
of the resolution of $B_s - \bar B_s$ oscillations. 

Nature actually provided us with an `encore' in $B^0$ oscillations. 
It had been recognized from the beginning that within the SM one predicts 
$\Delta M_{B_s} \gg \Delta M_{B_d}$, i.e. that $B_s$ mesons oscillate much faster than 
$B_d$ mesons. Both receive their dominant contributions from $t \bar t$ quarks in the quark box diagram making their ratio depend on the CKM parameters and the hadronic matrix element of 
the relevant four-quark operator only: 
\beq 
\frac{\Delta M_{B_s}}{\Delta M_{B_d}} \simeq 
\frac{B_sf^2_{B_s}}{B_df^2_{B_d}}\frac{|V(ts)|^2}{|V(td)|^2}\; . 
\eeq
This relation also exhibits the phenomenological interest in measuring $\Delta M_{B_s}$, namely to 
obtain an accurate value for $|V(td)|$. Lattice QCD is usually invoked to gain theoretical control over the first ratio of hadronic quantities. Taking its findings together with the  CKM constraints on 
$|V(ts)/V(ts)|$ yields the following SM prediction:  
\beq 
\left. \Delta M_{B_s}\right| _{SM} = \left(18.3 ^{+6.5}_{-1.5}\right) \, {\rm ps}^{-1} \; \; 
\hat = \;  \left( 1.20 ^{+0.43}_{-0.10}  \right) \cdot 10^{-2} \; {\rm eV}            \; \; \;  {\rm CKM\; fit} 
\eeq
Those rapid oscillations have been resolved now by CDF \cite{CDFBSPUB} and D0  \cite{D0BSPUB}: 
\bea 
\Delta M_{B_s} &=& 
\left\{   
\begin{array}{ll}
\left(19 \pm 2 \right) \, {\rm ps}^{-1} & {\rm D0}   
\\  
\left(17.77 \pm 0.10 \pm 0.07\right)\, {\rm ps}^{-1} & {\rm CDF}  
\end{array} 
\right. \\ 
x_s &=& \frac{\Delta M_{B_s}}{\Gamma _{B_s}} \simeq 25 
\label{DIRECTCPDATA} 
\eea 
These findings represent another triumph of CKM theory even more impressive 
than a mere comparison of the observed and predicted values of $\Delta M(B_s)$, as explained later. 

There is also marginal evidence for $\Delta \Gamma (B_s) \neq 0$ \cite{HFAG}
\beq 
\frac{\Delta \Gamma (B_s)}{\Gamma (B_s)} = 0.31 \pm 0.13 \; . 
\label{DGEXP}
\eeq
My heart wishes that $\frac{\Delta \Gamma _s}{\bar \Gamma_s}$ were indeed as large 
as 0.5 or even larger. For it would open up a whole new realm of \cp~studies in 
$B_s$ decays with a great potential to identify New Physics. Yet my head tells me that values 
exceeding 0.25 or so are very unlikely; it would point at a severe 
limitation in our theoretical understanding of $B$ lifetimes. For only on-shell intermediate 
states $f$ in $B^0 \to f \to \bar B^0$ can contribute to $\Delta \Gamma (B)$, and for 
$B^0 = B_s$ these are predominantly driven by $b\to c \bar c s$. Let $R(b\to c \bar cs)$ denote their fraction of all $B_s$ decays. 
If these transitions contribute only 
to $\Gamma (B_s(CP=+))$ one has $\Delta \Gamma_s/\bar \Gamma_s = 2R(b\to c \bar cs)$. 
Of course this is actually an upper bound quite unlikely to be even remotely saturated. 
With the estimate $R(b\to c \bar cs) \simeq 25\% $, which is consistent with the data on 
the charm content of $B_{u,d}$ decays this upper bound reads 50\%. More realistic calculations 
have yielded considerably smaller predictions: 
\beq 
\frac{\Delta \Gamma_s}{\bar \Gamma_s} = \left\{
\begin{array}{l} 0.22 \cdot \left(\frac{f(B_s)}{220\, \MeV}\right)^2 \\
0.12 \pm 0.05  
\end{array}
\right. \;  ; 
\label{DELTAPRED}
\eeq 
where the two predictions are taken from Refs.\cite{AZIMOV} and \cite{LENZNEW}, respectively.  
A value as high as $0.20 - 0.25$ \% is thus not out of the question theoretically, and Eq.(\ref{DGEXP}) is still consistent with it. One should note that invoking New Physics would actually `backfire' since it leads to a lower prediction.
If, however, a value exceeding 25\% were established experimentally, we had to draw at least one of the following conclusions: (i) $R(b\to c \bar cs)$ actually exceeds the estimate of 0.25 significantly. This would imply 
at the very least that the charm content is higher in $B_s$ than $B_{u,d}$ decays by a 
commensurate amount and the $B_s$ semileptonic branching ratio lower. 
(ii) Such an enhancement of  $R(b\to c \bar cs)$ would presumably -- though not necessarily -- imply 
that the average $B_s$ width exceeds the $B_d$ width by more than the predicted 1-2\% level. 
That means in analyzing $B_s$ lifetimes one should allow $\bar \tau (B_s)$ to float 
{\em freely}. (iii) If in the end one found the charm content of $B_s$ and $B$ decays to be quite 
similar and 
$\bar \tau (B_s)$ $\simeq$ $\tau (B_d)$, yet  
$\Delta \Gamma_s/\bar \Gamma_s$ to exceed 0.25, we had to concede a loss of theoretical control 
over $\Delta \Gamma$. This would be disappointing, yet not inconceivable: the a priori reasonable 
ansatz of evaluating both  
$\Delta \Gamma_B$ and $\Delta M_B$ from quark box diagrams -- with the only 
manifest difference being that the internal quarks are charm in the former and top in the 
latter case -- obscures the fact that the dynamical situation is actually different. In the latter case the 
effective transition operator is a local one involving a considerable amount of averaging over 
off-shell transitions; the former is shaped by on-shell channels with a relatively small amount 
of phase space: for the $B_s$ resides barely 1.5 GeV above the $D_s \bar D_s$ threshold. 
To say it differently: the observable $\Delta \Gamma_s$ is more vulnerable to limitations 
of quark-hadron duality than $\Delta M_s$ and even beauty lifetimes 
\footnote{These are all dominated by nonleptonic transitions, where duality violations can be 
significantly larger than for semileptonic modes.} . 

In summary: establishing $\Delta \Gamma_s \neq 0$ amounts to important qualitative progress in our 
knowledge of beauty hadrons; it can be of great practical help in providing us with novel probes of CP 
violations in $B_s$ decays, and it can provide us theorists with a reality check concerning the reliability 
of our theoretical tools for nonleptonic $B$ decays.

\subsubsection{Large \cp~Asymmetries in $B$ Decays With{\em out} 
 `Plausible Deniability'}
\label{PLAUS}

The above mentioned observation of a long $B$ lifetime pointed to 
$|V(cb)| \sim {\cal O}(\lambda ^2)$ with $\lambda = {\rm sin}\theta_C$. Together with the 
expected observation $|V(ub)| \ll |V(cb)|$ and coupled with the assumption 
of three-family unitarity this allows to expand the CKM matrix in powers of $\lambda$, which 
yields the following most intriguing result through order $\lambda ^5$, as first recognized by Wolfenstein: 
\beq 
{\bf V}_{CKM} = 
\left( 
\begin{array}{ccc} 
1 - \frac{1}{2} \lambda ^2 & \lambda & 
A \lambda ^3 (\rho - i \eta + \frac{i}{2} \eta \lambda ^2) \\
- \lambda & 1 - \frac{1}{2} \lambda ^2 - i \eta A^2 \lambda ^4 & 
A\lambda ^2 (1 + i\eta \lambda ^2 ) \\ 
A \lambda ^3 (1 - \rho - i \eta ) \\
& - A\lambda ^2 & 1 
\end{array}
\right) 
\label{WOLFKM}
\eeq  
The three Euler angles and one complex phase of the representation given in Eq.(\ref{PDGKM}) 
is taken over by the four real quantities $\lambda$, $A$, $\rho$ and $\eta$; 
$\lambda$ is the expansion parameter with $\lambda \ll 1$, whereas $A$, $\rho$ and $\eta$ 
are a priori of order unity, as will be discussed in some detail later on. I.e., the `long' 
lifetime of beauty hadrons of around 1 psec together with beauty's affinity to transform itself into charm 
and the assumption of only three quark families tell us 
that the CKM matrix exhibits a very peculiar hierarchical pattern in powers of $\lambda$: 
\beq 
V_{CKM} = 
\left( 
\begin{array}{ccc} 
1 & {\cal O}(\lambda ) & {\cal O}(\lambda ^3) \\ 
{\cal O}(\lambda ) & 1 & {\cal O}(\lambda ^2) \\ 
{\cal O}(\lambda ^3) & {\cal O}(\lambda ^2) & 1 
\end {array} 
\right) 
\; \; \; , \; \; \; \lambda = {\rm sin}\theta _C 
\eeq 
As explained in Sect.\ref{QMASSES}, we know this matrix has to be unitary. Yet in addition it is almost 
the identity matrix, almost symmetric and the moduli of its elements shrink with the distance from the diagonal. 
It has to contain a message from nature -- albeit in a highly encoded form. 

My view of the situation is 
best described by a poem by the German poet Joseph von Eichendorff from the late romantic 
period \footnote{I have been told that early romantic writers would have used the term `symmetry' 
instead of `song'.}: 

\vspace{3mm}

\begin{tabular} {ll}
Schl\"aft ein Lied in allen Dingen, &  There sleeps a song in all things \\
die da tr\"aumen fort und fort, & that dream on and on, \\
und die Welt hebt an zu singen, & and the world will start to sing, \\
findst Du nur das Zauberwort. & if you find the magic word.
\end{tabular}
 
\vspace{3mm}

\noindent 
The six triangle relations obtained from the unitarity condition fall into three categories:  
\begin{enumerate}
\item 
$K^0$ triangle: 
\beq 
\begin{array}{ccc} 
V^*(ud)V(us) + &V^*(cd)V(cs) + &V^*(td) V(ts) = 
\delta _{ds}= 0 \\
{\cal O}(\lambda ) & {\cal O}(\lambda ) & {\cal O}(\lambda ^5) 
\end{array} 
\label{TRI1} 
\eeq 
$D^0$ triangle:
\beq 
\begin{array}{ccc} 
V^*(ud)V(cd) + &V^*(us)V(cs) + &V^*(ub) V(cb) = 
\delta _{uc}= 0 \\
{\cal O}(\lambda ) & {\cal O}(\lambda ) & {\cal O}(\lambda ^5) \; , 
\end{array} 
\label{TRI2} 
\eeq 
where below each product of matrix elements I have noted 
their size in powers of $\lambda $. 
These two triangles are extremely `squashed': two sides are 
of order $\lambda $, the third one of order $\lambda ^5$ and their 
ratio of order $\lambda ^4 \simeq 2.3 \cdot 10^{-3}$; 
Eq.(\ref{TRI1}) and Eq.(\ref{TRI2}) control the situation in 
strange and charm decays; the relevant weak phases there 
are obviously tiny.

\item 
$B_s$ triangle: 
\beq 
\begin{array}{ccc} 
V^*(us)V(ub) + &V^*(cs)V(cb) + &V^*(ts) V(tb) = 
\delta _{sb}= 0 \\
{\cal O}(\lambda ^4) & {\cal O}(\lambda ^2) & {\cal O}(\lambda ^2) 
\end{array} 
\label{TRI3} 
\eeq 
$tc$ triangle: 
\beq 
\begin{array}{ccc} 
V^*(td)V(cd) + &V^*(ts)V(cs) + &V^*(tb) V(cb) = 
\delta _{ct}=0 \\
{\cal O}(\lambda ^4) & {\cal O}(\lambda ^2) & {\cal O}(\lambda ^2) 
\end{array} 
\label{TRI4} 
\eeq 
The third and fourth triangles are still rather squashed, yet less so: 
two sides are of order $\lambda ^2$ and the third one of order 
$\lambda ^4$. 

\item 
$B_d$ triangle: 
\beq 
\begin{array}{ccc} 
V^*(ud)V(ub) + &V^*(cd)V(cb) + &V^*(td) V(tb) = 
\delta _{db}=0 \\
{\cal O}(\lambda ^3) & {\cal O}(\lambda ^3) & {\cal O}(\lambda ^3) 
\end{array} 
\label{TRI5} 
\eeq 
$ut$ triangle: 
\beq 
\begin{array}{ccc} 
V^*(td)V(ud) + &V^*(ts)V(us) + &V^*(tb) V(ub) = 
\delta _{ut}=0 \\
{\cal O}(\lambda ^3) & {\cal O}(\lambda ^3) & {\cal O}(\lambda ^3) 
\end{array}
\label{TRI6}  
\eeq 
The last two triangles have sides that are all of the same 
order, namely $\lambda ^3$. All their angles are therefore 
naturally large, i.e. $\sim$ several $\times$ $10$ degrees! Since to 
leading order in $\lambda$ one has 
\beq 
V(ud) \simeq V(tb) \; , \; V(cd) \simeq - V(us) \; , \; 
V(ts) \simeq - V(cb) 
\eeq 
we see that the triangles of Eqs.(\ref{TRI5}, \ref{TRI6}) 
actually coincide to that order.  
\begin{figure}[t]
\vspace{6.0cm}
\includegraphics{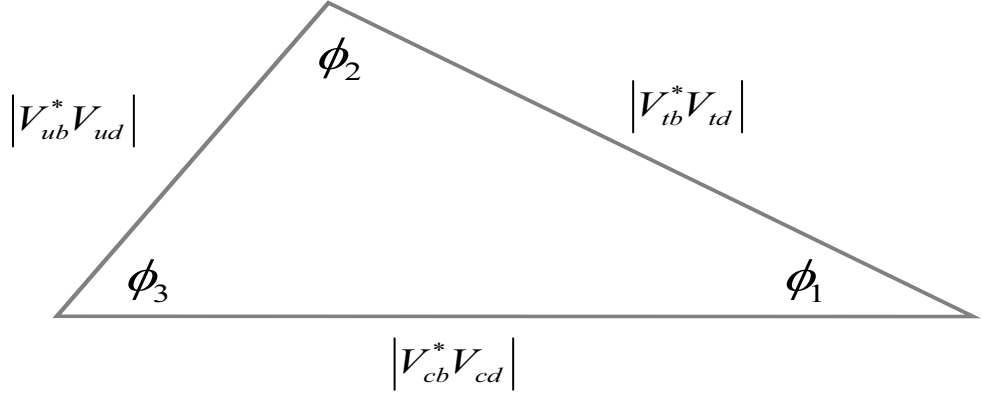}
\caption{The CKM Unitarity Triangle}
\label{CKMTRIANGLENOT} 
\end{figure}
The sides of this triangle having naturally large angles, see Fig.(\ref{CKMTRIANGLENOT}), are 
given by $\lambda \cdot V(cb)$, $V(ub)$ and 
$V^*(td)$; these are all quantities that control important 
aspects of $B$ decays, namely CKM favoured and disfavoured 
$B_{u,d}$ decays and $B_d - \bar B_d$ oscillations. The $B_d$ triangle of Eq.(\ref{TRI5}) 
is usually referred to as {\em `the' CKM unitarity triangle}.  

\end{enumerate} 
Let the reader be reminded that all six triangles, despite their very different shapes, have 
the same area, see Eq.(\ref{JARL}), reflecting the single CKM phase for three families.  

Some comments on notation might not be completely useless. The BABAR collaboration 
and its followers refer to the three angles of the CKM unitarity triangle as 
$\alpha$, $\beta$ and $\gamma$; the BELLE collaboration instead has adopted the 
notation $\phi_1$, $\phi_2$ and $\phi_3$. While it poses no problem to be 
conversant in both languages, the latter has not only historical priority on its side 
\cite{BJSANDA}, but is 
also more rational. For the angles $\phi_i$ in the 
`$bd$' triangle of Eq.(\ref{TRI5}) are always opposite the side defined by 
$V^*(id)V(ib)$.   Furthermore this classification scheme can readily be generalized to all six 
unitarity triangles;  those triangles can be labeled by $kl$ with 
$k \neq l= d,s,b$ or $k\neq l = u,c,t$, see Eqs.(\ref{TRI1}) -- (\ref{TRI6}). Its 18 angles 
can then be unambiguously denoted by $\phi_i^{kl}$: it is the angle in triangle $kl$ opposite 
the side $V^*(ik)V(il)$ or $V^*(ki)V(li)$, respectively. {\em Therefore I view the notation 
$\phi_i^{(kl)}$ as the only truly Cartesian one.}  

The discovery of $B_d - \bar B_d$ oscillations {\em defined} the 
`CKM Paradigm of Large \cp~Violation in $B$ Decays' that had been {\em anticipated} in 1980: 
\begin{itemize}
\item 
A host of nonleptonic $B$ channels has to exhibit sizable \cp~asymmetries. 
\item 
For $B_d$ decays to flavour-nonspecific final states (like \cp~eigenstates) the \cp~asymmetries 
depend on the time of decay in a very characteristic manner; their size should typically be measured 
in units of 10\% rather than 0.1\%. 
\item 
{\em There is no plausible deniability for the CKM description, if such asymmetries are 
not found.}  
\item 
For $m_t \geq 150$ GeV the SM prediction for $\epsilon_K$ is dominated by the top quark 
contribution like $\Delta M_{B_d}$. It thus drops out from their ratio, and sin$2\phi_1$ can be 
predicted within the SM irrespective of the (superheavy) top quark mass. In the early 1990's, i.e., 
before the direct discovery of top quarks, it was predicted \cite{BEFORETOP} 
\beq 
\frac{\epsilon_K}{\Delta M_{B_d}} \propto {\rm sin}2\phi_1 \sim 0.6 - 0.7
\eeq  
with values for $B_Bf_B^2$ inserted as now estimated by LQCD. 
\item 
The \cp~asymmetry in the Cabibbo favoured channels $B_s \to \psi \phi/\psi \eta$ is Cabibbo suppressed, i.e. below 4\%, for reasons very specific to CKM theory, as pointed out already in 
1980 \cite{BS80}. 

\end{itemize}

In 1974 finally top quarks were observed directly with a mass fully consistent 
with the indirect estimates given above; the most recent analyses from CDF \& D0 list  
\beq 
m_t = 172.7 \pm 2.9 \; {\rm GeV}
\eeq

\subsubsection{Data in 1998}
\label{DATA(*}

\cp~violation had been observed only in the decays of neutral kaons, and all its 
manifestations -- $K_L \to \pi^+\pi^-$, $\pi^0\pi^0$, 
$K^0 \to \pi^+\pi^-$ vs. $\bar K^0 \to \pi^+\pi^-$, $K_L\to l^+\nu \pi^-$ 
vs. $K_L \to l^- \bar \nu \pi^+$ -- could be described for 35 years with a {\em single real} number, 
namely $|\eta_{+-}|$ or $\Phi (\Delta S=2) = {\rm arg}(M_{12}/\Gamma_{12})$.  

There was intriguing, though not conclusive evidence for {\em direct} \cp~violation: 
\beq 
\frac{\epsilon ^{\prime}}{\epsilon_K} = 
\left\{ 
\begin{array}{ll} 
(2.30 \pm 0.65)\cdot 10^{-3} & {\rm NA31} \\
(0.74 \pm 0.59)\cdot 10^{-3} & {\rm E731}
\end{array}
\right.
\eeq
These measurements were made in the 1980's and had been launched by theory 
guestimates  
suggesting values for $\epsilon^{\prime}$ that would be within the reach of these experiments. 
Theory, however, had `moved on' favouring values $\leq 10^{-3}$ -- or so it was claimed.

\subsection{Completion of a Heroic Era}
\label{HERO}

{\em Direct} CP violation 
has been unequivocally established in 1999. The present world average dominated 
by the data from NA48 and KTeV reads as follows \cite{SOZZI}: 
\beq 
\langle \epsilon ^{\prime}/\epsilon _K \rangle = 
(1.63 \pm 0.22) \cdot 10^{-3} 
\eeq 
Quoting the result in this way does not do justice to the experimental 
achievement, since $\epsilon _K$ is a very small number itself. 
The sensitivity achieved becomes more obvious when quoted in terms of actual 
widths \cite{SOZZI}:
\beq 
\frac{\Gamma (K^0 \to \pi ^+ \pi ^-) - 
\Gamma (\bar K^0 \to \pi ^+ \pi ^-)}
{\Gamma (K^0 \to \pi ^+ \pi ^-) + 
\Gamma (\bar K^0 \to \pi ^+ \pi ^-)} = 
(5.04 \pm 0.82) \cdot 10^{-6} \; !
\label{DIRECTK}
\eeq
This represents a discovery of the very first rank 
\footnote{As a consequence of Eq.(\ref{DIRECTK}) I am not impressed by \cpt~tests falling short 
of the $10^{-6}$ level.}. Its significance does not depend on whether the  
SM can reproduce it or not -- which is the most concise confirmation of 
how important it is. 
The HEP community can take pride 
in this achievement; the tale behind it is a most fascinating one about imagination and perseverance. The two groups and their predecessors deserve our respect; they 
have certainly earned my admiration. 

The experimental findings are consistent with CKM theory on the qualitative level, since the latter 
does not represent a superweak scenario even for strange decays due to the existence of Penguin 
operators. It is not inconsistent with it even quantitatively. One should keep in mind that within the 
SM $\epsilon^{\prime}/\epsilon_K$ has to be considerably suppressed.  
$\epsilon^{\prime}$ requires interference between $\Delta I = 1/2\,  \& \, 3/2$ amplitudes and is thus 
reduced by the `$\Delta I = 1/2$ rule': $|T(\Delta I = 3/2)/T(\Delta I = 1/2)| \sim 1/20$. Furthermore 
$\epsilon^{\prime}$ is generated by loop diagrams -- as is $\epsilon_K$; yet the top quark mass 
enhances $\epsilon_K$ powerlike -- $|\epsilon_K| \propto m_t^2/M_W^2$ -- whereas 
$\epsilon^{\prime}$ only logarithmically. When there is only one weak phase -- as is the case 
for CKM theory -- one has $|\epsilon^{\prime}/\epsilon_K| \propto {\rm log}m_t^2/m_t^2$, i.e. 
greatly reduced again for superheavy top quarks (revisit Homework \# 4). 

CKM theory can go beyond such semiquantitative statements, but one should not expect a 
{\em precise} prediction from it in the near future. For the problem of uncertainties in the evaluation 
of hadronic matrix elements  is compounded by the fact that the two main contributions to 
$\epsilon^{\prime}$ are similar in magnitude, yet opposite in sign \cite{EPSPRIMETH}.

\subsubsection{CKM Theory at the End of the 2nd Millenium}

\begin{figure}[t]
\vspace{7.0cm}
\includegraphics{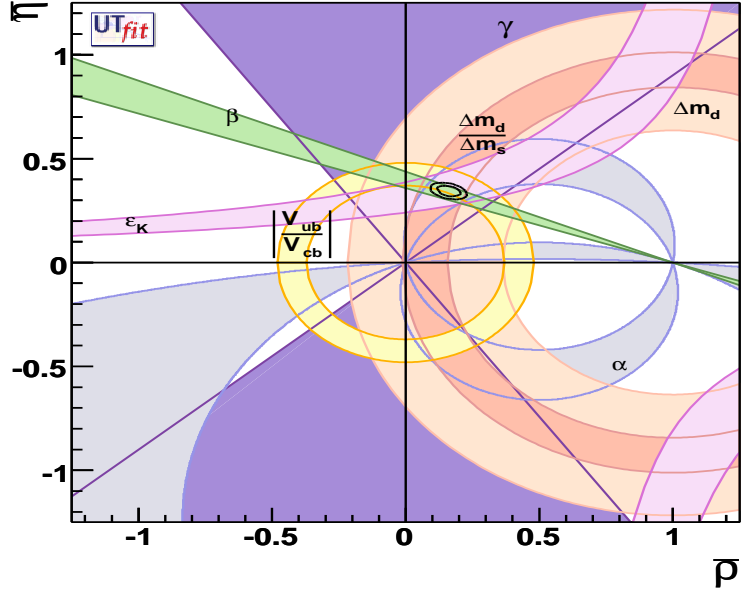}
\caption{The CKM Unitarity Triangle fit.}
\label{CKMTRIANGLEFIT} 
\end{figure}

It is indeed true that large fractions of the observed values for $\Delta M_K$, $\epsilon_K$ and 
$\Delta M_B$ and even most of $\epsilon^{\prime}$ could be due to New Physics given 
the limitations in our theoretical control over hadronic matrix elements. Equivalently constraints from 
these and other data translate into `broad' bands in plots of the unitarity triangle, see 
Fig.\ref{CKMTRIANGLEFIT}. 

The problem with this statement is that it is not even wrong -- it misses the real point. Let me illustrate 
it by a local example first. If you plot the whereabouts of the students at this school on a local map, you would find a seemingly broad band stretching from Aronsborg to Stockholm and Uppsala; however when you look at the `big' picture -- say a 
map of Europe -- you realize these students are very closely bunched together 
in one tiny spot on the map. 
This cannot be by accident, there has to be a good reason for it, which, I hope, is obvious in this specific case. Likewise for the problem at hand: Observables like 
$\Gamma (B \to l \nu X_{c,u})$, $\Gamma (K \to l \nu \pi)$, $\Delta M_K$, $\Delta M_B$, 
$\epsilon _K$ and sin$2\phi_1$ etc. represent very different dynamical regimes that proceed on 
time scales hat span several orders of magnitude.  The very fact that CKM theory 
can accommodate such diverse observables always within a factor two or better and 
relate them in such a manner that its parameters can be plotted as meaningful constraints on a 
triangle is highly nontrivial and -- in my view -- must reflect some underlying, yet unknown 
dynamical layer. Furthermore the CKM parameters exhibit an unusual hierarchical pattern -- 
$|V(ud)| \sim |V(cs)| \sim |V(tb)| \sim 1$, $|V(us)| \simeq |V(cd)| \simeq \lambda$, 
$|V(cb)| \sim |V(ts)| \sim {\cal O}(\lambda ^2)$, $|V(ub)| \sim |V(td)| \sim {\cal O}(\lambda ^3)$ -- 
as do the quark masses culminating in $m_t \simeq 175$ GeV. Picking such values for these parameters would have been seen as frivolous at best -- had they not been forced upon us by 
(independent) data. Thus I view it already as a big success for CKM theory that the experimental constraints on its parameters can be represented through triangle plots in a meaningful way. 

\begin{center}
{\bf Interlude: Singing the Praise of Hadronization}
\end{center}

\noindent 
Hadronization and nonperturbative dynamics in general are usually viewed as unwelcome 
complication, if not outright nuisances. A case in point was already mentioned: while 
I view the CKM predictions for $\Delta M_K$, $\Delta M_B$, $\epsilon_K$ to be in 
remarkable agreement with the data, significant contributions from New Physics could 
be hiding there behind the theoretical uncertainties due to lack of computational control 
over hadronization. Yet {\em without} hadronization bound states of quarks and antiquarks will not form; without 
the existence of kaons $K^0 - \bar K^0$ oscillations  obviously cannot occur. 
It is hadronization that provides the `cooling' of the (anti)quark degrees of freedom, which 
allows subtle quantum mechanical effects to add up coherently over macroscopic distances. 
Otherwise  
one would not have access to a super-tiny energy difference Im${\cal M}_{12} \sim 10^{-8}$ eV, 
which is very sensitive to different layers of dynamics, 
and indirect \cp~violation could not manifest itself. The same would hold for $B$ mesons and 
$B^0 - \bar B^0$ oscillations. 

\noindent 
To express it in a more down to earth way:  
\begin{itemize}
\item 
Hadronization leads to the formation of kaons and pions with masses exceeding 
greatly (current) quark masses.  
It is the {\em hadronic} phase space that suppresses the \cp~{\em conserving} rate for 
$K_L \to 3 \pi$ by a factor $\sim 500$, since the $K_L$ barely resides above the three pion threshold. 
\item 
It awards `patience'; i.e. one can `wait' for a pure $K_L$ beam to emerge after starting out with a 
beam consisting of $K^0$ and $\bar K^0$. 
\item 
It enables \cp~violation to emerge in the {\em existence} of a reaction, namely 
$K_L \to 2 \pi$ rather than an asymmetry; this greatly facilitates its observation. 
\end{itemize}
For these reasons alone we should praise hadronization as the hero in the tale of \cp~violation 
rather than the villain it is all too often portrayed. 

\begin{center}
{\bf End of Interlude}
\end{center}
 
Looking at the present CKM triangle fit shown in Fig. \ref{CKMTRIANGLEFIT} one realizes another 
triumph of CKM theory appears imminent: if one removed at present the constraints from $\epsilon_K$ 
and sin$2\phi_1$, i.e. \cp~constraints,  then a `flat' CKM `triangle' is barely 
compatible with the constraints on $|V(td)|$ from $\Delta M(B_d)$ and on 
$|V(ub)/V(cb)|$ from semileptonic $B$ decays -- processes not sensitive to \cp~violation per se.  However  
a measurement of 
$\Delta M(B_s)$ through resolving $B_s - \bar B_s$ oscillations in the near future would 
definitely require 
this triangle to be nontrivial: `\cp~insensitive observables would imply \cp~violation'! 

The first unequivocal manifestation of a Penguin contribution surfaced in radiative $B$ decays, 
first the exclusive channel $B \to \gamma K^*$ and subsequently the inclusive one 
$B\to \gamma X_s$. These transitions represent flavour changing neutral currents and as such 
represent a one-loop, i.e. quantum process. 

By the end of the second millenium a rich and diverse body of data on flavour dynamics had been 
accumulated, and CKM theory provided a surprisingly successful description of it. This prompted some daring spirits to perform detailed fits of the CKM triangle to infer a rather accurate prediction for 
the \cp~asymmetry in $B_d \to \psi K_S$ \cite{PARODI}: 
\beq 
{\rm sin}2\phi_1 = 0.72 \pm 0.07
\eeq

\subsection{Summary of Lecture I}
\label{SUMI}

The status of flavour dynamics in general and of CKM theory in particular in just before the turn of the 
millenium can be summarized as follows: 
\begin{itemize}
\item 
"Never underestimate Nature's ability to come up with an unexpected trick." Physicists 
had thought to have seen it all after the shock of parity violation in 1957 -- and then the 
`earthquake' of \cp~violation struck in 1964. 
\item 
"CKM theory -- all it does, it works." 
We have only an `engineering' solution for the generation of masses (the Higgs mechanism), yet 
no deeper understanding in particular of fermion masses and family replication. Yet 
CKM theory, which is based on a set of 
{\em mass related} parameters (fermion masses, CKM parameters) that any sober person 
would view as frivolous -- were they not forced upon us by data -- describes successfully a vast and very diverse array of transitions characterized by dynamical scales that a priori span several orders of magnitude. 
\item 
While no accurate CKM prediction for $\epsilon ^{\prime}/\epsilon_K$ is available now (and presumably for some time to come), it is highly nontrivial that the predictions match to data to better than an order 
of magnitude. It provides some understanding why direct \cp~violation is so feeble in kaon decays: 
it is greatly reduced by the $\Delta I =1/2$ rule -- i.e. 
$T(S=1; \Delta I =3/2)/T(S=1; \Delta I = 1/2) \simeq 1/20$ -- and the unexpectedly large top quark mass. 
\item 
The SM has to produce a host of truly large \cp~asymmetries in $B$ decays -- there is no plausible 
deniability. This is far from trivial: based on a tiny \cp~impurity in the $K^0 - \bar K^0$ system 
one predicts an almost maximal \cp~asymmetry in $B_d$ decays: 
\beq 
 {\rm few} \times 0.001 \; {\rm \cp~asymm. \; in \; } K^0 - \bar K^0 \hspace{1cm}  
\Longrightarrow \hspace{1cm} {\rm few} \times 0.1 \; {\rm \cp~asymm. \; in \; } B_d - \bar B_d \;  ; 
\eeq
i.e., an effect 
two orders of magnitude larger. 
\item 
While it is quite possible, or even likely, that New Physics will affect \cp~asymmetries in $B$ decays, 
we cannot expect it to create a numerically massive impact except for some special cases. 
\end{itemize}

\section{Lecture II:  "Flavour Dynamics 2000 - 2006" -- The `Expected' Triumph of a Peculiar Theory}
\label{LECT2}

As explained in the previous lecture, within CKM theory one is unequivocally lead to a paradigm 
of large \cp~violation in $B$ decays. This realization became so widely accepted that 
two $B$ factories 
employing $e^+e^- \to \Upsilon (4S) \to B \bar B$ were constructed -- one at KEK in Japan and one 
in Stanford in the US -- together with specialized detectors, around which two collaborations 
gathered, the BELLE and BABAR collaborations, respectively.

\subsection{Establishing the CKM Ansatz as a Theory -- \cp~Violation in $B$ Decays}
\label{ANSATZTOTHEORY}

The three angles $\phi_{1,2,3}$ in the CKM unitarity triangle 
(see Fig.\ref{CKMTRIANGLENOT}  for notation) can be determined 
through \cp~asymmetries in $B_d(t) \to \psi K_S, \pi^+\pi^-$ 
and $B_d \to K^+ \pi^-$ -- in principle. In practice the angle $\phi_3$ can be extracted from 
$B^{\pm} \to D^{\rm neut}K^{\pm}$ with better theoretical control, and $B \to 3\pi , \; 4 \pi$ 
offer various experimental advantages over $B \to 2 \pi$. 
These issues will be addressed in five acts plus two interludes.

\subsubsection{Act I: $B_d(t) \to \psi K_S$ and $\phi_1$ (a.k.a. $\beta$) }
\label{ACT1}

The first published result on the \cp~asymmetry in $B_d \to \psi K_S$ was actually obtained by 
the OPAL collaboration at LEP I \cite{OPAL}: 
\beq 
{\rm sin}2\phi_1 = 3.2 ^{+1.8}_{-2.0} \pm 0.5 \; , 
\eeq
where the `unphysical' value of sin$2\phi_1$ is made possible, since a large background subtraction 
has to be performed. 
The first value inside the physical range was obtained by CDF \cite{CDFPHI1} : 
\beq 
{\rm sin}2\phi_1 =0.79 \pm 0.44 
\eeq
In 2000 the two $B$ factory collaborations BABAR and BELLE presented their first measurements 
\cite{HFAG}: 
\beq 
{\rm sin}2\phi_1 =  
\left\{ 
\begin{array}{l} 
0.12 \pm 0.37 \pm 0.09 \; \; {\rm BABAR\;  '00} \\ 
0.45 \pm 0.44 \pm 0.09 \; \; 
{\rm BELLE\;  '00}
\end{array}  
\right. 
\eeq
Already one year later these inconclusive numbers turned into conclusive ones, and 
the first \cp~violation outside the $K^0 - \bar K^0$ complex was established: 
\beq 
{\rm sin}2\phi_1 =  
\left\{ 
\begin{array}{l} 
0.59 \pm 0.14 \pm 0.05 \; \; {\rm BABAR\; '01} \\ 
0.99 \pm 0.14 \pm 0.06 \; \; 
{\rm BELLE\;  '01}
\end{array}  
\right. 
\eeq
By 2003 the numbers from the two experiments had well converged  
\beq 
{\rm sin}2\phi_1 =  
\left\{ 
\begin{array}{l} 
0.741 \pm 0.067 \pm 0.03 \; \; {\rm BABAR\; '03} \\ 
0.733 \pm 0.057 \pm 0.028 \; \; 
{\rm BELLE\; '03}
\end{array}  
\right. 
\eeq
allowing one to state just the world averages, which is actually a BABAR/BELLE average \cite{HFAG}: 
\beq 
{\rm sin}2\phi_1 =  
\left\{ 
\begin{array}{l} 
0.726 \pm 0.037  \; \; {\rm WA\;  '04} \\ 
0.685 \pm 0.032  \; \; 
{\rm WA \; '05} \\
0.675 \pm 0.026 \; \; {\rm WA \; '06}
\end{array}  
\right. 
\label{WA040506} 
\eeq
{\bf The \cp~asymmetry in $B_d \to \psi K_S$ is there, is huge and as expected even quantitatively.}  
For CKM fits based on constraints from $|V(ub)/V(cb)|$, $B^0 - \bar B^0$ oscillations and -- as 
the only \cp~sensitive observable -- $\epsilon_K$ yield \cite{UTFIT} 
\beq 
{\rm sin}2\phi_1|_{\rm CKM}  = 0.755 \pm 0.039 \; . 
\label{PHI1CKM}
\eeq
The CKM prediction has stayed within the $\sim 0.72 - 0.75$ interval for the last several years. 
Through 2005 it has been in impressive agreement with the data. In 2006 a hint of a deviation has emerged. It is not more than that, since it is not (yet) statistically significant and furthermore depends 
very much on the value extracted for $|V(ub)/V(cb)|$ and its uncertainty. The latter might very well be underestimated, as discussed later. This is illustrated by Fig.\ref{CKMTRIANGLEFIT} showing these constraints. This figure actually obscures another impressive triumph of CKM theory: 
the \cp~{\em in}sensitive observables $|V(ub)/V(cb)|$ and $\Delta M(B_d)/\Delta M(B_s)$ -- 
i.e. observables that do {\em not} require \cp~violation for acquiring a non-zero value -- imply 
\begin{figure}[t]
\vspace{5.0cm}
\includegraphics{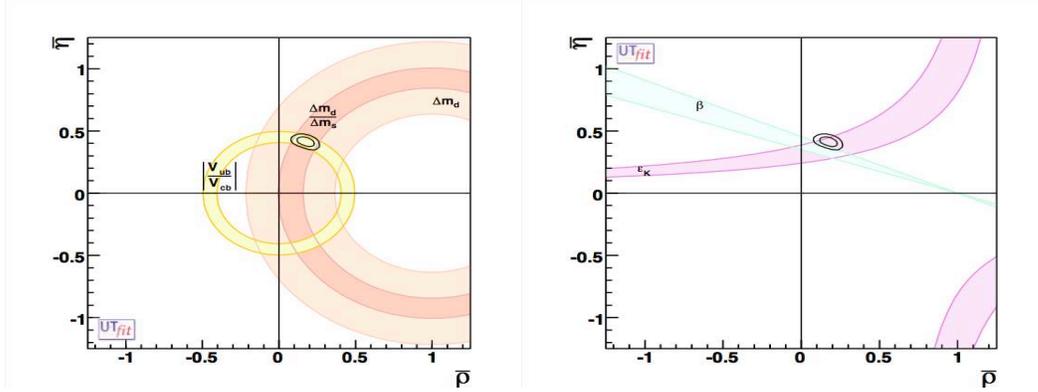}
\caption{CKM unitarity triangle from $|V(ub)/V(cb)|$ and $\Delta M(B_d)/\Delta M(B_s)$ on the left and  
       compared to constraints from 
$\epsilon_K$ and sin2$\phi_1/\beta$ on the right 
(courtesy of M. Pierini)
    \label{fig1} }
\end{figure}
\begin{itemize}
\item 
a non-flat CKM triangle and thus \cp~violation, see the left of Fig.~\ref{fig1} 
\item 
that is fully consistent with the observed \cp~sensitive observables $\epsilon_K$ 
and sin$2\phi_1$, see the right of Fig.~\ref{fig1}. 
\end{itemize}

\subsubsection{\cp~violation in $K$ and $B$ decays -- exactly the same, only different}
\label{AUSTRIAN}

\begin{figure}[t]
\vspace{8.0cm}
\includegraphics{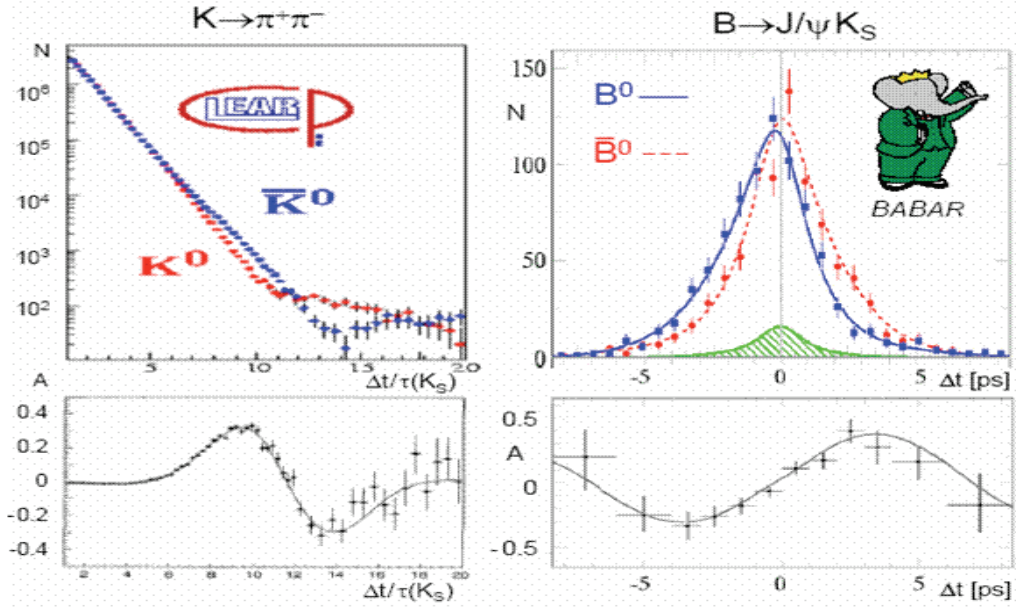}
\caption{
The observed decay time distributions for $K^0$ vs. $\bar K^0$ from CPLEAR on the left 
and for $B_d$ vs.$\bar B_d$ from BABAR on the right. 
\label{KBcompfig} }
\end{figure}

There are several similarities between $K^0 - \bar K^0$ and  
$B_d - \bar B_d$ oscillations even on the quantitative level. Their values for $x = \Delta M/\Gamma$ and 
thus for $\chi$ are very similar. It is even more intriguing that also their pattern of \cp~asymmetries in 
$K^0(t)/\bar K^0 (t) \to \pi^+\pi^-$ and $B_d(t)/\bar B_d (t) \to \psi K_S$ is very similar. Consider 
the two lower plots in Fig.\ref{KBcompfig}, which show the asymmetry directly as a function of 
$\Delta t$: it looks intriguingly similar qualitatively and even quantitatively. The lower left plot 
shows that the difference between $K^0 \to \pi^+\pi^-$ and $\bar K^0 \to \pi^+\pi^-$ is actually 
measured in units of 10 \% for $\Delta t \sim (8 - 16) \tau_{K_S}$, which is the $K_S-K_L$ interference region.  

Clearly one can find domains in $K \to \pi^+\pi^-$ that exhibit  a truly large \cp~asymmetry. 
Nevertheless it is an empirical fact that \cp~violation in $B$ decays is much larger than in 
$K$ decays. For the mass eigenstates of neutral kaons are 
very well approximated by \cp~eigenstates, as can be read off from the  upper left plot: it shows that the 
vast majority of $K \to \pi^+\pi^-$ events follow a single exponential decay law that coincides for 
$K^0$ and $\bar K^0$ transitions. This is in marked contrast to 
the $B_d \to \psi K_S$ and $\bar B_d \to \psi K_S$ transitions, which in no domain are well approximated by a single exponential law and do not coincide at all, except for 
$\Delta t =0$, as it has to be, see Sect.\ref{EPRIMPORT}.

\subsubsection{Interlude: "Praise the Gods Twice for EPR Correlations"} 
\label{EPRIMPORT} 

 The BABAR and BELLE analyses are based on a glorious application of quantum mechanics and in 
 particular EPR correlations\cite{EPR}.  The \cp~asymmetry in $B_d \to \psi K_S$ had been predicted to 
 exhibit a peculiar dependence on the time of decay, since it involves $B_d - \bar B_d$ oscillations 
 in an essential way: 
\beq 
{\rm rate} (B_d(t)[\bar B_d(t)] \to \psi K_S) \propto e^{-t/\tau _B} (1- [+] A {\rm sin}\Delta m_B t) \; , 
\label{ASYM}
\eeq  
At first it would seem that an asymmetry of the form given in Eq.(\ref{ASYM}) could not be measured for practical reasons. For in the reaction
\beq 
 e^+e^- \to \Upsilon (4S) \to B_d \bar B_d
\label{UPS4S}
\eeq
the point where the $B$ meson pair is produced is ill determined due to the finite size of the electron and positron beam spots: the latter amounts to about 1 mm in the longitudinal direction, while a $B$ meson typically travels only about a quarter of that distance before it decays. 
It would then seem that the length of the flight path of the $B$ mesons is poorly known and that averaging over this ignorance would greatly dilute or even eliminate the signal. 
 
It is here where the existence of a EPR correlation comes to the rescue. While the two $B$ mesons in the reaction of Eq.(\ref{UPS4S}) oscillate back and forth between a $B_d$ and $\bar B_d$, they change their flavour identity in a {\em completely correlated} way.  For the $B \bar B$ pair forms a \oc~{\em odd} state; Bose statistics then tells us that there cannot be two identical flavour hadrons in the final state: 
\beq 
 e^+e^- \to \Upsilon (4S) \to B_d \bar B_d \not \to B_d B_d, \; \bar B_d \bar B_d
 \label{NOTID}
 \eeq
 Once one of the $B$ mesons decays through a flavour specific mode, say $B_d \to l^+\nu X$ 
 [$\bar B_d \to l^- \bar \nu X$], then we know unequivocally that the other $B$ meson was a 
 $\bar B_d$ [$B_d$] at {\em that} time. The time evolution of $\bar B_d(t) [B_d(t)] \to \psi K_S$ as described by 
 Eq.(\ref{ASYM}) starts at {\em that} time as well; i.e., the relevant time parameter is the {\em interval between} 
 the two times of decay, not those times themselves. That time interval is related to -- and thus can be inferred from -- 
 the distance between the two decay vertices, which is well defined and can be measured.

\begin{figure}[t]
\vspace{8.0cm}
\includegraphics{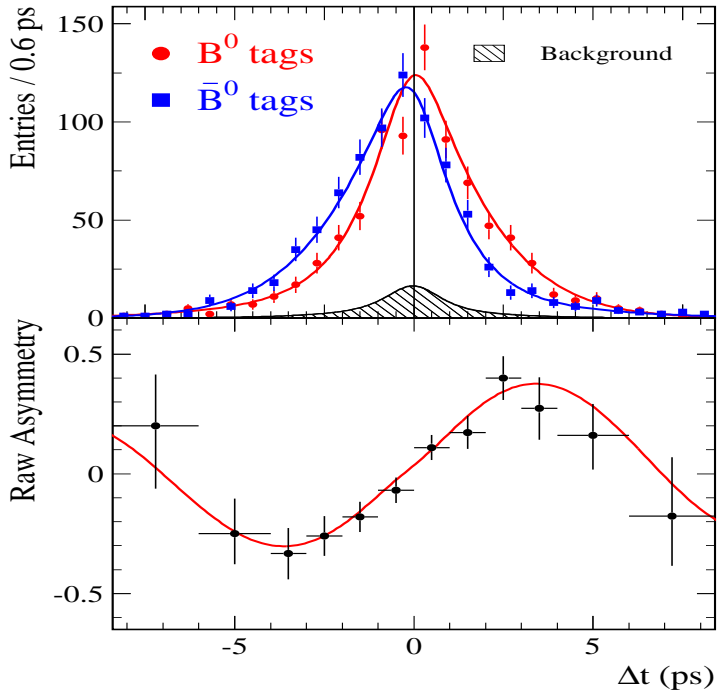}
\caption{The observed decay time distributions for $B^0$ (red) and
  $\bar B^0$ (blue) decays.}
\label{timeasymmfig} 
\end{figure}

 The great {\em practical} value of the EPR correlation is instrumental for another consideration as well, namely how to 
 see directly from the data that \cp~violation is matched by T violation. Fig.\ref{timeasymmfig} shows two distributions, one for the 
 interval  $\Delta t$ between the times of decays $B_d \to l^{+}X$ and $\bar B_d \to \psi K_S$ and the other one for the 
\cp~conjugate process $\bar B_d \to l^{-}X$ and $B_d \to \psi K_S$. They are clearly different proving that \cp~is broken. Yet they show more: the shape of the two distributions is actually the same 
(within experimental uncertainties) the only difference being that the average of $\Delta t$ is 
{\em positive} for $(l^-X)_{\bar B} (\psi K_S)$ and 
 {\em negative} for $(l^+X)_{B} (\psi K_S)$ events. I.e., there is a (slight) preference for 
 $B_d \to \psi K_S$ 
 [$\bar B_d \to \psi K_S$] to occur {\em after} [{\em before}] and thus more [less] slowly (rather than just more rarely) than $\bar B \to l^-X$ [$B \to l^+ X$]. Invoking \cpt~invariance merely for semileptonic $B$ decays -- yet not for nonleptonic 
 transitions -- synchronizes the starting point of the $B$ and $\bar B$ decay `clocks', and the 
 EPR correlation keeps them synchronized. We thus see that \cp~and 
\ot~violation are `just' 
 different sides of the same coin. 
 As explained above, EPR correlations are essential for 
 this argument! 
 
 The reader can be forgiven for feeling that this argument is of academic interest only, since 
 \cpt~invariance of all 
 processes is based on very general arguments. Yet the main point to be noted is that EPR correlations, which 
 represent some of quantum mechanics' most puzzling features, serve as an essential precision tool, which is routinely used in these measurements. I feel it is thus inappropriate to refer to EPR correlations as a paradox.

\subsubsection{Act II: $B_d(t) \to pions$ and $\phi_2$ (a.k.a. $\alpha$)}
\label{ACT2}

\subsubsubsection{$B \to 2 \pi$}
\label{2PION}

The situation is theoretically more complex than for $B_d (t) \to \psi K_S$ due to two 
reasons: 
\begin{itemize}
\item 
While both final states $\pi \pi$ and $\psi K_S$ are \cp~eigenstates, the former unlike the latter 
is not reached through an isoscalar transition. The two pions can form an $I=0$ or $I=2$ 
configuration (similar to $K\to 2\pi$), which in general will be affected differently by the strong interactions. 
\item 
For all practical purposes $B_d \to \psi K_S$ is described by two tree diagrams 
representing the two effective operators 
$(\bar c_L\gamma _{\mu}b_L)(\bar s_L\gamma ^{\mu}c_L)$ and 
$(\bar c_L\gamma _{\mu}\lambda _i b_L)(\bar s_L\gamma ^{\mu}\lambda _ic_L)$ with the 
$\lambda_i$ representing the $SU(3)_C$ matrices. Yet for $B\to \pi \pi$ we have 
effective operators 
$(\bar d_L\gamma _{\mu}\lambda _i b_L)(\bar q\gamma ^{\mu}\lambda _i q)$ generated by 
the Cabibbo suppressed Penguin loop diagrams in addition to the two tree 
operators $(\bar u_L\gamma _{\mu}b_L)(\bar d_L\gamma ^{\mu}u_L)$ and 
$(\bar u_L\gamma _{\mu}\lambda _i b_L)(\bar d_L\gamma ^{\mu}\lambda _iu_L)$. 
\end{itemize} 
This greater complexity manifests itself already in the phenomenological description of 
the time dependent \cp~asymmetry: 
\beq 
\frac{R_+(\Delta t) - R_-(\Delta t)}{R_+(\Delta t) + R_-(\Delta t)} = S {\rm sin}(\Delta M_d \Delta t) - 
C {\rm cos}(\Delta M_d \Delta t) \; , \;  S^2 + C^2 \leq 1 
\eeq
where $R_{+[-]}(\Delta t)$ denotes the rate for 
$B^{tag}(t)\bar B_d(t+\Delta)[\bar B^{tag}(t)B_d(t+\Delta)]$ and 
\beq 
S = \frac{2{\rm Im}\frac{q}{p}\bar \rho_{\pi^+\pi^-}}{1+\left|\frac{q}{p}\bar \rho_{\pi^+\pi^-}\right| ^2} \; , \; \; 
C = \frac{1 - \left|\frac{q}{p}\bar \rho_{\pi^+\pi^-}\right| ^2}
{1+\left|\frac{q}{p}\bar \rho_{\pi^+\pi^-}\right| ^2} 
\eeq
As before, due to the EPR correlation between the two neutral $B$ mesons, it is the 
{\em relative} time interval $\Delta t$ between the two $B$ decays that matters, not their 
lifetime. The new feature is that one has also a cosine dependence on $\Delta t$. 

BABAR and BELLE find 
\bea 
\label{SPIPI} 
S &=&  
\left\{ 
\begin{array}{l} 
- 0.53 \pm 0.14 \pm 0.02 \; \; {\rm BABAR\; '06} \\ 
- 0.61 \pm 0.10 \pm 0.04 \; \; 
{\rm BELLE\; '06} \\
- 0.59 \pm 0.09 \hspace{1.4cm} {\rm HFAG} 
\end{array}  
\right. \\
C &=&  
\left\{ 
\begin{array}{l} 
- 0.16 \pm 0.11 \pm 0.03 \; \; {\rm BABAR\; '06} \\ 
- 0.55 \pm 0.08 \pm 0.05 \; \; 
{\rm BELLE\; '06} \\
- 0.39 \pm 0.07 \hspace{1.4cm} {\rm HFAG}
\end{array}  
\right.
\label{CPIPI}
\eea
While BABAR and BELLE agree nicely on $S$ making the HFAG average straightforward, their 
findings on $C$ indicate different messages making the HFAG average more iffy. 

$S \neq 0$ has been established and thus \cp~violation also in this channel. While 
BELLE finds $C\neq 0$ as well, BABAR's number is still 
consistent with $C =0$.  
$C\neq 0$ obviously represents {\em direct} \cp~violation. Yet it is often overlooked that 
also $S$ can reveal such \cp~violation. For if one studies $B_d$ decays into two \cp~eigenstates 
$f_a$ and $f_b$ and finds   
\beq 
S(f_a) \neq \eta(f_a) \eta(f_b) S(f_b) 
\eeq
with $\eta_i$ denoting the \cp~parities of $f_i$, then one has established 
{\em direct} \cp~violation. For the case under study that means even if $C(\pi \pi )= 0$, 
yet $S(\pi^+\pi^-) \neq - S(\psi K_S)$, one has observed unequivocally {\em direct} 
\cp~violation. One should note that such direct \cp~violation might not necessarily induce 
$C\neq 0$. For the latter requires, as explained below in 
Sect. \ref{INT2} (see Eq.(\ref{PARTIAL})), that two different amplitudes 
contribute coherently to $B_d \to f_b$ with non-zero relative weak as well as strong phases. 
$S(f_a) \neq \eta(f_a) \eta(f_b) S(f_b)$ on the other hand only requires that the two 
overall amplitudes for $B_d \to f_a$ and $B_d \to f_b$ possess a relative phase. This can be 
illustrated with a familiar example from CKM dynamics: If there were no Penguin operators for 
$B_d \to \pi^+\pi^-$ (or it could be ignored quantitatively), one would have 
$C(\pi^+\pi^-) = 0$, yet at the same time $S(\psi K_S) = {\rm sin}(2\phi_1)$ together with 
$S(\pi^+\pi^-)  = {\rm sin}(2\phi_2) \neq - {\rm sin}(2\phi_1)$.  I.e.,  {\em without} direct CP violation one would have to 
find $C=0$ and $S = - {\rm sin}2\phi_1$ \cite{ELEFANT}. Yet since the measured value of $S$ is within one 
sigma of - sin$2\phi_1$ this distinction is mainly of academic interest at the moment.

Once the categorical issue of whether there is {\em direct} \cp~violation has been settled, one can take up the challenge of extracting a value for 
$\phi_2$ from the data 
\footnote{The complications due to the presence of the Penguin contribution are all too often referred to as `Penguin pollution'. Personally I find it quite unfair to blame our lack of theoretical control on water fowls rather than on the guilty party, namely us. }. This can be done in a model independent way by 
analyzing $B_d(t) \to \pi^+\pi^-$, $\pi^0\pi^0$ and $B^{\pm} \to \pi^{\pm}\pi^0$ transitions 
and performing an isospin decomposition. For the Penguin contribution cannot affect 
$B_d(t) \to [\pi \pi]_{I=2}$ modes. Unfortunately there is a serious experimental bottle neck, 
namely to study $B_d (t) \to \pi^0\pi^0$ with sufficient accuracy. Therefore alternative 
decays have been suggested, in particular $B \to \rho \pi$ and $\rho \rho$. 

\subsubsubsection{$B \to 3\pi /4 \pi$}
\label{34PION}

The final states in $B\to 3\pi$ and $4\pi$ are largely of the $\rho \pi$ and $\rho \rho$ form, respectively. 
For those one can also undertake an isospin decomposition to disentangle the Penguin contribution. 
These channels are less challenging {\em experimentally} than $B_{d,u} \to 2\pi$, yet they pose some complex  
{\em theoretical} problems. 

For going from the experimental starting point 
$B \to 3 \pi$ to $B \to \pi \rho$ configurations is quite nontrivial. There are other contributions to the three-pion final state like $\sigma \pi$, and cutting on the dipion mass provides a rather imperfect filter 
due to the large $\rho$ width. It hardly matters in this context whether the $\sigma$ is a bona fide resonance or some 
other dynamical enhancement. This actually leads to a further complication, namely that the $\sigma$ structure cannot be 
described adequately by a Breit-Wigner shape. As analyzed first in \cite{ANTONELLO} and then 
in more detail in \cite{ULF} ignoring such complications can induce a significant systematic uncertainty in the extracted value of $\phi_2$.

The modes $B_{d,u} \to \rho \rho$ contain even more theoretical 
complexities, since they have to be extracted from $B \to 4 \pi$ final states, where one has to 
allow for $\sigma \rho$, $2\sigma $, $\rho 2\pi$ etc. in addition to $2\rho$.

My point here is one of caution rather than of agnosticism. The concerns sketched above might 
well be more academic than practical with the present statistics. My main conclusions are the 
following: (i) I remain unpersuaded that {\em averaging} over the values for 
$\phi_2$ obtained {\em so far} from the three methods listed above provides a reliable value, since I do 
not think that the systematic uncertainties have sufficiently been analyzed. (ii) It will be 
mandatory to study those in a comprehensive way, before we  can make full use of the even larger data sets that will become available in the next few years. As I have emphasized repeatedly, our aim has 
to be to reduce the uncertainty down to at least the 5\% level in a way that can be {\em defended}. 
(iii) In the end we will need 
\begin{itemize}
\item
to perform time {\em dependent} Dalitz plot analyses (and their generalizations) and 
\item 
involve the expertise that already exists or can obtained concerning low-energy 
hadronization processes like final state interactions among low energy pions and kaons; 
valuable information can be gained on those issues from $D_{(s)} \to \pi$'s, kaons etc. as well 
as $D_{(s)} \to l \nu K\pi/\pi \pi /KK$, in particular when analyzed with state-of-the-art tools of  
chiral dynamics. 
\end{itemize}

\subsubsection{Act III, $1^{st}$ Version: $B_d \to K^+\pi^-$}
\label{ACT3}
It was pointed out in a seminal paper \cite{SONIBKPI} that (rare) transitions like 
$\bar B_d \to K^- + \pi$'s have the ingredients for sizable direct \cp~asymmetries: 
\begin{itemize}
\item 
Two different amplitudes can contribute coherently, namely the highly CKM 
suppressed tree diagram with $b \to u \bar u s$ and the Penguin diagram 
with $b \to s \bar qq$. 
\item 
The tree diagram contains a large weak phase from $V(ub)$. 
\item 
The Penguin diagram with an internal charm quark loop exhibits an imaginary part, which can be 
viewed -- at least qualitatively -- as a strong phase generated by the production and subsequent 
annihilation of a $c \bar c$ pair (the diagram with an internal $u$ quark loop acts merely as a 
subtraction point allowing a proper definition of the operator). 
\item 
While the Penguin diagram with an internal top quark loop is actually not essential, the 
corresponding effective operator can 
be calculated quite reliably, since integrating out first the top quarks and then the $W$ boson leads to a truly local operator. Determining its matrix elements is however another matter. 

\end{itemize}
To translate these features into accurate numbers represents a formidable task, we have not mastered yet. In Ref. \cite{CECILIABOOK} an early and detailed effort was made to treat $\bar B_d \to K^-\pi^+$ theoretically with the following results: 
\beq 
{\rm BR}(\bar B_d \to K^- \pi ^+) \sim 10^{-5} \; , \; \; 
A_{\cp} \sim  - 0.10
\label{BKPIPRED}
\eeq 
Those numbers turn out to be rather prescient, since they are in gratifying agreement with the data 
$$
{\rm BR}(\bar B_d \to K^- \pi ^+) = (1.85 \pm 0.11) \cdot 10^{-5} 
$$
\beq 
A_{\cp} =
\left \{  
\begin{array}{l} 
 - 0.133 \pm 0.030 \pm 0.009  \; \; \; {\rm BABAR} \\
- 0.113 \pm 0.021   \; \; \; {\rm BELLE} 
\end{array}  
\right. 
\label{BKPIDATA}
\eeq 
Cynics might point out that the authors in  \cite{CECILIABOOK} did not give a specific estimate of the 
theoretical uncertainties in Eq.(\ref{BKPIPRED}). More recent authors have been more ambitious -- 
with somewhat mixed success. I list the predictions inferred from pQCD \cite{pQCD} and 
QCD Factorization \cite{QCDFACT} and the data for the three modes 
$\bar B_d \to K^-\pi^+$ and $B^- \to K^- \pi^0, \, \bar K^0 \pi^-$: 
\beq 
A_{\cp} (B_d \to K^-\pi^+)=
\left \{  
\begin{array}{l} 
 - 0.133 \pm 0.030 \pm 0.009 \; \; \; {\rm BABAR} \\
- 0.113 \pm 0.021 \; \; \; {\rm BELLE}\\
-0.09 ^{+0.05+0.04}_{-0.08-0.06} \; \; \; {\rm pQCD} \\
+ 0.05 \pm 0.09 \; \; \; {\rm QCD\; Fact.} 
\end{array}  
\right. 
\label{K-PI+}
\eeq 
\beq 
A_{\cp} (B^- \to K^-\pi^0)=
\left \{  
\begin{array}{l} 
 + 0.06 \pm 0.06 \pm 0.01 \; \; \; {\rm BABAR} \\
+ 0.04 \pm 0.04 \pm 0.02 \; \; \; {\rm BELLE}\\
-0.01 ^{+0.03+0.03}_{-0.05-0.05}\; \; \;  {\rm pQCD}\\
+ 0.07 \pm 0.09 \; \; \; {\rm QCD\; Fact.}  
\end{array}  
\right. 
\label{K-PI0}
\eeq 
\beq 
A_{\cp} (B^- \to \bar K^0\pi^-)=
\left \{  
\begin{array}{l} 
 - 0.09 \pm 0.05 \pm 0.01  \; \; \; {\rm BABAR} \\
+ 0.05 \pm 0.05 \pm 0.01   \; \; \; {\rm BELLE}\\
+ 0.00 \; \; \; {\rm pQCD}\\
+ 0.01 \pm 0.01\; \; \; {\rm QCD\; Fact.} 
\end{array}  
\right. 
\label{K0PI-}
\eeq
As explained next the size of these asymmetries depends very much on hadronization effecs, namely 
hadronic matrix elements and strong phase shifts. While the observed asymmetry in $B_d \to K\pi$ 
with CKM expectations, we do not have an accurate predictions. 

\subsubsection{Interlude: On Final State Interactions and \cpt~Invariance}
\label{INT2}

Due to \cpt~invariance  
\cp~violation can be implemented only through a  
complex phase in some effective couplings.  For it to become  
observable two different, yet coherent amplitudes have to  
contribute to an observable. There are two types of scenarios for 
implementing this requirement: 
\begin{enumerate}
\item 
When studying a final state $f$ that can be reached by a $\Delta B = 1$ transition 
from $B^0$ as well as $\bar B^0$, then $B^0-\bar B^0$ oscillations driven by $\Delta B =2$ dynamics 
provide the second amplitude, the weight of 
which varies with time. This is what happens in $B_d \to \psi K_S$, $\pi^+\pi^-$. 
\item 
Two different $\Delta B =1$ amplitudes ${\cal M}_{a,b}$ of fixed ratio --
distinguished by,  say, their isospin content -- exist leading 
{\em coherently} to the same final state: 
\beq  
T(B \to f) = \lambda_a {\cal M}_a  + \lambda_b {\cal M}_b   
\label{T2M}
\eeq 
I have factored out the {\em weak} couplings $\lambda_{a,b}$ while allowing 
the amplitudes ${\cal M}_{a,b}$ to be still complex due to strong or electromagnetic FSI.   
For the \cp~conjugate reaction one has  
\beq  
T(\bar B \to \bar f) = \lambda_a^* {\cal M}_a + 
\lambda_b^* {\cal M}_b  
\eeq 
It is important to note that the reduced amplitudes 
${\cal M}_{a,b}$ remain 
unchanged, since strong and electromagnetic forces conserve \cp .  
Therefore we find   
\beq  
\Gamma (\bar B \to \bar f) - \Gamma (B \to  f) = 
\frac{2{\rm Im}\lambda_a \lambda_b^* \cdot {\rm Im}{\cal M}_a{\cal M}_b^*}
{|\lambda_a|^2|{\cal M}_a|^2 + |\lambda_b|^2|{\cal M}_b|^2 
+ 2{\rm Re}\lambda_a \lambda_b^* \cdot {\rm Re}{\cal M}_a{\cal M}_b^*} 
\label{PARTIAL} 
\eeq 
i.e. for a \cp~asymmetry to become observable, two  
conditions have to satisfied simultaneously irrespective of the underlying dynamics: 
\begin{itemize}
\item 
Im $\lambda_a \lambda_b^* \neq 0$, i.e.  there has to be a relative phase between the weak 
coulings $\lambda_{a,b}$. 
\item 
Im${\cal M}_a{\cal M}_b^* \neq 0$, i.e. final state interactions (FSI) have to induce a phase shift 
between ${\cal M}_{a,b}$. 
\end{itemize}

\end{enumerate} 
It is often not fully appreciated that \cpt~invariance places constraints on 
the phases of the ${\cal M}_{a,b}$. For it implies much more than equality of masses and lifetimes of particles and antiparticles. It tells us that the 
widths for {\em sub}classes of transitions for particles and 
antiparticles have to coincide already, either identically or 
at least practically. Just writing down strong phases in an equation 
like Eq.(\ref{T2M}) does {\em not automatically} satisfy \cpt~constraints.

I will illustrate this feature first with two 
simple examples and then express it in more general terms. 
\begin{itemize}
\item 
\cpt~invariance already implies $\Gamma (K^- \to \pi ^- \pi ^0) = 
\Gamma (K^+ \to \pi ^+ \pi ^0)$ up to small electromagnetic 
corrections, since in that case there are no other channels it 
can rescatter with. 
\item 
While 
$\Gamma (K^0 \to \pi^+\pi^-) \neq \Gamma (\bar K^0 \to \pi^+\pi^-)$ 
and $\Gamma (K^0 \to \pi^0\pi^0) \neq \Gamma (\bar K^0 \to \pi^0\pi^0)$ 
one has 
$\Gamma (K^0 \to \pi^+\pi^- + \pi^0\pi^0) = 
\Gamma (\bar K^0 \to \pi^+\pi^- + \pi^0\pi^0)$. 
\item 
Let us now consider a scenario where a particle $P$ and its antiparticle 
$\bar P$ can each decay into two final states only, namely $a,b$ and 
$\bar a, \bar b$, respectively 
\cite{WOLFFSI,URIFSI}. Let us further assume that strong (and
electromagnetic) forces drive transitions among $a$ and $b$ -- and 
likewise for $\bar a$ and $\bar b$ -- as described by an S matrix 
${\cal S}$. The latter can then be decomposed into two parts  
\beq 
{\cal S} = {\cal S}^{diag} + {\cal S}^{off-diag} \; , 
\eeq
where ${\cal S}^{diag}$ contains the diagonal transitions 
$a \Rightarrow a$, $b \Rightarrow b$ 
\beq 
{\cal S}^{diag}_{ss} = e^{2i\delta _s} \; , s=a,b 
\eeq
and 
${\cal S}^{off-diag}$ the off-diagonal ones 
$a \Rightarrow b$, $b \Rightarrow a$:
\beq 
{\cal S}^{off-diag}_{ab} = 
2i{\cal T}^{resc}_{ab} e^{i(\delta _a + \delta _b)} 
\eeq 
with 
\beq 
{\cal T}^{resc}_{ab} = {\cal T}^{resc}_{ba} = 
({\cal T}^{resc}_{ab})^* \; , 
\eeq
since the strong and electromagnetic forces driving the 
rescattering 
conserve \cp~and \ot . The resulting S matrix is unitary to first 
order in ${\cal T}^{resc}_{ab}$. 
\cpt~invariance implies the following relation between the 
weak decay amplitude of $\bar P$ and $P$: 
\bea 
T(P \to a) &=& e^{i\delta _a}\left[ T_a + T_b i{\cal T}^{resc}_{ab}\right] \\
T(\bar P \to \bar a) &=& e^{i\delta _a}\left[ T^*_a +T^*_bi{\cal T}^{resc}_{ab}\right]
\eea
and thus 
\beq 
\Delta \gamma (a) \equiv |T(\bar P \to \bar a)|^2 - 
|T(P \to  a)|^2 = 4 {\cal T}^{resc}_{ab} {\rm Im}T^*_a T_b \; ; 
\label{DELTAA}
\eeq
likewise
\beq 
\Delta \gamma (b) \equiv |T(\bar P \to \bar b)|^2 - 
|T(P \to  b)|^2 = 4 {\cal T}^{resc}_{ab} {\rm Im}T^*_b T_a 
\eeq
and therefore as expected 
\beq 
\Delta \gamma (b) = - \Delta \gamma (b)
\eeq
Some further features can be read off from Eq.(\ref{DELTAA}):
\begin{enumerate} 
\item 
If the 
two channels that rescatter have comparable widths -- 
$\Gamma (P \to a) \sim \Gamma (P \to b)$ -- one would like the
rescattering 
$b \leftrightarrow a$ to proceed via the usual strong forces; for
otherwise  the asymmetry $\Delta \Gamma $ is suppressed relative to these 
widths by the
electromagnetic coupling. 
\item 
If on the other hand the channels 
command very different widths -- say 
$\Gamma (P \to a) \gg \Gamma (P \to b)$ -- then a large {\em relative} 
asymmetry in $P \to b$ is accompagnied by a tiny one in 
$P \to a$. 
\end{enumerate} 
This simple scenario can easily be extended to two sets $A$ and $B$ of 
final states s.t. for all states $a$ in set $A$ the transition 
amplitudes have the same weak coupling and likewise for states 
$b$ in set $B$. One then finds 
\beq 
\Delta \gamma (a) = 4 \sum _{b\, \in \, B}{\cal T}^{resc}_{ab}{\rm Im}T_a^*T_b 
\eeq
The sum over all CP asymmetries for states $a \, \in \, A$ cancels the 
correponding sum over $b\, \in \, B$: 
\beq 
\sum _{a\, \in \, A} \Delta \gamma (a) 
= 4 \sum _{b\, \in \, B}{\cal T}^{resc}_{ab}{\rm Im}T_a^*T_b = 
- \sum _{b\, \in \, B} \Delta \gamma (b)
\eeq
\end{itemize}
These considerations tell us that the \cp~asymmetry averaged over certain 
classes of channels defined by their quantum numbers has to 
vanish. Yet these channels can still be very heterogenous, 
namely consisting of two- and quasi-two-body modes, 
three-body channels and other multi-body decays. 
Hence we can conclude: 
\begin{itemize}
\item 
If one finds a direct \cp~asymmetry in one channel, 
one can infer -- based on rather general grounds -- which other channels
have to exhibit the compensating  asymmetry as required by \cpt~invariance.
Observing them would enhance  the significance of the measurements very
considerably.  
\item 
Typically there can be several classes of rescattering channels. 
The SM weak dynamics select a subclass of those where the 
compensating asymmetries have to emerge. QCD frameworks like 
generalized factorization can be invoked to estimate the 
relative weight of the asymmetries in the different classes. 
Analyzing them can teach us important lessons about the 
inner workings of QCD. 
\item 
If New Physics generates the required weak phases (or at least 
contributes significantly to them), it can induce 
rescattering with novel classes of channels. The pattern in the 
compensating asymmetries then can tell us something about the 
features of the New Physics 
involved.  
\end{itemize}

I want to end this Interlude by adding that Penguins are rather smart beings: they know about these \cpt~constraints. For when one considers the imaginary parts of the Penguin diagrams, 
which are obtained by cutting the internal quark lines, namely the up and charm quarks (top quarks do not contribute there, since they cannot reach their mass shell in $b$ decays), one realizes that 
\cp~asymmetries in $B \to K +\pi$'s are compensated by those in $B\to D \bar D_s + \pi$'s.

\subsubsection{Act III, $2^{nd}$ Version: 
$\phi_3$  from $B^+ \to D_{neut}K^+$ vs. $B^- \to D_{neut} K^-$}
\label{DNEUTK}

As first mentioned in 1980 \cite{CS80}, then explained in more detail in 1985 \cite{BS85} 
and further developed in \cite{GRONWYL},  
the modes $B^{\pm} \to D_{neut}K^{\pm}$ should exhibit direct \cp~violation driven by the 
angle $\phi_3$, if the neutral $D$ mesons decay to final states that are {\em common} to 
$D^0$ and $\bar D^0$. Based on simplicity the original idea was to rely on two-body modes like 
$K_S\pi^0$, $K^+K^-$, $\pi^+\pi^-$, $K^{\pm}\pi^{\mp}$. One drawback of that method are the small 
branching ratios and low efficiencies. 

A new method was pioneered by BELLE and then implemented also by BABAR, namely to employ 
$D_{neut} \to K_S \pi^+\pi^-$ and perform a full Dalitz plot analysis. This requires a very 
considerable analysis effort -- yet once this initial investment has been made, it will pay handsome profit in the long run. For obtaining at least a decent description of the full Dalitz plot population 
provides  considerable cross checks concerning systematic uncertainties and thus a high degree of 
confidence in the results. BELLE and BABAR find:  
\beq 
\phi_3 = 
\left\{
\begin{array}{ll} 53^o  \pm 18^o (stat) \pm 3^o(syst) \pm 9^o ({\rm model}) & {\rm BELLE}\\
92^o  \pm 41^o (stat) \pm 11^o(syst) \pm 12^o ({\rm model}) & {\rm BABAR}
\end{array}
\right.
\eeq 
At present these studies are severely statistics limited; one should also note that with more statistics 
one will be able to reduce in particular the model dependence. I view this method as the best one to 
extract a reliable value for $\phi_3$, where the error estimate can be defended due to the many constraints inherent in a Dalitz plot analysis. It exemplifies how the complexities of 
hadronization can be harnessed to establish confidence in the accuracy of our results. 

\subsubsection{Act IV: $\phi_1$ from \cp~Violation in $B_d \to$ 3 Kaons -- Snatching Victory from the Jaws of Defeat or Defeat from the Jaws of Victory}
\label{KAONS}

Analysing \cp~violation in $B_d \to \phi K_S$ decays is a most promising way to search 
for New Physics. For the underlying quark-level transition $b \to s \bar s s$ represents a pure 
loop-effect in the SM, it is described by a {\em single} $\Delta B=1$\& $\Delta I=0$ operator (a `Penguin'), a reliable 
SM prediction exists for it \cite{GROSS} -- 
sin$2\phi_1(B_d \to \psi K_S) \simeq {\rm sin}2\phi_1(B_d \to \phi K_S)$ -- 
and the $\phi$ meson represents a {\em narrow} resonance.  

Great excitement was created when BELLE reported a large discrepancy between the predicted and 
observed \cp~asymmetry in $B_d \to \phi K_S$ in the summer of 2003:  
\beq 
{\rm sin}2\phi_1(B_d \to \phi K_S) = 
\left\{
\begin{array}{ll} - 0.96 \pm 0.5 \pm 0.10 & {\rm BELLE ('03)}\\
0.45 \pm 0.43 \pm 0.07 & {\rm BABAR ('03)}
\end{array}
\right. \;  ; 
\eeq
Based on more data taken, this discrepancy has 
shrunk considerably: the BABAR/BELLE average for 2005 yields \cite{LANCERI}
\beq 
{\rm sin}2\phi_1(B_d \to \psi K_S) = 0.685 \pm 0.032
\eeq
versus  
\beq 
{\rm sin}2\phi_1(B_d \to \phi K_S) = 
\left\{
\begin{array}{ll} 0.44 \pm 0.27 \pm 0.05 & {\rm BELLE ('05)} \\
0.50 \pm 0.25 ^{+0.07}_{-0.04} & {\rm BABAR ('05)} 
\end{array}
\right. \;  ; 
\eeq
while the 2006 values read as follows: 
\beq 
{\rm sin}2\phi_1(B_d \to \psi K_S) = 0.675 \pm 0.026
\label{PHI06}
\eeq
compared to 
\beq 
{\rm sin}2\phi_1(B_d \to \phi K_S) = 
\left\{
\begin{array}{ll} 0.50 \pm 0.21 \pm 0.06 & {\rm BELLE ('06)}\\
0.12 \pm 0.31 \pm 0.10 & {\rm BABAR ('06)}\\
0.39 \pm 0.18 & {\rm HFAG ('06)}
\end{array}
\right. \;  ; 
\eeq
I summarize the situation as follows: 
\begin{itemize}
\item 
Performing dedicated \cp~studies in channels driven mainly or even predominantly by 
$b \to s q \bar q$ to search for New Physics signatures makes eminent sense since the SM 
contribution, in particular from the one-loop Penguin operator, is greatly suppressed. 
\item 
The experimental situation is far from settled, as can be seen also from how the central 
value have moved over the years. It is tantalizing to see that the $S$ contribution for all the 
modes in this category -- $B_d \to \pi^0 K_S, \rho^0 K_S, \omega K_S, f_0 K_S$ -- are all low 
compared to the SM expectation Eq.(\ref{PHI06}). Yet none of them is significantly lower; 
for none of these modes a non-zero \cp~asymmetry has been established except for 
\beq 
{\rm sin}2\phi_1(B_d \to \eta ^{\prime} K_S) = 
\left\{
\begin{array}{ll} 0.64 \pm 0.10 \pm 0.04 & {\rm BELLE ('06)}\\
0.58 \pm 0.10 \pm 0.03 & {\rm BABAR ('06)}\\
0.61 \pm 0.07 & {\rm HFAG ('06)}
\end{array}
\right. \;  ; 
\label{CPETAKS}
\eeq
\item 
Obviously there is considerable space still for significant deviations from SM predictions. 
It 
is ironic that such a smaller deviation, although not significant, is actually more believable as 
signaling an incompleteness of the SM than 
the large one originally reported by BELLE. 
While it is tempting to average over all these hadronic transitions, I would firmly resist this temptation 
for the time being, till several modes exhibit a significant asymmetry. 
\item 
One complication has to be studied, though, in particular if the observed 
value of sin$2\phi_1(B_d \to \phi K_S)$ falls below the predicted one by a moderate amount only. 
For one is actually observing $B_d \to K^+K^-K_S$. If there is a single weak phase like in the SM one finds 
\beq 
{\rm sin}2\phi_1(B_d \to \phi K_S) = - {\rm sin}2\phi_1(B_d \to `f_0(980)' K_S) \; , 
\eeq 
where $`f_0(980)'$ denotes any {\em scalar} $K^+K^-$ configuration with a mass close to that of the 
$\phi$, be it a resonance or not. A smallish pollution by such a $`f_0(980)' K_S$ -- by, say, 
10\% {\em in amplitude} --  
can thus reduce the asymmetry assigned to $B_d \to \phi K_S$ significantly -- by 20\% in this example. 
\item 
In the end it is therefore mandatory to perform a {\em full time dependent Dalitz plot analysis} 
for $B_d \to K^+K^-K_S$ and compare it with that for $B_d \to 3 K_S$ and 
$B^+ \to K^+K^-K^+, \, K^+K_SK_S$ and also with $D \to 3K$. BABAR has presented a preliminary such 
study. This is a very challenging task, but 
in my view essential. There is no `royal' way to fundamental insights. 
\footnote{The ruler of a Greek city in southern Italy once approached the resident sage with the request 
to be educated in mathematics, but in a `royal way', since he was very busy with many 
obligations. Whereupon the sage replied with admirable candor: 
"There is no royal way to mathematics."} 
\item 
An important intermediate step in this direction is given by one application of 
{\em Bianco's Razor} \cite{RIO}, namely to analyze the \cp~asymmetry in $B_d \to [K^+K^-]_MK_S$ as a 
function of the cut $M$ on the $K^+K^-$ mass. 
\end{itemize}
All of this might well lead to another triumph of the SM, when its predictions agree with accurate data in the future even for these rare transition rates dominated by loop-contributions, i.e pure quantum effects.  
It is equally possible -- personally I think it actually more likely -- that future precision data will expose 
New Physics contributions. In that sense the SM might snatch victory from the jaws of defeat -- or defeat from the jaws of victory. For us that are seeking indirect manifestations of New Physics it is the other way around.

In any case the issue has to be pursued with vigour, since these reactions provide such a natural portal to New Physics on one hand and possess such an accurate yardstick from 
$B_d \to \psi K_S$.

\subsubsection{The Beginning of Act V -- \cp~Violation in Charged $B$ Decays}
\label{CPCHARGED}

So far \cp~violation has not been established yet in the decays of {\em charged} mesons, which is not 
surprising, since meson-antimeson oscillations cannot occur there and it has to be purely {\em direct} 
\cp~violation. Now BELLE \cite{CHARGEDCPV} has found strong evidence for a large \cp~asymmetry in charged $B$ decays 
with a 3.9 sigma significance, namely in $B^{\pm} \to K^{\pm}\rho^0$ observed in 
$B^{\pm} \to K^{\pm}\pi^{\pm}\pi^{\mp}$ : 
\beq 
A_{\cp} (B^{\pm} \to K^{\pm}\rho^0) = \left( 30 \pm 11 \pm 2.0 ^{+11}_{-4} \right) \% 
\label{CPCHARGED}
\eeq 
I find it a most intriguing signal since a more detailed inspection of the mass peak region shows 
a pattern as expected for a genune effect. 
Furthermore a similar signal is seen in BABAR's data, and it would make sense to make to undertake a careful 
average over the two data sets. 

I view BELLE's and BABAR's analyses of the Dalitz plot for $B^{\pm} \to K^{\pm}\pi^{\pm}\pi^{\mp}$ 
as important pilot studies, from which one can infer important lessons about the strengths and pitfalls 
of such studies in general.

\subsection{Loop Induced Rare $B_{u,d}$ Transitions}
\label{LOOPDEC}

Processes that require a loop diagram to proceed -- i.e. are classically forbidden -- provide 
a particularly intriguing stage to probe fundamental dynamics. 

It marked a tremendous  success for the SM, when radiative $B$ decays were measured, first in the 
exclusive mode $B \to \gamma K^*$ and subsequently also inclusively: $B \to \gamma X$. 
Both the rate and the photon spectrum are in remarkable agreement with SM prediction; they have 
been harnessed to extracting heavy quark parameters, as explained below. 

More recently the next, i.e. even rarer level has been reached with transitions to final states 
containing a pair of charged leptons:   
\beq 
{\rm BR}(B \to l^+l^- X) = 
\left\{
\begin{array}{ll} (6.2 \pm 1.1\pm 1.5)\cdot 10^{-6} & {\rm BABAR/BELLE}\\
(4.7 \pm 0.7)\cdot 10^{-6} & {\rm SM}
\end{array}
\right. \;   . 
\eeq
Again the data are consistent with the SM prediction \cite{ALIETAL1}, yet the present experimental uncertainties 
are very sizable. We are just at the beginning of studying $B \to l^+l^-X$, and it has to be 
pursued in a dedicated and comprehensive manner as discussed in Lect. III. 

The analogous decays with a $\nu \bar \nu$ instead of the $l^+l^-$ pair is irresistibly attractive to theorists -- although quite resistibly so to experimentalists:
\beq 
{\rm BR}(B \to \nu \bar \nu X) 
\left\{
\begin{array}{ll} \leq 7.7 \cdot 10^{-4} & {\rm ALEPH}\\
= 3.5 \cdot 10^{-5} & {\rm SM}
\end{array}
\right. \;   ; 
\eeq
\beq 
{\rm BR}(B \to \nu \bar \nu K) 
\left\{
\begin{array}{ll} \leq 7.0 \cdot 10^{-5} & {\rm BABAR}\\
= (3.8^{+1.2}_{-0.6}) \cdot 10^{-6} & {\rm SM} 
\end{array}
\right. \;   , 
\eeq
where the SM predictions are taken from Refs.\cite{BUBU} and  \cite{BUHIISI}, respectively.

\subsection{Other Rare Decays}
\label{OTHERRARE}

There are some relatively rare $B$ decays that could conceivably reveal New Physics, although they 
proceed already on the tree level. Semileptonic decays involving $\tau$ leptons are one example and 
will be discussed in Lect. III. The most topical example is $B^+ \to \tau \nu$ which has been pursued 
vigorously since it provides information on the decay constant $f_B$ and is sensitive to contributions from charged Higgs fields. A first signal  has been found by BELLE with a 3.5 sigma significance: 
\beq 
{\rm BR} (B^- \to \tau ^- \bar \nu) = \left( 1.79 ^{+0.56\; +0.46} _{- 0.49\; -0.51}\right) \cdot 10^{-4}
\eeq
Hence one extracts 
\beq 
f_B |V(ub)| = \left(  10.1 ^{+1.6\; +1.3}_{-1.4\; - 1.4}\right) \cdot 10^{-4} \; {\rm GeV}
\eeq

\subsection{Adding High Accuracy to High Sensitivity}
\label{ADDING}

As mentioned before and addressed in more detail in Lect. III, we can{\em not count} on a 
{\em numerically} massive impact of New Physics in heavy flavour transitions. 
Therefore it no longer suffices to rely on the high sensitivity that loop processes like $B^0 - \bar B^0$ oscillations or radiative $B$ decays possess to New Physics; we have to strive also for high accuracy. 

The spectacular success of the $B$ factories and the emerging successes of CDF and D0 to obtain 
high quality data on beauty transitions in a hadronic environment give us confidence that even greater 
experimental precision can be achieved in the future. However this would be of little help if it could 
not be matched by a decrease in the theoretical uncertainties. I will describe now why I think that theory will be able to hold up its side of the bargain as well and what the required elements for such an undertaking have to be. 

The question is: "Can we answer the challenge of $\sim \%$ accuracy?" One guiding principle 
will be in Lenin's concise words:  
\begin{center}
`Trust is good -- control is better!'
\end{center} 

 \begin{table}
\begin{center}
\begin{tabular}{lll} 
\hline
Heavy Quark Parameter      & value as of 2005 & relative uncertainty    \\
\hline
      $m_b$(1 GeV)   & $ = (4.59 \pm 0.04)$ GeV & $\hat = \; \; 1.0\, \%$ \\ 
      $m_c$(1 GeV)   & = $(1.14 \pm 0.06)$ GeV & $\hat = \; \; 5.3\, \%$ \\ 
      $m_b$(1 GeV) $-0.67 m_c$(1GeV)  & = $(3.82 \pm 0.017)$ GeV & $\hat = \; \; 0.5\, \%$ \\    
      $|V(cb)|$ & = $(41.58 \pm 0.67) \cdot 10^{-3}$ & $\hat = \; \; 1.6\, \%$ \\
    \hline
    $|V(us)|_{KTeV}$ & = $0.2252 \pm 0.0022$ & $\hat = \; \; 1.1\, \%$ \\
    \hline 
\end{tabular}  
\caption{The 2005 values \cite{FLAECHER} 
of $b$ and $c$ quark masses and of $|V(cb)|$ compared to the Cabibbo angle}
\label{tab:STATUS05} 
\end{center} 
\end{table}

Table \ref{tab:STATUS05} provides a sketch of the theoretical control we have achieved over 
some aspects of $B$ decays. I hope it will excite the curiosity of the reader and fortify her/him to 
read the following more technical discussion; let me add that one can skip this Sect.\ref{ADDING} can 
be at the first reading and continue with Sect.\ref{SUMII}.

\subsubsection{Heavy Quark Theory}
\label{HQTH}

While QCD is the only candidate among {\em local} quantum field theories to describe the 
strong interactions, as explained in Lecture I in Sects. \ref{QCD} \& \ref{SU(2)}, 
$SU(2)_L \times U(1)$ is merely the minimal theory for the electroweak 
forces. Obtaining reliable information about the latter is, however, limited by our lack of full 
calculational control over the former.  

It had been 
conjectured for more than thirty years that 
the theoretical treatment of heavy flavour hadrons should be facilitated, when the 
heavy quark mass greatly exceeds greatly the nonperturbative scale of QCD 
\footnote{A striking prediction has been that super-heavy top quarks -- i.e. with 
$m_t \geq 150$ GeV -- would decay, {\em before} they could hadronize \cite{RAPALLO} 
thus bringing top quarks under full theoretical control. For then the decay width of top quarks is of order 
1 GeV and provides an infrared cutoff for QCD corrections.  
This feature comes with a price, though, in so far 
as \cp~studies are concerned: without hadronization as a `cooling' mechanism, the degree 
of coherence between different transition amplitudes -- a necessary condition for \cp~violation to become observable -- will be rather tiny.}: 
\beq 
m_Q \gg \Lambda _{QCD} \; . 
\eeq
This conjecture has been transformed into a reliable theoretical framework only in the last 
fifteen years, as far as beauty hadrons are concerned. I refer to it as Heavy Quark Theory (mentioned already in Sect. \ref{TECH}); comprehensive reviews with references to the original 
literature can be found in Refs. \cite{HQREV} and 
\cite{URALTSEV}. Its goal is to treat nonperturbative dynamics {\em quantitatively}, as it affects heavy flavour {\em hadrons}, in full conformity with QCD and without model assumptions. It has achieved this goal already for several classes of 
beauty meson transitions with a reliability and accuracy that before would 
have seemed unattainable.

Heavy Quark Theory is based on a two-part strategy analogous to the one adopted in 
chiral perturbation theory -- another theoretical technology to deal reliably with nonperturbative dynamics 
in a special setting. Like there Heavy Quark Theory combines two basic elements, namely an 
{\em asymptotic symmetry principle} and a {\em dynamical treatment}  
telling us how the asymptotic limit is approached: 
\begin{enumerate}
\item 
The symmetry principle is {\em Heavy Quark Symmetry} 
stating that all sufficiently heavy quarks behave identically 
under the strong interactions without sensitivity to their spin. 
This can easily be illustrated with the Pauli Hamiltonian describing the 
interaction of a quark of mass $m_Q$ with a gauge field $A_{\mu}= (A_0, \vec A)$: 
\beq 
{\cal H}_{\rm Pauli} = - A_0 + \frac{(i\vec \partial - \vec A)^2}{2m_Q} + 
\frac{\vec \sigma \cdot \vec B}{2m_Q} \Longrightarrow - A_0 
\; \; {\rm as} \; \; m_Q \to \infty \; ; 
\label{PAULI} 
\eeq 
i.e., in the infinite mass limit, quarks act like {\em static} objects 
{\em without} spin dynamics and subject only to colour Coulomb fields. 

This simple consideration illustrates a general feature of heavy quark theory, namely that 
the spin of the heavy quark $Q$ decouples from the dynamics in the heavy quark limit. 
Hadrons $H_Q$ can therefore be labeled by the angular momentum 
$j_q$ carried by its `light' components 
-- light valence quarks, gluons and sea quarks --  in addition to its total spin $S$. 
The S wave pseudoscalar and vector mesons  -- $B$ \& $B^*$ and $D$ \& $D^*$ -- then form the ground state doublet of heavy quark 
symmetry with $[S,j_q] = [0,\frac{1}{2}], [1,\frac{1}{2}]$; a quartet of P wave configurations 
form the first excited states with $[S,j_q] = [0,\frac{1}{2}], [1,\frac{1}{2}], [1,\frac{3}{2}], [2,\frac{3}{2}]$. 

Heavy quark symmetry can be understood 
in an intuitive way: consider a hadron $H_Q$ containing a 
heavy quark $Q$ with mass $m_Q \gg \Lambda _{QCD}$ surrounded 
by a "cloud" of light degrees of freedom carrying quantum 
numbers of an antiquark $\bar q$ or diquark $qq$ 
\footnote{This cloud is often referred to -- 
somewhat disrespectfully -- as `brown muck', a phrase coined by the late Nathan Isgur.}. 
This cloud has 
a rather complex structure: in addition to $\bar q$ 
(for mesons) or $qq$ (for baryons) it 
contains an indefinite number of $q \bar q$ pairs and gluons 
that are strongly coupled to and constantly 
fluctuate into each other. There is, however, 
one thing we know: since typical frequencies 
of these fluctuations are $\sim {\cal O}({\rm few}) \times 
\Lambda _{QCD}$, the normally dominant {\em soft} dynamics 
allow the heavy quark to exchange momenta of order few times 
$\Lambda _{QCD}$ only with its surrounding medium. 
$Q \bar Q$ pairs then cannot play a significant role, 
and the heavy quark can be treated as a quantum mechanical 
object rather than a field theoretic entity 
requiring second quantization. This provides a 
tremendous computational simplification even while 
maintaining a field theoretic description for the light 
degrees of freedom. Furthermore techniques developed 
long ago in QED can profitably be adapted here. 
\item 
We can go further and describe the interactions between 
$Q$ and its surrounding light degrees of freedom through 
an expansion in powers of $1/m_Q$ -- the Heavy 
Quark Expansion (HQE). This allows us to 
analyze {\em pre}-asymptotic effects, i.e. effects 
that fade away like powers of $1/m_Q$ as 
$m_Q \ra \infty$.
\end{enumerate}   
Let me anticipate the lessons we have learnt: we have 
\begin{itemize}
\item 
identified the sources of the non-perturbative corrections;  
\item 
found them to be smaller than they could have been; 
\item 
succeeded in relating the basic quantities of the Heavy Quark 
Theory -- KM paramters, masses and kinetic energy of 
heavy quarks, 
etc. -- to various a priori independant observables with a considerable  
amount of  redundancy; 
\item 
developed a better understanding of incorporating perturbative 
and nonperturbative corrections without double-counting.  
\end{itemize} 
In the following I will sketch the concepts 
on which the Heavy Quark Expansions are based, the 
techniques employed, the results obtained and 
the problems encountered. It will not constitute 
a self-sufficient introduction into this vast 
and ever expanding field. My intent is to provide a guide through the literature 
for the committed student. 

\subsubsection{H(eavy) Q(uark) E(xpansions), Fundamentals}
\label{HQEXP} 

In describing weak decays of heavy flavour {\em hadrons} 
one has to incorporate perturbative as well as 
nonperturbative contributions in a self-consistent 
and complete  
way. The only known way to 
tackle such a task invokes the 
{\em Operator Product Expansion a la Wilson} 
involving an {\em effective} Lagrangian. Further 
conceptual insights as well as practical results can be 
gained by analysing {\em sum rules}; in particular they 
shed light on various aspects and formulations of 
{\em quark-hadron duality}. 

\subsubsubsection{Operator Product Expansion (OPE) 
for Inclusive Weak Decays}
\label{OPE}

Similar to the well-known case of 
$\sigma (e^+ e^- \to had)$ one invokes the optical 
theorem to describe the decay into a sufficiently 
{\em inclusive} final state $f$ through the 
imaginary part of the forward scattering operator 
evaluated to second order in the weak interactions  
\beq 
\hat T( Q \to Q) = 
{\rm Im}  \int d^4x 
\, i\{ {\cal L}_W(x) {\cal L}_W^{\dagger}(0) \} _T
\label{TRANSOP}
\eeq 
with the subscript $T$ denoting the time-ordered 
product and ${\cal L}_W$ the relevant weak Lagrangian  
\footnote{There are two qualitative differences to the 
case of $e^+ e^- \to had$: in describing weak decays 
of a hadron $H_Q$ (i) one employs the weak rather than the 
electromagnetic Lagrangian, and (ii) one takes the 
expectation value between the $H_Q$ state rather than the hadronic 
vacuum.}.   
The expression in Eq.(\ref{TRANSOP}) represents in general 
a non-local operator with the space-time separation 
$x$ being fixed by the inverse of the {\em energy release}. If 
the latter is large compared to typical hadronic 
scales, then the product is dominated by short-distance 
physics, and one can apply a Wilsonian OPE, which yields an infinite 
series of {\em local} operators of increasing dimension 
\footnote{
I will formulate the expansion in powers of 
$1/m_Q$, although it has to be kept in mind that it is 
really controlled by the inverse of the {\em energy release}.  
While there is no fundamental difference between 
the two for $b \to c/u l \bar \nu$ or 
$b \to c/u \bar ud$, since $m_b, \, 
m_b - m_{c,u} \gg  
\Lambda _{QCD}$, the expansion becomes of 
somewhat dubious reliability for $b \to c \bar cs$. 
It actually would break down for a scenario 
$Q_2 \to Q_1 l \bar \nu$ with $m_{Q_2} \simeq 
m_{Q_1}$ -- in contrast to HQET! 
}. 
The width for the decay of a hadron $H_Q$ containing 
$Q$ is then obtained by taking the $H_Q$ expectation value 
of the operator $\hat T$:  
$$ 
\frac{\matel{H_Q}{{\rm Im}\hat T(Q \to f \to Q)}{H_Q}}
{2M_{H_Q}} \propto 
\Gamma (H_Q \to f) = \frac{G_F^2m_Q^5(\mu )}{192 \pi ^3}
|V_{CKM}|^2 \cdot 
$$
$$
\cdot \left[ c_3^{(f)}(\mu )
\frac{\matel{H_Q}{\bar QQ}{H_Q}_{(\mu )}}{2M_{H_Q}} + 
\frac{c_5^{(f)}(\mu )}{m_Q^2}
\frac{\matel{H_Q}{\bar Q\frac{i}{2}\sigma \cdot GQ}
{H_Q}_{(\mu )}}{2M_{H_Q}} + \right. 
$$
\beq
\left. + \sum _i \frac{c_{6,i}^{(f)}(\mu )}{m_Q^3} 
\cdot \frac{
\matel{H_Q}{(\bar Q\Gamma _iq)(\bar q\Gamma _iQ)}
{H_Q}_{(\mu )}}{2M_{H_Q}} + {\cal O}(1/m_Q^4) 
\right] 
\label{MASTER}
\eeq  
Eq.(\ref{MASTER}) exhibits the following important features:
\begin{itemize}
\item 
An {\em auxiliary} scale $\mu$ has been introduced to consistently separate 
short and long distance dynamics: 
\beq 
{\rm short \; distance} \; \; < \; \; 
\mu ^{-1} \; \; < \; \; 
{\rm long \; distance} 
\eeq 
with the former entering through the coefficients and 
the latter through the effective operators; their 
matrix elements will thus depend on $\mu$. 

{\em In principle} the value of $\mu$ does not 
matter: it reflects merely our computational procedure 
rather than how nature goes about its business. The 
$\mu$ dependance of the coefficients thus has to cancel 
against that of the corresponding matrix elements. 

{\em In practise} however there are competing 
demands on the choice of $\mu$: 
\begin{itemize}
\item 
On one hand one has to choose 
\beq 
\mu \gg \Lambda _{QCD} \; ; 
\label{MULARGE} 
\eeq 
otherwise radiative corrections cannot be treated 
within {\em perturbative} QCD. 
\item 
On the other hand many computational techniques 
for evaluating {\em matrix elements}  
-- among them the Heavy Quark Expansions -- 
require 
\beq 
\mu \ll m_b 
\label{MUSMALL} 
\eeq 
\end{itemize} 
The choice 
\beq 
\mu \sim 1\; {\rm GeV}
\eeq
satisfies both of these requirements. It is important to check that the obtained 
numerical results do not exhibit a significant sensitivity to the exact value of 
$\mu$ when varying the latter in a reasonable range. 
\item 
{\em Short-distance} dynamics 
shape the c number coefficients $c_i^{(f)}$. 
{\em In practise} they are evaluated in 
{\em perturbative} QCD. It is 
quite conceivable, though, that also {\em nonperturbative}  
contributions arise there; yet they are believed to be 
fairly small in beauty decays \cite{CHIBISOV}. 

By the same token these short-distance coefficients provide also the portals, 
through which New Physics can enter in a natural way. 
\item  
Nonperturbative contributions on 
the other hand enter through the {\em expectation values} 
of operators of dimension higher than three -- 
$\bar Q\frac{i}{2}\sigma \cdot GQ$ etc. -- and higher order 
corrections to the expectation 
value of the leading operator   
$\bar QQ$, see below. 
\item 
In practice we cannot go beyond evaluating the first few terms in this expansions. 
More specifically we are limited to contributions through order 
$1/m_Q^3$; those are described in terms of six heavy quark parameters, namely 
two quark masses -- $m_{b,c}$ --, two expectation values of dimension-five 
operators -- $\mu_{\pi}^2$ and $\mu_G^2$ -- and of dimension-six operators -- 
the Darwin and `LS' terms, $\rho^3_D$ and $\rho^3_{LS}$, respectively 
\footnote{For simplicity I ignore here socalled `Intrinsic Charm' contributions, see 
\cite{IC}.}.  
\begin{itemize}
\item
This small and universal set of nonperturbative quantities describes a host of 
observables in $B$ transitions. Therefore their values can be determined from some of these 
observables and still leave a large number of predictions. 
\item 
It opens the door to a novel symbiosis of different theoretical technologies for heavy flavour 
dynamics -- in particular between HQE and lattice QCD. For the HQP can be inferred from lattice 
studies. This enhances the power of and confidence in both technologies by 
\begin{itemize}
\item 
increasing the range of applications and 
\item 
providing more validation points. 

\end{itemize}
I will give some examples later on.

\end{itemize} 
\item 
Expanding the expectation value of the leading operator 
$\bar QQ$ for a pseudoscalar meson $P_Q$ with quantum number $Q$  in powers 
of $1/m_Q$ yields 
\beq 
\frac{1}{2M_{P_Q}} 
\matel{P_Q}{\bar QQ}{P_Q} = 
1 - \frac{\mu _{\pi}^2}{2m_Q^2} + 
\frac{\mu _G^2}{2m_Q^2} 
+{\cal O}(1/m_Q^3) \; ; 
\label{QQ}
\eeq 
$\mu ^2_{\pi }(\mu )$ and 
$\mu ^2_G(\mu )$ denote the expectation values of the kinetic and 
chromomagetic operators, respectively: 
\beq 
\mu ^2_{\pi }(\mu ) \equiv \frac{1}{2M_{H_Q}} 
\matel{H_Q}{\bar Q \vec \pi ^2 Q}{H_Q}_{(\mu )} 
\; , \; 
\mu ^2_G (\mu ) \equiv \frac{1}{2M_{H_Q}} 
\matel{H_Q}{\bar Q \frac{i}{2} 
\sigma \cdot G Q}{H_Q}_{(\mu )} \; ; 
\eeq 
for short they are often called the kinetic and chromomagnetic moments. 

Eq.(\ref{QQ}) implies that one has 
$\matel{H_Q}{\bar QQ}{H_Q}_{(\mu )}/2M_{H_Q} = 1$ 
for $m_Q \to \infty$; i.e., the free quark model expression emerges 
asymptotically for the total width.

\item 
The {\em leading} nonperturbative corrections 
arise at order $1/m_Q^2$ only. That means they are 
rather small in beauty decays since 
$(\mu /m_Q)^2 \sim $ few \% for $\mu \leq 1$ GeV. 
\item 
This smallness of nonperturbative contributions explains a posteriori, why 
partonic expressions when coupled with a `smart' perturbative treatment 
often provide a decent approximation. 
\item 
These nonperturbative contributions which are 
power suppressed can be described only if considerable 
care is applied in treating the {\em parametrically 
larger} perturbative corrections. 
\item 
Explicitely flavour dependant effects arise in order 
$1/m_Q^3$. They mainly drive the differences in the 
lifetimes of the various mesons of a given heavy 
flavour.
\item 
An important practical distinction to the OPE treatment of $e^+e^- \to had$ or 
deep-inelastic lepton nucleon scattering is the fact that the weak width depends on the 
fifth power of the heavy quark mass, see Eq.(\ref{MASTER}), and thus requires particular 
care in dealing with the delicate concept of quark masses.  
\end{itemize} 
One general, albeit subtle point has to be kept in mind here: while everybody these 
days invokes the OPE it is often not done employing Wilson's prescription with the  
auxilliary scale $\mu$, and different definitions of the relevant operators have 
been suggested. While results from one prescription can be translated into another one 
order by order, great care has to be applied. I will adopt here the socalled `kinetic scheme' 
with $\mu \simeq 1$ GeV. It should be noted that the quantities $\mu ^2_{\pi }(\mu )$ and 
$\mu ^2_{G}(\mu )$ are quite distinct from the socalled HQET parameters $\lambda_1$ and 
$\lambda_2$ although the operators look identical. Furthermore the fact that perturbative 
corrections are rather smallish in the kinetic scheme does generally {\em not} hold in other schemes.

The absence of corrections of order 
$1/m_Q$ \cite{BUV} 
is particularly noteworthy and intriguing since 
such corrections do exist for hadronic masses --  
$M_{H_Q} = m_Q (1+ \bar \Lambda/m_Q + 
{\cal O}(1/m_Q^2) )$ -- and those control 
the phase space. Technically this follows from the 
fact that there is no {\em independant} 
dimension-four operator that could emerge in the OPE 
\footnote{The operator $\bar Q i \not D Q$ can be reduced 
to the leading operator $\bar QQ$ through the equation 
of motion.}. This result can be illuminated in more 
physical terms as follows. Bound-state effects in the 
initial state like mass shifts do generate corrections 
of order $1/m_Q$ to the total width; yet so does 
hadronization in the final state. {\em Local} 
colour symmetry demands that those effects cancel 
each other out. {\em It has to be emphasized that the 
absence of corrections linear in $1/m_Q$ is an 
unambiguous consequence of the OPE description.} 
{\em If} their presence were forced upon us, we would have 
encountered a {\em qualitative} change in our QCD paradigm. 
A discussion of this point has arisen recently phrased 
in the terminology of quark-hadron duality. I will return 
to this point later.   

\subsubsubsection{Sum Rules}
\label{SUMRULES}

There are classes of sum rules derived from QCD proper that  relate the 
heavy quark parameters appearing in the OPE for inclusive $B\to l \nu X_c$ -- like 
$\mu_{\pi}^2$, $\mu _G^2$ etc. -- with restricted sums over exclusive channels. They 
provide rigorous definitions, inequalities and experimental constraints \cite{HQSR};  
e.g.:
\bea 
\mu_{\pi}^2 (\mu)/3 &=& \sum _n^{\mu} \epsilon_n^2\left| \tau_{1/2}^{(n)}(1)\right| ^2 +
2\sum _m^{\mu} \epsilon_m^2\left| \tau_{3/2}^{(m)}(1)\right| ^2 
\\
\mu_{G}^2 (\mu)/3 &=& 
-2\sum _n^{\mu} \epsilon_n^2\left| \tau_{1/2}^{(n)}(1)\right| ^2 +
2\sum _m^{\mu} \epsilon_m^2\left| \tau_{3/2}^{(m)}(1)\right| ^2
\label{SUMRULES}
\eea
where $\tau_{1/2}$,  $\tau _{3/2}$ are the amplitudes for 
$B\to l \nu D(j_q)$ with $D(j_q)$ a hadronic system beyond the $D$ and $D^*$,  
$j_q = 1/2 \& 3/2$ the angular momentum carried by the light degrees of freedom in 
$D(j_q)$, as explained in the paragraph below Eq.(\ref{PAULI}), 
and $\epsilon_m$ the excitation energy of the $m$th such system above the $D$  
with $\epsilon _m \leq \mu$.

These sum rules have become of great practical value. I want to emphasize here 
one of their conceptual features: they show that {\em the heavy quark parameters in the 
kinetic scheme are observables themselves}.

\subsubsubsection{Quark-hadron Duality}
\label{DUALITY}

The concept of quark-hadron duality (or duality for short), which  goes back to the early days of the quark model, refers to the notion that a {\em quark}-level description 
should provide a good description of transition rates that involve {\em hadrons}, if one sums over a 
sufficient number of channels. This is a rather vague formulation: How many channels are "sufficiently" many? How good an approximation can one expect? How process dependent is it? Yet it is 
typical in the sense that no precise definition of duality had been given for a long time, and the concept has been used in many different 
incarnations. A certain lack of intellectual rigour can be of great euristic value in the `early going' -- 
but not forever. 

A precise definition requires theoretical control over perturbative as well as nonperturbative dynamics. 
For limitations to duality have to be seen as effects 
{\em over and beyond} uncertainties due to truncations in 
the perturbative and nonperturbative expansions. To be more explicit: duality violations 
are due to corrections {\em not} accounted for due to 
\begin{itemize}
\item 
truncations in the expansion and 
\item limitations in the algorithm employed. 
\end{itemize}
One important requirement is to have an OPE treatment of the process under study, since otherwise we have no unambiguous and systematic inclusion of nonperturbative corrections.  This is certainly the 
case for inclusive semileptonic and radiative $B$ decays. 

While we have no complete theory for duality and its limitations, we have certainly moved beyond the 
folkloric stage in the last few years. We have developed a better understanding of the physics effects 
that can generate duality violations -- the presence of production thresholds for example -- and 
have identified mathematical portals through which duality violations can enter. The fact that we construct the OPE in the Euclidean range and then have to extrapolate it to the Minkowskian domain 
provides such a gateway. 

The problem with the sometimes heard statement that duality represents an additional ad-hoc assumption is that it is not even wrong -- it just misses the point. 

More details on this admittedly complex subject can be found in Ref.\cite{DUALMANNEL} and for the 
truly committed student in Ref.\cite{VADE}. Suffice it here to say that it had been predicted that duality 
violation in $\Gamma _{SL}(B)$ can safely be placed below 0.5 \% \cite{VADE}. 
The passion in the arguments over the potential size of duality violations in 
$B \to l \nu X$ has largely faded away, since, as I discuss later on, the experimental studies of it have 
shown no sign of such limitations.

\subsubsubsection{Heavy Quark Parameters}
\label{HEAVIES}

Through order $1/m_Q^3$ there are six heavy quark parameters (HQP), which fall into two different 
classes:  
\begin{enumerate}
\item 
The heavy quark masses $m_b$ and $m_c$; they are `external' to QCD; i.e. they can never be calculated by lattice QCD with{\em out} experimental input. 
\item 
The expectation values of the dimension five and six operators: $\mu_{\pi}^2$, $\mu_G^2$, 
$\rho _D^3$ and $\rho _{LS}^3$. They are `intrinsic' to QCD, i.e. can be calculated 
by lattice QCD with{\em out} experimental input. 
\end{enumerate}

Since weak decay widths depend on the fifth power of the heavy quark mass, great care has to be 
applied in defining this somewhat elusive entity in a way that can pass full muster by quantum field theory. To a numerically lesser degree this is true for the other HQP as well. Their dependance 
on the auxiliary scale $\mu$ has to be carefully tracked. 

$\bullet$ {\em Quark masses:} There is no quark mass {\em per se} -- one has to specify the renormalization scheme used and the scale, at which the mass is to be evaluated. 
The {\em pole} mass -- i.e. the position of the pole in the perturbative Green function -- has the  
convenient features that it is gauge invariant and infrared finite in perturbation theory. Yet in the 
complete theory it is infrared unstable \cite{HQREV} due to `renormalon' effects. Those introduce 
an {\em irreducible intrinsic} uncertainty into the quark mass: 
$m_Q(1+\delta (m_Q)/m_Q)$, with 
$\delta (m_Q)$ being roughly $\sim  \Lambda_{QCD}$. 
For the weak width it amounts to an uncertainty $\delta (m_Q^5) \sim 5 \delta (m_Q)/m_Q$; i.e., it   
is parametrically larger than the power suppressed terms $\sim {\cal O}(1/m_Q^2)$ one is 
striving to calculate. The pole mass is thus ill suited when including nonperturbative contributions. 
Instead one needs a running mass with an infrared cut-off $\mu$ to `freeze out' renormalons. 

\noindent 
The $\overline{MS}$ mass, which is a rather ad hoc expression convenient in perturbative 
computations rather than a parameter in an effective Lagrangian, would satisfy this requirement. 
It is indeed a convenient tool for treating reactions where the relevant scales exceed 
$m_Q$ in {\em production} processes like $Z^0 \to b \bar b$. Yet in {\em decays}, where 
the relevant scales are necessarily below $m_Q$ the $\overline{MS}$ mass is actually 
inconvenient or even inadequate. For it has a hand made infrared instability: 
\beq 
\overline{m}_Q (\mu) = \overline{m}_Q(\overline{m}_Q) \left[ 
1+ \frac{2\alpha _S}{\pi} {\rm log} \frac{\overline{m}_Q}{\mu} 
\right]  \to \infty \; \; {\rm as} \; \; \frac{\mu}{\overline{m}_Q} \to 0
\eeq

\noindent 
It is much more advantageous to use the `kinetic' mass instead with 
\beq 
\frac{dm_Q(\mu)}{d\mu} = - \frac{16 \alpha_S(\mu)}{3\pi} - 
\frac{4 \alpha_S(\mu)}{3\pi}\frac{\mu}{m_Q} + ...  , 
\label{KINMASSDEF}
\eeq
which has a linear scale dependence in the infrared. It is this kinetic mass I will use in the 
following. Its value had been extracted from 
\beq 
e^+e^- \to \Upsilon (4S) \to H_b H_b^{\prime} X 
\eeq
before 2002 by different authors with better than about 2\% accuracy 
\cite{MBUPSILON} based on an original idea of M. Voloshin. Their findings expressed 
in terms of the kinetic mass can be summarized as follows: 
\beq 
\langle m_b (1\; {\rm GeV})\rangle |_{\Upsilon (4S) \to b \bar b} = 4.57 \pm 0.08 \; {\rm GeV}
\label{MB4S}
\eeq
Charmonium sum rules yield 
\beq 
m_c(m_c) \simeq 1.25 \pm 0.15 \; {\rm GeV} \; . 
\label{MCONIUM}
\eeq
The HQE allows to relate the difference $m_b - m_c$ to the `spin averaged' beauty and charm meson 
masses and the higher order HQP \cite{HQREV}: 
\beq 
m_b - m_c = \langle M_B \rangle - \langle M_D\rangle + 
\left( \frac{1}{2m_c} - \frac{1}{2m_b}   \right)  \mu _{\pi}^2 + ... \simeq 
3.50 \; {\rm GeV} + 40 \; {\rm MeV} \cdot \frac{\mu_{\pi}^2 - 0.5\; ({\rm GeV})^2}{0.1 \; ({\rm GeV})^2} ...  
\label{MBMCDIFF}
\eeq 
Yet this relation is quite vulnerable since it is dominantly an expansion in $1/m_c$ rather than $1/m_b$ and {\em non}local correlators appear in order $1/m_c^2$. Therefore one is ill-advised to 
impose this relation a priori. One is of course free to consider it a posteriori. 

$\bullet$ {\em Chromomagnetic moment:} Its value can be inferred quite reliably from the 
hyperfine splitting in the $B^*$ and $B$ masses: 
\beq 
\mu_{G}^2 (1\; {\rm GeV}) \simeq \frac{3}{2}\left[ M^2(B^*) - M^2(B)  \right]   \simeq 0.35 \pm 0.03 \; ({\rm GeV})^2 
\label{CHROMOMAG}
\eeq

$\bullet$ {\em Kinetic moment:} The situation here is not quite so definite. We have a rigorous 
lower bound from the SV sum rules \cite{OPTICAL}:  
\beq 
\mu_{\pi}^2 (\mu ) \geq \mu_{G}^2 (\mu )
\label{MUPIBOUND}
\eeq
for any $\mu$; QCD sum rules yield 
\beq 
\mu_{\pi}^2 (1\; {\rm GeV}) \simeq 0.45 \pm 0.1\; ({\rm GeV})^2 
\label{MUPISR}
\eeq

$\bullet$ {\em Darwin and LS terms:} The numbers are less certain still for those. The saving 
grace is that their contributions are reduced in weight, since they represent 
${\cal O}(1/m_Q^3)$ terms. 
\beq 
\rho^3_D(1\; {\rm GeV}) \sim 0.1\; ({\rm GeV})^3 \; , \; 
- \rho^3_{LS}(\mu ) \leq \rho^3_{D}(\mu ) 
\label{DARWIN}
\eeq

\subsubsection{First Tests: Weak Lifetimes and SL Branching Ratios}
\label{FIRSTTESTS}

Let me begin with three general statements: 
\begin{itemize}
\item 
Within the SM  the semileptonic widths have to coincide for $D^0$ and $D^+$ mesons and 
for $B_d$ and $B_u$ mesons up to small isospin violations, since the 
semileptonic transition operators for $b \to l \nu c$ and $c \to l \nu s$ are isosinglets. The 
ratios of their semileptonic branching ratios are therefore equal to their lifetime ratios to a very 
good approximation: 
\beq 
\frac{{\rm BR}_{SL}(B^+)}{{\rm BR}_{SL}(B_d)} = \frac{\tau (B^+)}{\tau (B_d)} + 
{\cal O}\left( \left| \frac{V(ub)}{V(cb)}\right|^2 \right) \; , \; 
\frac{{\rm BR}_{SL}(D^+)}{{\rm BR}_{SL}(D^0)} = \frac{\tau (D^+)}{\tau (D^0)} + 
{\cal O}\left( \left| \frac{V(cd)}{V(cs)}\right|^2 \right)
\eeq 
For dynamical rather than symmetry reasons such a relation can be extended to $B_s$ and $D_s$ mesons \cite{DSSL}: 
\beq 
\frac{{\rm BR}_{SL}(B_s)}{{\rm BR}_{SL}(B_d)} \simeq \frac{\overline{\tau} (B_s)}{\tau (B_d)} \; , 
\eeq
where $\overline{\tau} (B_s)$ denotes the average of the two $B_s$ lifetimes. 
\item 
Yet the semileptonic widths of heavy flavour baryons will {\em not} be universal for a given 
flavour. The ratios of their semileptonic branching ratios will therefore not reflect their lifetime ratios. 
In particular for the charmed baryons one predicts large differences in their semileptonic widths 
\cite{VOLOSHINSL}. 

\item 
It is more challenging for theory to predict the absolute value of a semileptonic branching ratio than the 
ratio of such branching ratios.  

\end{itemize}

\subsubsubsection{Charm lifetimes}
\label{CHARMLIFES}

The lifetimes of all seven $C=1$ charm hadrons have been measured now with the 
FOCUS experiment being the only one that has contributed to all seven lifetimes. In 
Table \ref{tab:TABLECHARM} the predictions based on the HQE (together 
with brief theory comments) are juxtaposed to the data 
\cite{CICERONE}.  While a priori the HQE might be expected to fail even on the semiquantitative level since $\mu_{had}/m_c \sim 1/2$ is an uncomfortably large expansion parameter, it works surprisingly well in describing the lifetime ratios even for baryons 
except for $\tau (\Xi_c^+)$ being about 50 \% longer than predicted. This agreement should be viewed as quite nontrivial, since these lifetimes span more than an order 
of magnitude between the shortest and longest: $\tau (D^+)/\tau (\Omega_c) \simeq 14$. 
It provides one of the better arguments for charm acting like a heavy quark at least in cases, 
when the leading nonperturbative correction is of order $1/m_c^2$ rather than $1/m_c$. 

\begin{table}
\begin{center}
\small{\begin{tabular}{llll}
\hline
 & $1/m_c$ expect. & theory comments & data   \\
\hline
$\frac{\tau (D^+)}{\tau (D^0)}$ &  
$\sim 1+\left( \frac{f_D}{200\; \MeV} \right)^2 \sim 2.4$  
& PI dominant               & $2.54 \pm 0.01$  \\    
$\frac{\tau (D_s^+)}{\tau (D^0)}$ & 0.9 - 1.3[1.0 - 1.07] & 
{\em with} [{\em without}] WA  & $1.22 \pm 0.02$ \\  
$\frac{\tau (\Lambda _c^+)}{\tau (D^0)}$ & $\sim 0.5$  
& quark model matrix elements       & $0.49 \pm 0.01$ \\   
$\frac{\tau (\Xi _c^+)}{\tau (\Lambda _c^+)}$ & $\sim 1.3 - 1.7$ &  
ditto                                  &  $2.2 \pm 0.1$\\
$\frac{\tau (\Lambda _c^+)}{\tau (\Xi _c^0)}$ & $\sim 1.6 - 2.2$ &  
ditto                                  & $2.0 \pm 0.4$ \\
$\frac{\tau (\Xi _c^+)}{\tau (\Xi _c^0)}$ & $\sim 2.8$ &  
ditto                                  & $4.5 \pm 0.9$ \\  
$\frac{\tau (\Xi _c^+)}{\tau (\Omega _c)}$ & $\sim 4$ &  
ditto                                  & $5.8 \pm 0.9$ \\
$\frac{\tau (\Xi _c^0)}{\tau (\Omega _c)}$ & $\sim 1.4$ &  
ditto                                  & $1.42 \pm 0.14$ \\
\hline
\end{tabular}}
\caption{The weak lifetime ratios of $C=1$ hadrons}
\label{tab:TABLECHARM}
\end{center}
\end{table}

The SELEX collaboration has reported candidates for weakly decaying double charm baryons. It is my judgment that those candidates cannot be $C=2$ baryons since their reported 
lifetimes are too short and do not show the expected hierarchy \cite{CICERONE}.

\subsubsubsection{Beauty lifetimes}
\label{BEAUTYLIFES}

Theoretically one is on considerably safer ground when applying the HQE to lifetime ratios 
of beauty hadrons, since the expansion parameter $\mu_{had}/m_b \sim 1/7$ is small compared to 
unity. The HQE provided predictions in the old-fashioned sense; i.e., it produced them {\em before} 
data with significant accuracy were known.  

\begin{table}
\begin{center}
\small{\begin{tabular}{llll}
\hline 
 & $1/m_b$ expect. & theory comments & data   \\
 \hline 
\hline
$\frac{\tau (B^+)}{\tau (B_d)}$ &  
$\sim 1+ 0.05\left( \frac{f_B}{200\; \MeV} \right)^2$    '92 \cite{MIRAGE}  
& PI in $\tau (B^+)$               & $1.076 \pm 0.008$  \cite{WINTER05} \\    
 & $1.06 \pm 0.02$    \cite{LENZNEW} & fact. at low scale 1GeV & \\ 
 \hline 
$\frac{\overline{\tau} (B_s)}{\tau (B_d)}$ & $1 \pm {\cal O}(0.01)$   '94 \cite{DSSL} 
& & $0.920 \pm 0.030$ \cite{WINTER05} \\
\hline 
 $\frac{\tau (\Lambda _b^-)}{\tau (B_d)}$ & $\geq 0.9 $    '93 \cite{STONEBOOK}
& quark model           & $0.806 \pm 0.047$   '04 \cite{WINTER05} \\   
 & $\simeq 0.94$ \& $\geq 0.88$  '96 \cite{BOOST,FAZIO} & matrix elements& $0.944 \pm 0.089$   '05 \cite{CDFNOTE} \\
 \hline
$\tau (B_c)$ & $\sim (0.3 - 0.7)$ psec   '94ff  \cite{MICHEL}  & largest lifetime diff. & 
$0.45 \pm 0.12$ psec  
\cite{WINTER05}\\
 & & no $1/m_Q$ term crucial & \\
 \hline   
$\frac{\Delta \Gamma (B_s)}{\overline{\Gamma}(B_s)}$ & 
$22\% \cdot  \left(\frac{f(B_s)}{220\, \MeV}\right) ^2$  '87 \cite{AZIMOV} & less reliable & 
$0.65 \pm 0.3$  CDF \\
 & $12 \pm 5\% $   '04 \cite{LENZNEW} & than $\Delta M(B_s)$ & $0.23 \pm 0.17$  D0 \\
\hline
\end{tabular}}
\caption{The weak lifetime ratios of $B=1$ hadrons}
\label{tab:TABLEBEAUTY}
\end{center}
\end{table}

Several comments are in order to interpret the results: 
\begin{itemize}
\item 
The $B^+ - B_d$ lifetime ratio has been measured now with better than 1\% accuracy -- and the very first prediction based on the HQE was remarkably on target \cite{MIRAGE}. 
\item 
The most dramatic deviation from a universal lifetime for $B=1$ hadrons has emerged in 
$B_c$ decays. Their lifetime is only a third of the other beauty lifetimes -- again in full agreement with 
the HQE {\em pre}diction.  That prediction is actually less obvious than it might seem. 
For the observed $B_c$ lifetime is close to the charm lifetime as given by $\tau (D^0)$, and that is what 
one would expect already in a naive parton model treatment, where 
$\Gamma (b \bar c) \simeq \Gamma (c)\cdot [1 + \Gamma(b)/\Gamma(c)]$. However it had been 
argued that inside such a tightly bound state the $b$ and $c$ quark masses had to be replaced by 
effective masses reduced by the (same) binding energy: 
$m_b ^{eff} = m_b - B.E.$, $m_c ^{eff} = m_c - B.E.$ with $B.E. \sim {\cal O}(\Lambda _{QCD})$. This would prolong the weak lifetimes of the two quark greatly, since those depend on the fifth power of the quark masses and would do so much more for the charm transition than for the beauty one. Yet such an effect would amount to a correction of order 
$1/m_Q$, which is not allowed by the OPE, as explained above at the end of Sect.\ref{OPE}; 
the more detailed argument can be found in Ref. \cite{BELLINI}.  
\item 
A veritable saga is emerging with respect to $\tau (\Lambda_b)$. The first prediction stated 
\cite{STONEBOOK} 
that $\tau (\Lambda_b)/\tau (B_d)$ could not fall below 0.9. A more detailed analysis led to two 
conclusions \cite{BOOST}, namely that the HQE most likely leads to 
\beq 
\frac{\tau (\Lambda_b)}{\tau (B_d)} \simeq 0.94 
\label{CENTRAL}
\eeq
with an uncertainty of a few percent, while a lower bound had to hold 
\beq 
\frac{\tau (\Lambda_b}{\tau (B_d)} \geq 0.88 \; .  
\label{LOWER}
\eeq
A violation of this bound would imply that we need a new paradigm for evaluating at least baryonic 
matrix elements. 

There are actually two questions one can ask concerning
$\tau(\Lambda_b)/\tau (B_d)$:
\begin{enumerate}
\item 
What is theoretically  the most likely value for $\tau(\Lambda_b)/\tau (B_d)$? 
\item 
How much lower can one reasonably push it? 
\end{enumerate}
While there is a connection between those two questions, they
clearly should be distinguished. Most theoretical analyses -- employing quark models, QCD 
sum rules or lattice studies -- agree on the first question, namely that the ratio is predicted 
to lie above 0.90. Yet the data  have for many years pointed to a significantly lower 
value $\sim 0.80$. This apparent discrepancy has given rise to the second question listed above. 
Ref.\cite{BOOST} provided a carefully reasoned answer to it. Ref.\cite{petrov} stated a value 
of $0.86 \pm 0.05$, which is sometimes quoted as the theory prediction. I object to 
viewing this value as the answer to the first question above; one might consider it as a response 
to the second question, although even then I remain skeptical of it. 

The new CDF result seems to reshuffle the cards. The question is whether it is just a high fluctuation 
-- implying a worrisome discrepancy between theory and experiment -- or represents a new trend to be confirmed in the future, which would represent an impressive `comeback' success for the HQE. 

No matter what the final verdict will be on $\tau (\Lambda_b)$, it is important to measure 
also $\tau (\Xi_b^0)$ and $\tau (\Xi_b^-)$ -- either to confirm success or diagnose failure. 
One expects \cite{VOLOSHINXIB}: 
\beq 
\tau(\Xi_b^0) \simeq \tau (\Lambda_b) < \tau (B_d) < \tau (\Xi_b^-) \; , 
\label{XIBLIFES}
\eeq
where the `$<$' signs indicate an about 7\% difference. 
{\em If} the $\Lambda_b$-$B_d$ lifetime difference were larger than predicted, one would 
like to know whether the whole lifetime hierarchy of Eq.(\ref{XIBLIFES}) is stretched out -- say 
`$<$' in 
$\tau (\Lambda_b) < \tau (B_d) < \tau (\Xi_b^-)$ represents differences of 10 \% or even more -- 
or whether the splittings in the baryon lifetimes are as expected, yet  their overall values reduced relative to $\tau (B_d)$.   

\item 
The original prediction that $\tau (B_d)/\overline{\tau}(B_s)$ is unity within 1 - 2 \% 
\cite{DSSL,STONEBOOK} has been confirmed by subsequent authors. Yet the data have stubbornly remained somewhat low. This measurement deserves great attention and effort. While I consider 
the prediction to be on good footing, it is based on an evaluation of a complex dynamical situation 
rather than a theorem or even symmetry. Establishing a discrepancy between theory and experiment here would raise some very intriguing questions.

\item 
The theoretical evaluation of $\Delta \Gamma _{B_s}$ and the available data has already been given 
in Sect. \ref{HOT}.

\end{itemize}

\subsubsection{The $V(cb)$ `Saga' -- A Case Study in Accuracy}
\label{VCBEXTRAC}

\subsubsubsection{Inclusive Semileptonic $B$ Decays}
\label{INCLSL}
The value of $|V(cb)|$ is extracted from $B\to l \nu X_c$ in two steps. 

{\bf A:} 
One expresses $\Gamma (B \to l \nu X_c)$ in terms of the HQP -- quark masses 
$m_b$, $m_c$ and the expectation values of local operators $\mu_{\pi}^2$, $\mu_G^2$, 
$\rho_D^3$ and $\rho_{LS}^3$ -- as accurately as possible, namely through 
${\cal O}(1/m_Q^3)$ and to all orders in the BLM treatment for the partonic contribution. 
Having 
precise values for these HQP is not only of obvious use for extracting $|V(cb)|$ and $|V(ub)|$, 
but also yields benchmarks for how much numerical control lattice QCD provides us over 
nonperturbative dynamics. 

{\bf B:}  
The numerical values of these HQP are extracted from the {\em shapes} of inclusive 
lepton distributions as encoded in their {\em normalized} moments.  Two types of moments have 
been utilized, namely lepton energy and hadronic mass moments. While the former are dominated by the contribution from the `partonic' term $\propto \matel{B}{\bar bb}{B}$, the latter are more sensitive to higher nonperturbative terms $\mu_{\pi}^2$ \& $\mu_G^2$ and thus have to form an integral part of the analysis. 
  
Executing the first step in the so-called kinetic scheme and inserting the experimental number for 
$\Gamma (B\to l \nu X_c)$ one arrives at \cite{BENSON1} 
\bea
\nonumber
\frac{|V(cb)|}{0.0417} &=& D_{exp}\cdot (1+\delta _{th})  [1+0.3 (\alpha_S(m_b) - 0.22)]  
\left[ 1 - 0.66(m_b - 4.6) + 0.39(m_c - 1.15)  \right. \\ 
\nonumber 
&& \left. +  0.013(\mu_{\pi}^2 - 0.4) + 0.05 (\mu_G^2 - 0.35) + 0.09 (\rho_D^3 - 0.2) + 
0.01 (\rho_{LS}^3 + 0.15 )\right]  \; , \\
&& D_{exp} = \sqrt{\frac{\rm BR_{SL}(B)}{0.105}}\sqrt{\frac{1.55\, {\rm ps}}{\tau_B}}
\label{VCBHQP}
\eea
where all the HQP are taken at the scale 1 GeV and their `seed' values are given in the 
appropriate power of GeV; the theoretical error at this point is  given by 
\beq 
\delta _{th} = \pm 0.5 \%|_{pert} \pm 1.2 \% |_{hWc} \pm 0.4 \% |_{hpc} \pm 0.3 \% |_{IC} 
\eeq 
reflecting the remaining uncertainty in the Wilson coefficient of the leading operator 
$\bar bb$, as yet uncalculated  perturbative corrections to the Wilson coefficients of the 
chromomagnetic and Darwin operators, higher order power corrections including 
duality violations in $\Gamma _{SL}(B)$ and nonperturbative effects due to operators containing 
charm fields, respectively.  Concerning the last item, in Ref.\cite{BENSON1} an error 
of 0.7 \% was stated. A dedicated analysis of such IC effects allowed to reduce 
this uncertainty down to 0.3 \% \cite{IC}. 

BaBar has performed the state-of-the-art analysis of several lepton energy 
and hadronic mass moments \cite{BABARVCB} obtaining 
an impressive fit with the following HQP in the kinetic scheme \cite{SCHEMES}: 
\bea 
m_b(1 \, \GeV)  = (4.61 \pm 0.068) \GeV , \, m_c(1 \, \GeV) = (1.18 \pm 0.092) \GeV  
\label{MB}\\
m_b(1 \, \GeV) - m_c(1 \, \GeV) = (3.436 \pm 0.032) \GeV  
\label{MBMMC}\\ 
\mu_{\pi}^2 (1\, \GeV) = (0.447 \pm 0.053) \GeV ^2 , \, 
\mu_{G}^2 (1\, \GeV) = (0.267 \pm 0.067) \GeV ^2 
\label{MUPI} \\
\rho_{D}^3 (1\, \GeV) = (0.195 \pm 0.029) \GeV ^3 
\label{RHOD}\\
|V(cb)|_{incl} = 41.390 \cdot (1 \pm 0.021) \times 10^{-3} 
\label{VCBBABAR}
\eea
The DELPHI collab. has refined their pioneering study of 2002 \cite{DELPHI02} 
obtaining \cite{DELPHI05}: 
\beq
|V(cb)|_{incl} = 42.1 \cdot (1 \pm 0.014|_{meas} \pm 0.014|_{fit} \pm 0.015|_{th}) \times 10^{-3} 
\label{DELPHI}
\eeq
A comprehensive analysis of all relevant data from $B$ decays, including from $B \to \gamma X$ yields the results listed in Table \ref{tab:STATUS05COMPLET} \cite{FLAECHER}, where they are compared 
to their predicted values. Some had already been 
given in Table \ref{tab:STATUS05}.  With these HQP one arrives at 
\beq 
\langle |V(cb)|_{incl}\rangle = 41.96 \cdot (1 \pm 0.0055|_{exp}\pm 0.0083|_{HQE}
\pm 0.014|_{\Gamma_{SL}}) \times 10^{-3} 
\label{VCBCOMPREHENSIVE}
\eeq

\begin{table}
\begin{center}
\small{\begin{tabular}{lll}
\hline
Heavy Quark Parameter      & value from $B\to l\nu X_c/\gamma X$ & 
predict. from other observ.    \\
\hline
      $m_b$(1 GeV)   &$ = (4.59 \pm 0.025|_{exp}\pm 0.030|_{HQE})$GeV&
      $=(4.57 \pm 0.08$)GeV,{\small Eq.(\ref{MB4S})}\\ 
      $m_c$(1 GeV)   &= $(1.142 \pm 0.037|_{exp}\pm 0.045|_{HQE})$GeV&=$(1.25 \pm 0.15)$GeV, 
      {\small Eq.(\ref{MCONIUM})}\\ 
      $[m_b - m_c]$(1GeV)  &= $(3.446 \pm 0.025)$GeV& = $(3.46 \pm X)$GeV,  
      Eq.(\ref{MBMCDIFF})\\
      $[m_b -0.67 m_c]$(1GeV)  & = $(3.82 \pm 0.017)$ GeV &  \\    
      $\mu_G^2$(1GeV) & = $(0.297 \pm 0.024|_{exp}\pm 0.046|_{HQE}){\rm GeV}^2$ 
      &=$(0.35 \pm 0.03)\GeV^2$,{\small Eq.(\ref{CHROMOMAG})}\\
      $\mu_{\pi}^2$(1GeV) & = $(0.401 \pm 0.019|_{exp}\pm 0.035|_{HQE}){\rm GeV}^2$ & 
      $\geq \mu_G^2$(1GeV), Eq.(\ref{MUPIBOUND}) \\
       & 
      &= $(0.45 \pm 0.1) \GeV^2$,Eq.(\ref{MUPISR}) \\
      $\rho_D^3$(1GeV) & = $(0.174 \pm 0.009|_{exp}\pm 0.022|_{HQE}){\rm GeV}^3$ & 
      $\sim + 0.1\; {\rm GeV}^3$, Eq.(\ref{DARWIN}) \\ 
        $\rho_{LS}^3$(1GeV) & = -$(0.183 \pm 0.054|_{exp}\pm 0.071|_{HQE}){\rm GeV}^3$ & 
      $\sim -0.1\; {\rm GeV}^3$, Eq.(\ref{DARWIN}) \\
\hline
\end{tabular}}
\caption{The 2005 values of the HQP obtained from a comprehensive analysis of 
$B \to l \nu X_c$ and $B \to \gamma X$ \cite{FLAECHER} and compared to predictions}
\label{tab:STATUS05COMPLET}
\end{center}
\end{table}

For a full appreciation of these results one has to note the following: 
\begin{itemize}
\item 
With just these six parameters one obtains an excellent fit to several energy and hadronic 
mass moments even for different values of the lower cut on the lepton or photon energy. Varying those lower cuts also provides more direct information on the respective energy spectra beyond the moments. 
\item 
Even better the fit remains very good, when one `seeds' two of these HQP to their predicted values, namely 
$\mu_G^2 (1\, \GeV) = 0.35 \pm 0.03 \GeV^2$ as inferred from the $B^*-B$ hyperfine mass splitting 
and $\rho_{LS}^3 = - 0.1 \GeV^3$ allowing only the other four HQP to float. 
\item 
These HQP are treated as free fitting parameters. It could easily have happened that they assume 
unreasonable or even unphysical values. Yet they take on very special values fully consistent with all constraints that can be placed on them by theoretical means as well as other experimental input. 
To cite but a few examples: 
\begin{itemize}
\item 
The value for 
$m_b$ inferred from the {\em weak decay} of a $B$ meson agrees completely within the 
stated uncertainties with what has been derived from the {\em electromagnetic} 
and {\em strong production} of $b$ hadrons just above threshold.
\item 
The rigorous inequality $\mu_{\pi}^2 > \mu_G^2$, which had {\em not} been imposed as a 
constraint, is satisfied. 
\item 
$\mu_G^2$ indeed emerges with the correct value, as does $\mu_{\pi}^2$.  

\end{itemize}

\item
$m_b$-$m_c$ agrees very well with what one infers from the spin-averaged $B$ and $D$ meson masses. 
However this {\em a posteriori} agreement does {\em not} justify imposing it as an {\em a priori}  constraint. For the mass relation involves an expansion in $1/m_c$, which is of less than sterling 
reliability. Therefore I have denoted its uncertainty by $X$. 
\item 
The $1$ \% error in $m_b$ taken at face value might suggest that it alone would generate more than a 2.5 \% uncertainty in $|V(cb)|$, i.e. by itself saturating
the total error given in Eq.(\ref{VCBCOMPREHENSIVE}). The resolution of this apparent contradiction is as follows.  
The dependance of the total semileptonic width and also of the lowest lepton energy 
moments on $m_b$ \& $m_c$ can be approximated 
by $m_b^2(m_b - m_c)^3$ for the actual quark masses; for the leading contribution this can 
be written as $\Gamma _{SL}(B) \propto (m_b - \frac{2}{3}m_c)^5$. From the values for 
$m_b$ and $m_c$, Eq.(\ref{MB}), and their correlation given in \cite{BABARVCB} one derives 
\beq 
m_b(1 \, \GeV) - 0.67 m_c(1 \, \GeV) = (3.819 \pm 0.017) \GeV = 3.819\cdot (1\pm 0.45\%)\GeV . 
\label{MBM23MC}
\eeq
I.e., it is basically this peculiar combination that is measured directly through $\Gamma_{SL}(B)$, 
and thus its error is so tiny. 
It induces an uncertainty of 1.1 \% into the value for $|V(cb)|$. 

Eq.(\ref{MBM23MC}) has another important use in the future, namely to provide a very stiff 
validation challenge to lattice QCD's determinations of $m_b$ and $m_c$. 

\end{itemize}
With all these cross checks we can defend the smallness of the stated uncertainties. The analysis of Ref.\cite{GLOBAL} arrives at similar 
numbers (although I cannot quite follow their error analysis). 

More work remains to be done: (i) The errors on the hadronic mass moments are still 
sizable; decreasing them will have a significant impact on the accuracy of $m_b$ and $\mu_{\pi}^2$. 
(ii) As discussed in more detail below, imposing high cuts on the lepton energy degrades the 
reliability of the theoretical description. Yet even so it would be instructive to analyze 
at which cut theory and data part ways. I will return to this point below.  (iii) As another preparation for $V(ub)$ extractions one 
can measure $q^2$ moments or mass moments with a $q^2$ cut to see how well one can 
reproduce the known $V(cb)$. 

\subsubsubsection{Exclusive Semileptonic $B$ Decays}
\label{EXCLSL}
While it is my judgment that the most precise value for $|V(cb)|$ can be extracted from 
$B\to l \nu X_c$, this does not mean that there is no motivation for analyzing exclusive modes. On the contrary: the fact that one extracts a value for $|V(cb)|$ from $B\to l \nu D^*$ at zero recoil fully consistent within a smallish uncertainty represents a great success since the systematics experimentally as well as theoretically are very different: 
\beq 
|V(cb)|_{B\to D^*} =  0.0416 \cdot (1\pm 0.022|_{exp} \pm 0.06|_{th}  )  
  \; \; \; \; {\rm for} \; \; \; \; F_{B\to D^*}(0) = 0.90 \pm 0.05
\eeq
It has been suggested \cite{BPS} to treat $B\to l \nu D$ with the 
`BPS expansion' based on $\mu_{\pi}^2 \simeq \mu_G^2$ and extract 
$|V(cb)|$ with a theoretical error not larger than $\sim 2\%$. It would be most instructive to compare the formfactors and their slopes found in this approach with those of LQCD \cite{OKA}.

\subsubsection{The Adventure Continues: $V(ub)$}
\label{VUBSECT}

There are several lessons we can derive from the $V(cb)$ saga: 
(i) Measuring various moments of $B\to l \nu X_u$ and extracting HQP from them is a 
powerful tool to strengthen confidence in the analysis. Yet it 
is done for validation purposes only. For there is no need to `reinvent the wheel':
{\em When calculating the width and (low) moments of 
$B\to l \nu X_u$ one has to use the values of the HQP as determined in 
$B\to l \nu X_c$}. 
(ii) $\Gamma (B \to l \nu X_u)$ is actually under better theoretical control than 
$\Gamma (B \to l \nu X_c)$ since the expansion parameter is smaller -- 
$\frac{\mu_{had}}{m_b}$ vs. $\frac{\mu_{had}}{m_b-m_c}$ -- and 
${\cal O}(\alpha_S^2)$ corrections are known exactly. 

\noindent {\bf On the Impact of Cuts:} 
In  practice there arises a formidable complication: to distinguish 
$b\to u$ from the huge $b\to c$ background, one applies  
cuts on variables like lepton energy $E_l$, 
hadronic mass $M_X$, the lepton-pair invariant mass $q^2$. As a general rule the 
more severe the cut, the less reliable the theoretical calculation becomes. 
More specifically 
the imposition of a cut introduces a new dimensional scale called `hardness' {\cal Q} \cite{MISUSE}.  
Nonperturbative contributions emerge scaled by powers of $1/{\cal Q}$ rather than 
$1/m_b$. If {\cal Q} is much smaller than $m_b$ such an expansion becomes unreliable. 
Furthermore the OPE cannot capture terms of the form 
$e^{-{\cal Q}/\mu}$. While these are irrelevant for ${\cal Q} \sim m_b$, they quickly gain relevance 
when {\cal Q} approaches $\mu$. Ignoring this effect would lead to a `bias', i.e. a {\em systematic} 
shift of the HQP away from their true values. 

This impact has been studied for radiative $B$ decays with their simpler kinematics in a pilot study 
\cite{MISUSE} and a detailed analysis \cite{BENSON2} 
of the average photon energy  and its variance. The first provides a measure mainly of 
$m_b/2$, the latter of $\mu_{\pi}^2/12$. These biases were found to be relevant down to 
$E_{\rm cut} = 1.85\, \GeV$ and increasing quickly above 2 GeV. While the existence of such 
effects is of a general nature, the estimate of their size involves model dependent elements. 
Yet as long as those corrections are of moderate size, they can be considered reliable. Once they 
become large, we are losing theoretical control. 
\begin{figure}[t]
\vspace{8.0cm}
\includegraphics{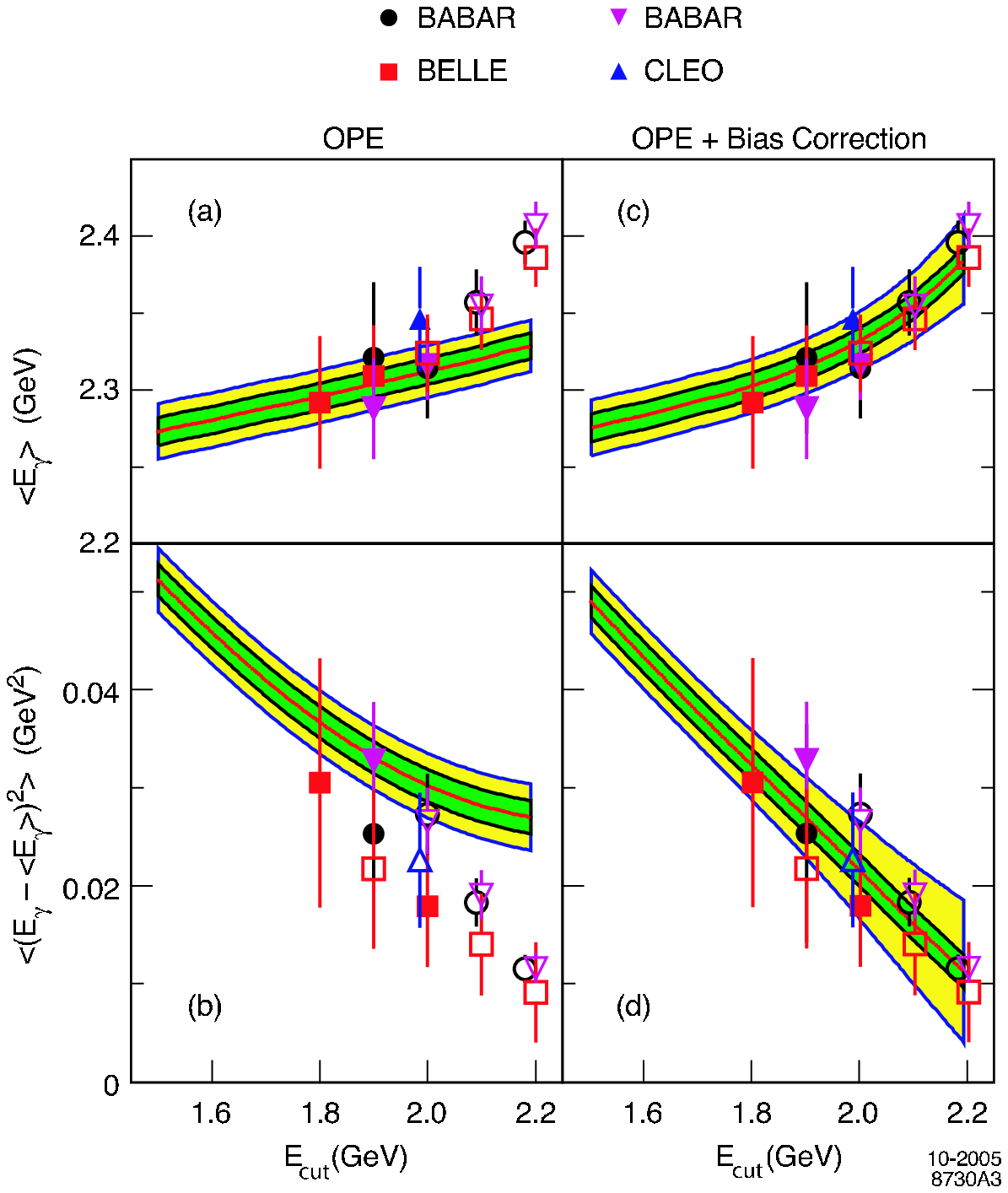}
\caption{The first and second moments of the photon energy in $B\to \gamma X$ compared to 
OPE predictions without and with bias corrections. The inner band indicates the experimental uncertainties only; the outer bands add the theoretical ones; from Ref.\cite{FLAECHER}}
\label{BIASIMPACT} 
\end{figure}

Fig.\ref{BIASIMPACT} shows data for the average photon energy and its variance for different lower cuts on the photon energy from CLEO, BABAR and BELLE compared to the OPE predictions without and with bias corrections on the left and right, respectively. The comparison shows the need for those 
bias corrections and them being under computational control over a sizable range of $E_{cut}$. 
Even more important than providing us with possibly more accurate values for $m_b$ and 
$\mu_{\pi}^2$,  these studies enhance confidence in our theoretical 
tools.  

These findings lead to the following conclusions: (i) As far as theory is concerned there is a high premium on keeping the cuts as low as possible. (ii) Such cuts introduce biases in the HQP values extracted from the truncated moments; yet within a certain range of the cut variables those biases can be corrected for and thus should not be used to justify inflating the theoretical uncertainties. 
(iii) In any case measuring the moments as functions of the cuts provides powerful cross checks for our theoretical control. 

\noindent{\bf `Let a Thousand Blossoms Bloom':} 
Several suggestions have been made for cuts to suppress the $b\to c$ background to 
managable proportions. None provides a panacea. 
The most straightforward one is to focus on the lepton energy endpoint region; however it captures 
merely a small fraction of the total $b\to u$ rate, which can be estimated only with considerable 
model dependance. This model sensitivity can be moderated with information on the heavy quark 
distribution function inferred from $B\to \gamma X$. Furthermore 
weak annihilation contributes only in the endpoint region and with different weight in $B_d$ and 
$B_u$ decays \cite{DSSL}. Thus the lepton spectra have to be measured {\em separately} for charged and neutral $B$ decays. 

Measuring the hadronic recoil mass spectrum up to a maximal value 
$M_X^{\rm max}$ captures the lion share of the $b\to u$ rate if $M_X^{\rm max}$ is above 1.5 GeV; yet it is still vulnerable to theoretical uncertainties in the very low $q^2$ region. This problem can be 
addressed in two different ways: adopting Alexander the Great's treatment of the Gordian knot 
one can  
impose a lower cut on $q^2$ or one can describe the low $q^2$ region with  the help of the measured  
energy spectrum in $B\to \gamma X$ for 1.8 GeV $\leq E_{\gamma} \leq$ 2.0 GeV. Alternatively 
one can apply a combination of cuts. Studying $B_d$ and $B_u$ decays is still desirable, yet not as 
essential as for the previous case. 

In any case one should not restrict oneself to a fixed cut, but vary the latter over some reasonable range 
and analyze to which degree theory can reproduce this cut dependence to demonstrate 
control over the uncertainties. 

There is not a single `catholic' path to the promised land of a precise value for 
$|V(ub)|$; presumably many paths will have to be combined 
\cite{BAUERPUERTO}. Yet it seems quite realistic that the 
error can be reduced to about 5 \% over the next few years. 

\subsection{Summary of Lect. II}
\label{SUMII}

As explained in Lect. I while CKM forces are generated by the exchange of gauge bosons, its couplings involve elements of the CKM matrix. Yet those originate in the elements of the up- and down-type 
quark {\em mass matrices}. Thus the CKM parameters are intrinsically connected with one of the central 
mysteries of the SM, namely the generation in particular of fermion masses and family replication. 
Furthermore the hierarchy in the quark masses and the likewise hierarchical pattern of the CKM matrix 
elements strongly hints at some deeper level of dynamics about which we are quite ignorant.  
Nevertheless CKM theory with its mysterious origins has proved itself to be highly successful in describing even quantitatively a host of phenomena occurring over  a wide array of scales. It 
lead to the `Paradigm of Large \cp~Violation in $B$ Decays as a prediction in the old-fashioned sense; 
i.e., predictions were made well before data of the required sensitivity existed. From the observation 
of a tiny and shy phenomenon -- \cp~violation in $K_L$ decays on the ${\cal O}(10^{-3})$ level -- it 
predicted without `plausible deniability' almost ubiquitous manifestations of \cp~violation about two orders of magnitude larger in $B$ decays. This big picture has been confirmed now in qualitative as well as impressively quantitative agreement with SM predictions: 
\begin{itemize}
\item 
Two \cp~insensitive observables, namely $|V(ub)/V(cb)|$ and $\Delta M_{B_d}/\Delta M_{B_s}$, imply that \cp~violation has to exist and in a way that at present is fully consistent with the measurements 
of $\epsilon$ and sin$2\phi_1$ and others. 
\item 
Time dependent \cp~asymmetries in the range of several $\times$ 10 \% have been established in 
$B_d \to \psi K_S$, $\pi^+\pi^-$ and $\eta^{\prime} K_S$ with several others on the brink of being found. 
\item 
{\em Direct} \cp~violation of about 10\% or even larger have been discovered in $B_d \to \pi^+\pi^-$ 
and $K^- \pi^+$. 
\item 
The first significant sign of \cp~violation in a charged meson has surfaced in 
$B^{\pm} \to K^{\pm} \rho^0$. 
\item 
The optimists among us might discern the first signs of tension between data and the predictions 
of CKM theory in  $|V(ub)/V(cb)|$ \& $\Delta M_{B_d}/\Delta M_{B_s}$ vs. sin$2\phi_1$ and in the 
\cp~asymmetries in $b \to s q \bar q$ vs. $b\to c \bar c s$ driven transitions.

\end{itemize}
For all these successes it is quite inappropriate to refer anymore to CKM theory as an `ansatz' with the 
latter's patronizing flavour \footnote{The German `ansatz' refers to an educated guess.}.  Instead I would characterize these developments as "the expected triumph 
of peculiar theory". However, as explained in Lect. III, it makes great sense -- actually it is mandatory to search for its phenomenological limitations in future even more sensitive data sets. This will require great advances in experimental sensitivity -- I have no doubt about their feasibility -- and further 
progress in out quantitative theoretical control over heavy flavour decays. I have presented some case studies, which give reason for optimism in this area as well. An essential element there is the availability 
of a comprehensive set of high quality data: among other things they provide the motivation for theorists to sharpen their tools, and they allow us to defend our estimates of uncertainties rather than merely state them.   

I will indulge myself in three more `cultural' conclusions: 
\begin{itemize}
\item 
The aforementioned "CKM Paradigm of Large \cp~Violation in $B$ Decays" is due to the confluence 
of several favourable, yet a priori less than likely factors that must be seen as gifts from nature who had 
\begin{itemize}
\item 
arranged for a huge top mass, 
\item 
a "long" $B$ lifetime, 
\item 
the $\Upsilon (4S)$ resonance being above the $B \bar B$, yet below the $B\bar B^*$ thresholds and 
\item 
regaled us previously with charm hadrons, which prompted the development detectors with an 
effective resolution that is needed to track $B$ decays.  
\end{itemize}
\item 
`Quantum mysteries' like EPR correlations with their intrinsic non-local features were essential  
for observing \cp~violation involving $B_d- \bar B_d$ oscillations in 
$\Upsilon (4S) \to B_d \bar B_d$ and to establish that indeed there is \ot~violation commensurate 
with \cp~violation. 
\item 
While hadronization is not easily brought under quantitative theoretical control it enhances greatly 
observable \cp~asymmetries and can provide most valuable cross checks for our interpretation of data.  
 
\end{itemize}   

\section{Lecture III: Probing the Flavour Paradigm of the {\em Emerging New} Standard Model}
\label{LECTIII}
 
\subsection{On the Incompleteness of the SM}
\label{INCOMPLET}

As described in the previous lectures the SM has scored novel -- i.e., qualitatively new -- successes in the last few years in the realm of flavour dynamics. Due to the very peculiar structure of the latter they have to be viewed as amazing. Yet even so the situation can be characterized with a slightly modified quote from Einstein: 
\begin{center} 
"We know a lot -- yet understand so little." 
\end{center} 
I.e., these successes do {\em not} invalidate  the general arguments in favour of the SM being 
{\em incomplete} -- the search for New Physics is as mandatory as ever. 

You have heard about the need to search for New Physics before and what the outcome has been of such efforts so far, have you not?  And it reminds you of a quote by Samuel Beckett: 
\begin{center}
"Ever tried? Ever failed? \\
No matter. \\
Try again. Fail again. Fail better."
\end{center}
Only an Irishman can express profound skepticism concerning the world in such a poetic way. 
Beckett actually spent most of his life in Paris, since Parisians like to listen to someone expressing such a world view, even while they do not share it. Being in the service of Notre Dame du Lac, the home of the `Fighting Irish', I cannot just ignore such advice. 

My colleague and friend Antonio Masiero likes to say: "You have to be lucky to find New Physics." True enough -- yet 
let me quote someone who just missed by one year being a fellow countryman of Masiero, namely 
Napoleon, who said: "Being lucky is part of the job description for generals." Quite seriously I think 
that if you as  
an high energy physicist do not believe that someday somewhere you will be a general -- maybe not 
in a major encounter, but at least in a skirmish -- then you are frankly in the wrong line of business.

My response to these concerns is: "Cheer up -- we know there is New Physics -- 
we will not fail forever!" I will marshall the arguments -- compelling ones in my judgment -- that point to the existence of New Physics.    

\subsubsection{Theoretical Shortcomings}
\label{THARG}

These arguments have been given already in the beginning of Lecture I. 
\begin{itemize}
\item 
{\em Quantization of electric charge}: While electric charge quantization 
\beq 
Q_e = 3 Q_d = - \frac{3}{2} Q_u 
\eeq
is an essential ingredient of the SM -- it allows to vitiate the ABJ anomaly -- it does not offer any 
understanding. It would naturally be explained through Grand Unification at very high energy scales 
implemented through, e.g., $SO(10)$ gauge dynamics. 
I call this the `guaranteed New Physics'. 

\item 
{\em Family Replication and CKM Structure}: We infer from the observed width of $Z^0$ decays that there are  three (light) neutrino species. The hierarchical pattern of CKM parameters as revealed by the data is so peculiar as to suggest that some other dynamical 
layer has to underlie it. I refer to it as `strongly suspected New Physics' or {\bf ssNP}. 
We are quite in the dark about its relevant scales. 
Saying we pin our hopes for explaining the family replication on Super-String or M theory is a scholarly way of saying 
we have hardly a clue what that {\bf ssNP} is. 

\item 
{\em Electroweak Symmetry Breaking and the Gauge Hierarchy}: What are the dynamics driving the electroweak symmetry breaking of 
$SU(2)_L\times U(1) \to U(1)_{QED}$. How can we tame  the instability of Higgs dynamics with its quadratic mass divergence? 
I find the arguments compelling that point to New Physics at the 
$\sim 1$ TeV scale -- like low-energy SUSY; therefore I call it the `confidently predicted' New Physics 
or {\bf cpNP}. 

\item 
Furthermore the more specific `Strong \cp~Problem' of QCD has not been resolved. 
Similar to the other shortcomings it is a purely theoretical problem in the sense that the offending coefficient for the \op~and \cp~odd operator $\tilde G\cdot G$ can be 
fine-tuned to zero , see Sect.\ref{FLY}, -- yet in my eyes that is not a flaw. 
\end{itemize}

\subsubsection{Experimental Signs}
\label{EXPARG}

Strong, albeit not conclusive (by itself) evidence for neutrino oscillations comes from the 
KAMLAND and K2K experiments in Japan studying the evolution of neutrino beams on earth. 

Yet compelling experimental evidence for the SM being incomplete comes from `heavenly signals', 
namely from astrophysics and cosmology.  
\begin{itemize}
\item 
{\em The Baryon Number of the Universe}: one finds only about one 
baryon per $10^9$ photons with the latter being mostly in the cosmic background radiation; there is 
no evidence for {\em primary} antimatter. 

$\ominus$ We know standard CKM dynamics is irrelevant for the Universe's baryon number. 

$\oplus$ Therefore New Physics has to exist. 

$\oplus$ The aforementioned New \cp~Paradigm tells us that \cp~violating phases can be large. 

\item 
{\em Dark Matter}: Analysis of the rotation curves of stars and galaxies reveal that there is a lot more 
`stuff' -- i.e. gravitating agents -- out there than meets the eye. About a quarter of the gravitating agents in the Universe are such dark matter, and they have to be mostly nonbaryonic. 

$\oplus$ The SM has {\em no} candidate for it. 

\item 
{\em Solar and Atmospheric $\nu$ Anomalies}: The sun has been `seen' by Super-Kamiokande in the 
light of neutrinos, as shown in Fig.\ref{SUNINNEUT}. Looking carefully one realizes that the sun looks paler than it should: more than half of the originally 
produced $\bar \nu _e$ disappear on the way to the earth by changing their identity. Muon neutrinos 
produced in the atmosphere perform a similar disappearance act. 
\begin{figure}[t]
\vspace{7.0cm}
\includegraphics{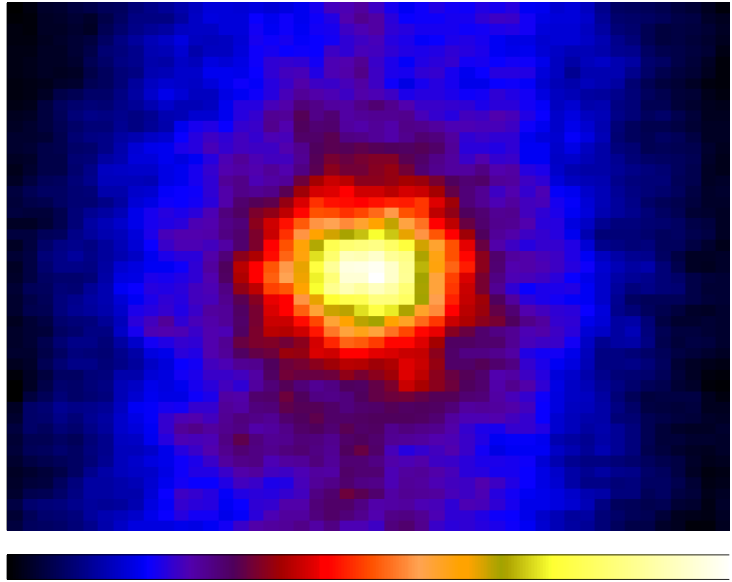}
\caption{The sun in the light of its neutrino emission as seen by the Super-Kamiokande detector; 
from Ref.\cite{SVOBODA}}
\label{SUNINNEUT} 
\end{figure}

These disappearances have to be attributed predominantly to neutrino oscillations (rather than 
neutrino decays). This requires neutrinos to carry {\em non}degenerate masses. 

\item 
{\em Dark Energy}: Type 1a supernovae are considered `standard candles'; i.e. considering 
their real light output known allows to infer their distance from their apparent brightness. 
When in 1998 two teams of researchers studied them at distance scales of about 
five billion light years, they found them to be fainter as a function of their redshift than what the 
conventional picture of the Universe's decelerating expansion would yield. Unless gravitational 
forces are modified over cosmological distances, one has to conclude the Universe is filled  with 
an hitherto completely unknown agent {\em accelerating} the expansion. A tiny, yet non-zero 
cosmological constant would apparently `do the trick' -- yet it would raise more fundamental 
puzzles.

\end{itemize}
These heavenly signals are unequivocal in pointing to New Physics, yet leave wide open 
the nature of this New Physics.

Thus we can be assured that New Physics exists `somehow' `somewhere', and quite likely 
even `nearby', namely around the TeV scale; above I have called the latter
{\bf cpNN}. The LHC program and the Linear Collider 
project are justified -- correctly -- to conduct campaigns for 
{\bf cpNP}. That is unlikely to shed light on the {\bf ssNP}, though it might. Likewise I would not 
{\em count} on 
a comprehensive and detailed program of heavy flavour studies to  
shed light on the {\bf ssNP} behind the flavour puzzle of the SM. 
Yet the argument is reasonably turned around: such a program 
will be essential to elucidate salient features of the 
{\bf cpNP} by probing the latter's flavour structure and having sensitivity to scales of order 10 TeV. 
One should keep in mind the following: one very popular example of 
{\bf cpNP} is supersymmetry; {\em yet it represents an organizing principle much more than even a class of theories}. I find it unlikely we can infer all required lessons by studying only flavour diagonal 
transitions. Heavy flavour decays provide a powerful and complementary probe of 
{\bf cpNP}. Their potential to reveal something about the {\bf ssNP} is a welcome extra not required 
for justifying efforts in that direction. 

Accordingly I see a dedicated heavy flavour program as an essential complement to the 
studies pursued at the high energy frontier at the TEVATRON, LHC and hopefully ILC. 
I will illustrate this assertion in the remainder of this lecture. 

\subsection{$\Delta S \neq 0$ -- the `Established Hero'}
\label{STRANGEHERO}

The chapter on $\Delta S \neq 0$ transitions is a most glorious one in the history of particle physics, 
as sketched in Table \ref{tab:STRANGELESSONS}.
\begin{table}
\begin{center}
\small{\begin{tabular}{ll}
\hline
Observation      & Lesson learnt     \\
\hline
      $\tau - \theta $ Puzzle   & \op~violation \\ 
      production rate $\gg$ decay rate  & concept of families \\ 
      suppression of flavour changing neutral currents  & GIM mechanism \& existence of charm \\
    $K_L \to \pi \pi$ & \cp~violation \& existence of top \\
    \hline 
\end{tabular}}
\caption{On the History of $\Delta S \neq 0$ Studies}
\label{tab:STRANGELESSONS}
\end{center}
\end{table} 
We should note that all these features, which now are pillars of the SM, were New Physics 
{\em at that time}! 

\subsubsection{Future `Bread \& Butter' Topics}

Detailed studies of radiative decays like $K\to \pi \gamma \gamma$ and 
$K \to \pi  \pi \gamma$ will allow deeper probes of chiral perturbation theory. The lessons thus 
obtained might lead to a better treatment of long distance dynamics' impact on the 
$\Delta I = 1/2$ rule, $\Delta M_K$, $\epsilon_K$ and $\epsilon ^{\prime}$. 

\subsubsection{The `Dark Horse'} 

The \ot~odd moment (see Sect.\ref{BASICSDISC}) 
\beq 
{\rm Pol}_{\perp}(\mu) \equiv \frac{\langle \vec s(\mu) \cdot (\vec p(\mu) \times \vec p(\pi))\rangle}
{|\vec p(\mu) \times \vec p(\pi)|} 
\eeq
measured in $K^+ \to \mu^+ \nu \pi^0$ would  
\begin{itemize}
\item 
represent genuine \ot~violation (as long as it exceeded the order $10^{-6}$ level) and 
\item 
constitute prima facie evidence for \cp~violation in {\em scalar} dynamics. 
\end{itemize}

\subsubsection{`Heresy'}
\label{HER}

The large \ot~odd correlation found in $K_L \to \pi^+\pi^- e^+e^-$ for the relative orientation 
of the $\pi^+\pi^-$ and $e^+e^-$ decay planes, see the discussion below in 
Sect.\ref{CPDIST}, is fully consistent with a \ot~violation as inferred from the \cp~violation expressed 
through $\epsilon_K$ -- yet it does not prove it \cite{BSTODD}. In an unabashedly contrived scenario 
-- something theorists usually avoid at great pains -- one could reconcile the data on 
$K_L \to \pi^+\pi^- e^+e^-$ with \ot~invariance without creating a conflict with known data. Yet the 
\cpt~violation required in this scenario would have to surface through \cite{BSTODD}
\beq 
\frac{\Gamma (K^+ \to \pi ^+ \pi^0) - \Gamma (K^- \to \pi ^- \pi^0)}
{\Gamma (K^+ \to \pi ^+ \pi^0) + \Gamma (K^- \to \pi ^- \pi^0)} > 10^{-3}
\eeq 

\subsubsection{The `Second Trojan War': $K \to \pi \nu \bar \nu$ }  
\label{SECTROJAN}

According to Greek Mythology the Trojan War described in Homer's Iliad was actually 
the second war over Troja. In a similar vein I view the heroic campaign over 
$K^0 - \bar K^0$ oscillations -- $\Delta M_K$, $\epsilon_K$ and $\epsilon ^{\prime}$ -- as 
a first one to be followed by a likewise epic struggle over the two ultra-rare modes 
$K^+ \to \pi^+ \nu \bar \nu$ and 
$K_L \to \pi^0 \nu \bar \nu$. This campaign has already been opened through the observation 
of the first through three events very roughly as expected within the SM. The second one, which 
requires \cp~violation for its mere existence, so far remains unobserved at a level well above 
SM predictions. These reactions 
are like `standard candles' for the SM: their rates are functions of 
$V(td)$ with a theoretical uncertainty of about 5\% and 2\% respectively, which is mainly due to 
the uncertainty in the charm quark mass. 

While their rates could be enhanced by New Physics greatly over their SM expectation, 
I personally find that somewhat unlikely for various reasons. Therefore I suggest one should aim for 
collecting ultimately about 1000 events of these modes to extract the value of 
$V(td)$ and/or identify likely signals of New Physics.

\subsection{The `King Kong' Scenario for New Physics Searches}
\label{KONG}

This scenario can be formulated as follows: "One is unlikely to encounter King Kong; yet 
once it happens one will have no doubt that one has come across something quite out of the ordinary!" 

What it means can be best illustrated with the historical precedent of $\Delta S \neq 0$ studies sketched 
above: the existence of New Physics can unequivocally be inferred if there is a 
{\em qualitative} conflict between data and expectation; i.e., if a theoretically `forbidden' process is 
found to proceed nevertheless -- like in $K_L \to \pi \pi$ -- or the discrepancy between expected and 
observed rates amounts to several orders of magnitude -- like in $K_L \to \mu ^+\mu^-$ or 
$\Delta M_K$. This does not mean that the effects are large or straightforward to discover -- only that they are much larger than the truly minute SM effects. 

History might repeat itself in the sense that future measurements might reveal such {\em qualitative} 
conflicts, where the case for the manifestation of New Physics is easily made. This does not mean that such measurements will be easy -- far from it,  as will become obvious. 

I have already mentioned one potential candidate for revealing such a qualitative conflict, namely 
the muon transverse polarization in $K_{\mu 3}$ decays. 

\subsubsection{Electric Dipole Moments}
\label{EDMS}

The energy shift of a system placed inside a weak 
electric field can be expressed through an expansion in terms of the components of that 
field $\vec E$: 
\beq 
\Delta {\cal E} = d_i E_i + d_{ij}E_iE_j + {\cal O}(E^3) 
\eeq
The coefficients $d_i$ of the term linear in the electric field form a vector $\vec d$, 
called an electric dipole moment (EDM). For a {\em non-}degenerate system -- it does not have to be elementary -- one infers from symmetry considerations that this vector has to be proportional to 
that systems spin: 
\beq 
\vec d \propto \vec s
\eeq
Yet, since 
\beq
E_i \stackrel{\ot} \to E_i \; \; , \; \; s_i \stackrel{\ot} \to - s_i 
\eeq
under time reversal \ot , a non-vanishing EDM constitutes \ot~violation. 

No EDM has been observed yet; the upper bounds of the neutron and electron EDM read 
as follows \cite{PDG06}: 
\bea 
d_N &<& 5 \cdot 10^{-26} \; {\rm e\, cm} \; \; \;  {\rm [from \; ultracold\; neutrons]}\\
d_e &<& 1.5 \cdot 10^{-27} \; {\rm e\, cm} \; \; \; {\rm [from\; atomic \; EDM]}
\eea
The experimental sensitivity achieved can be illustrated as follows: (i) An 
neutron EDM of $5\cdot 10^{-26}$ e cm of an object with a radius 
$r_N \sim 10^{-13}$ cm scales to a displacement of about 7 micron, i.e. less than 
the width of human hair, for an object of the size of the earth. (ii) Expressing the uncertainty 
in the measurement of the electron's magnetic dipole moment -- 
$\delta ((g-2)/2) \sim 10^{-11}$ in analogy to its EDM, one finds a sensitivity level of 
$\delta(F_2(0)/2m_e) \sim 2 \cdot 10^{-22}$ e cm compared to 
$d_e< 2\cdot 10^{-26}$ e cm. 

Despite the tremendous sensitivity reached these numbers are still several orders of magnitude above what is expected in CKM theory: 
\bea 
d_N^{CKM} &\leq &  10^{-30} \; {\rm e\, cm}  \\
d_e^{CKM} &\leq &  10^{-36} \; {\rm e\, cm} \; , 
\eea
where in $d_N^{CKM}$ I have ignored any contribution from the strong \cp~problem. 
These numbers are so tiny for reasons very specific to CKM theory, namely its chirality structure 
and the pattern in the quark and lepton masses. Yet New Physics scenarios with right-handed 
currents, flavour changing neutral currents, a non-minimal Higgs sector, heavy neutrinos etc. are 
likely to generate considerably larger numbers: $10^{-28} - 10^{-26}$ e cm represents a 
very possible range there quite irrespective of whether these new forces contribute to 
$\epsilon_K$ or not. This range appears to be within reach in the foreseeable future. 
There is a vibrant multiprong program going on at several places. 
Such experiments while being of the `table top' variety require tremendous efforts,  persistence and ingenuity -- yet the insights to be gained by finding a nonzero 
EDM somewhere are tremendous.

\subsubsection{Charm Decays}
\label{CHARMDEC}  

Charm dynamics is often viewed as physics with a great past -- it was instrumental in 
driving the paradigm shift from quarks as mathematical entities to physical objects and in 
providing essential support for accepting QCD as the theory of the strong interactions -- 
yet one without a future since the electroweak phenomenology for $\Delta C \neq 0$ transitions 
is decidedly on the `dull' side: `known' CKM parameters, slow $D^0 - \bar D^0$ oscillations, small 
\cp~asymmetries and extremely rare loop driven decays. 

Yet more thoughtful observers have realized that the very `dullness' of the SM phenomenology for charm 
provides us with a dual opportunity, namely to 
\begin{itemize}
\item
probe our quantitative understanding of QCD's nonperturbative dynamics thus calibrating our theoretical tools for $B$ decays and 
\item 
perform almost `zero-background' searches 
for New Physics. 
\end{itemize}
However the latter statement of `zero-background' has to be updated carefully since experiments over the last ten years have bounded the oscillation parameters $x_D$, $y_D$ to fall below  very few \% and direct \cp~asymmetries below several \%. While New Physics signals can still exceed SM predictions on 
\cp~asymmetries by orders of magnitude, they might not be large in absolute terms, as specified later 
\cite{GROSS2}. 

{\bf One should take note that charm is the only {\em up-}type quark allowing the full range of 
probes for New Physics, including flavour changing neutral currents}: while top quarks do not hadronize \cite{RAPALLO}, in the $u$ quark sector 
you cannot have $\pi^0 - \pi^0$ oscillations and many \cp~asymmetries are already ruled out by 
\cpt~invariance. My basic contention  is the following: {\em Charm transitions are a unique 
portal for obtaining a novel access to flavour dynamics with the experimental situation 
being a priori favourable (except for the lack of Cabibbo suppression)!} 

I will sketch such searches for New Physics in the context of $D^0 - \bar D^0$ oscillations and 
\cp~violation. 
\begin{enumerate}
\item 
Like for $K^0$ and $B^0$ mesons the oscillations of $D^0$ mesons represent a subtle quantum mechanical phenomenon of  
practical importance: it provides a probe for New Physics, albeit an ambiguous one, 
and constitutes an important ingredient for \cp~asymmetries arising in $D^0$ decays due to New Physics. 

In qualitative analogy to the $K^0$ and $B^0$ cases these phenomena can be characterized by two quantities, namely 
$x_D = \frac{\Delta M_D}{\Gamma_D}$ and $y_D =\frac{\Delta \Gamma_D}{2\Gamma_D}$.  
Oscillations  are slowed down in the SM due to GIM suppression and $SU(3)_{fl}$ symmetry. 
Comparing a {\em conservative} SM bound with the present data  
\beq 
x_D(SM), y_D(SM) < {\cal O}(0.01)  \; \; vs. \; \; 
\left. x_D\right|_{exp}  < 0.03 \; , \; \;  \left. y_D\right|_{exp} = 0.01 \pm 0.005 
\label{DOSC}
\eeq 
we conclude that the search has just now begun. There exists a considerable literature -- yet 
typically with several ad-hoc assumptions concerning the nonperturbative dynamics. It is widely understood that the usual quark box diagram is utterly irrelevant due to its untypically severe 
GIM suppression $(m_s/m_c)^4$. 
A systematic 
analysis based on an OPE treatment has been given in Ref.\cite{BUDOSC} in terms of powers of 
$1/m_c$ and $m_s$. Contributions from higher-dimensional operators with a much softer 
GIM reduction of $(m_s/\mu_{had})^2$ (even $m_s/\mu_{had}$ terms could arise) due to `condensate'  terms in the OPE  yield 
\beq 
\left. x_D (SM)\right|_{OPE}, \; \left. y_D (SM)\right|_{OPE} \sim {\cal O}(10^{-3}) \; . 
\eeq 
Ref.\cite{FALK} finds very similar numbers, albeit in a quite different approach. 

While one predicts similar numbers for $x_D(SM)$ and $y_D(SM)$, one should keep in mind 
that they arise in very different dynamical environments. $\Delta M_D$ is generated from 
{\em off}-shell intermediate states and thus is sensitive to New Physics, which could produce 
$x_D \sim {\cal O}(10^{-2})$. $\Delta \Gamma_D$ on the other hand is shaped by 
{\em on}-shell intermediate 
states; while it is hardly sensitive to New Physics, it involves much less averaging or `smearing' than 
$\Delta M_D$ making it thus much more vulnerable to violations of quark-hadron duality. Observing 
$y_D \sim 10^{-3}$ together with $x_D \sim 0.01$ would provide intriguing, though not conclusive 
evidence for New Physics, while $y_D \sim 0.01 \sim x_D$ would pose a true conundrum for its 
interpretation. 
\item 
Since the baryon number of the Universe implies the existence of New Physics in \cp~violating dynamics, it would be unwise not to undertake dedicated searches for \cp~asymmetries in 
charm decays, where the `background' from known physics is small: within the SM the effective weak phase is highly diluted, namely $\sim {\cal O}(\lambda ^4)$, and it can 
arise only in singly Cabibbo suppressed transitions, where one  
expects them to reach the 0.1 \% level; significantly larger values would signal New Physics.  
{\em Any} asymmetry in Cabibbo 
allowed or doubly suppressed channels requires the intervention of New Physics -- except for 
$D^{\pm}\to K_S\pi ^{\pm}$ \cite{CICERONE}, where the \cp~impurity in $K_S$ induces an asymmetry of 
$3.3\cdot 10^{-3}$. Several facts actually favour such searches: strong phase shifts 
required for direct \cp~violation to emerge in partial widths are in general large as are the branching ratios into relevant modes;  
finally \cp~asymmetries can be linear in New Physics amplitudes thus enhancing sensitivity to the 
latter.  As said above, the benchmark scale for KM asymmetries in singly Cabibbo suppressed 
partial widths is 
$0.1\%$. This does not exclude the possibility that CKM dynamics might exceptionally generate an \
asymmetry as `large' as 1\% in some special cases. It is therefore essential to analyze a host of 
channels. 

Decays to final states of {\em more than} two pseudoscalar or one pseudoscalar and one vector meson contain 
more dynamical information than given by their  widths; their distributions as described by Dalitz plots 
or \ot{\em -odd} moments can exhibit \cp~asymmetries that can be considerably larger than those for the 
width. Final state interactions while not necessary for the emergence of such effects, can fake a signal; 
yet that can be disentangled by comparing \ot{\em -odd} moments for \cp~conjugate modes. I view this as a very promising avenue, where we still have to develop the most effective analysis tools for small 
asymmetries.

\cp~violation involving $D^0 - \bar D^0$ oscillations can be searched for in final states common to $D^0$ 
and $\bar D^0$ decays like \cp~eigenstates -- $D^0 \to K_S\phi$, $K^+K^-$, $\pi^+\pi^-$ -- or 
doubly Cabibbo suppressed modes -- $D^0 \to K^+\pi^-$. The \cp~asymmetry is controlled by  
sin$\Delta m_Dt$ $\cdot$ Im$(q/p)\bar \rho (D\to f)$; within the SM both factors are small, namely 
$\sim {\cal O}(10^{-3})$, making such an asymmetry unobservably tiny -- unless there is New Physics! 
One should note 
that this observable is linear in $x_D$ rather than quadratic as for \cp~insensitive quantities.  
$D^0 - \bar D^0$ oscillations, \cp~violation and New Physics might thus be discovered simultaneously in a transition. 

One wants to reach the level at which SM effects are 
likely to emerge, namely down to time-{\em dependent} \cp~asymmetries 
in $D^0 \to K_S\phi$, $K^+K^-$, $\pi^+\pi^-$ [$K^+\pi^-$] down to $10^{-5}$ [$10^{-4}$] and 
{\em direct} \cp~asymmetries in partial widths and Dalitz plots down to $10^{-3}$.

\end{enumerate}
\subsubsubsection{\cp~Asymmetries in Final State Distributions}
\label{CPDIST}

So far \cp~violation has surfaced in time-integrated or time-dependent partial widths with 
one notable exception. A large \ot~odd moment was found in the rare $K_L$ mode --
BR$(K_L \ra \pi ^+ \pi ^- e^+ e^-) = 
(3.32 \pm 0.14 \pm 0.28 ) \cdot 10^{-7}$:    
With $\phi$ defined as the angle between the planes 
spanned by 
the two pions and the two leptons in the $K_L$ 
restframe:  
$$   
\phi \equiv \angle ( \vec n_l, \vec n_{\pi})
$$ 
\beq  
\vec n_l = \vec p_{e ^+}\times \vec p_{e ^-}/
|\vec p_{e ^+}\times \vec p_{e ^-}| \; , \;  
\vec n_{\pi} = \vec p_{\pi ^+}\times \vec p_{\pi ^-}/ 
|\vec p_{\pi ^+}\times \vec p_{\pi ^-}|
\label{PHISEHGAL}
\eeq    
one analyzes 
the decay rate as a function of $\phi$: 
\beq 
\frac{d\Gamma}{d\phi} = \Gamma _1 {\rm cos}^2\phi + 
\Gamma _2 {\rm sin}^2\phi + 
\Gamma _3 {\rm cos}\phi \, {\rm sin} \phi 
\eeq 
Since  
\beq 
{\rm cos}\phi \, {\rm sin} \phi = 
(\vec n_l \times \vec n_{\pi}) \cdot 
(\vec p_{\pi ^+} + \vec p_{\pi ^-}) 
(\vec n_l \cdot \vec n_{\pi})/
|\vec p_{\pi ^+} + \vec p_{\pi ^-}| 
\eeq
one notes that 
\beq 
{\rm cos}\phi \, {\rm sin} \phi \; \; \; 
\stackrel{{\bf T},{\bf CP}}{\longrightarrow} \; \; \; 
- \; {\rm cos}\phi \, {\rm sin} \phi 
\eeq    
under both \ot~ and \cp~transformations; i.e. the observable  
$\Gamma _3$ represents a \ot~- and \cp~-odd correlation. 
It can be projected out by comparing the $\phi$ 
distribution integrated over two quadrants: 
\beq 
A = 
\frac{\int _0^{\pi/2} d\phi \frac{d\Gamma}{d\phi} - 
\int _{\pi /2}^{\pi} d\phi \frac{d\Gamma}{d\phi}}
{\int _0^{\pi} d\phi \frac{d\Gamma}{d\phi}} = 
\frac{2\Gamma _3}{\pi (\Gamma _1 + \Gamma _2)} 
\eeq
It was first measured by KTEV and then confirmed by NA48 \cite{PDG06}:  
\beq 
A = (13.7 \pm 1.5)\% \, .
\label{KTEVSEHGAL2}
\eeq 
$A\neq 0$ is induced by $\epsilon_K$, the \cp~violation in the $K^0 - \bar K^0$ mass matrix, 
leading to the prediction \cite{SEGHALKL}
\beq 
A = (14.3 \pm 1.3)\% \, .
\eeq
The observed value for the \ot~odd moment $A$ is fully consistent with \ot~violation. Yet 
$A\neq 0$ {\em by itself} does not establish 
\ot~violation \cite{BSTODD}. 

One should note that this sizable forward-backward asymmetry is driven by the tiny quantity 
$|\eta_{+-}| \simeq 0.0023$, which can be understood. For 
$K_L \to \pi ^+ \pi ^- e^+ e^-$ is driven by the two sub-processes 
\bea 
K_L &\stackrel{\not {\cp} \&\Delta S =1}{\longrightarrow} \pi^+\pi^- 
\stackrel{E1}{\longrightarrow} \pi^+\pi^- \gamma ^* \to \pi ^+ \pi ^- e^+ e^- 
\\
K_L &\stackrel{M1\& \Delta S =1}{\longrightarrow} \pi^+\pi^- \gamma ^* \to \pi ^+ \pi ^- e^+ e^- \; , 
\eea
where the first reaction is suppressed, since it requires \cp~violation in  
$K_L \to 2\pi$, and the second one, since it involves an $M1$ transition.  Those two a priori very 
different suppression mechanisms happen to yield comparable amplitudes, which thus generate 
sizable interference. The price one pays is the small branching ratio.  

$D$ decays can be treated in an analogous way.  Consider the Cabibbo suppressed channel 
\footnote{This mode can exhibit direct \cp~violation even within the SM.}
\beq 
\stackrel{(-)}D \to K \bar K \pi^+\pi^-
\eeq
and define by $\phi$ now the angle between the $K \bar K$ and $\pi^+\pi^-$ planes. Then 
one has 
\bea 
\frac{d\Gamma}{d\phi}(D \to K \bar K\pi^+\pi^-) &=& \Gamma_1 {\rm cos}^2 \phi + 
\Gamma_2 {\rm sin}^2 \phi + \Gamma_3 {\rm cos} \phi {\rm sin}\phi \\
\frac{d\Gamma}{d\phi}(\bar D \to K \bar K\pi^+\pi^-) &=& \bar \Gamma_1 {\rm cos}^2 \phi + 
\bar \Gamma_2 {\rm sin}^2 \phi + \bar \Gamma_3 {\rm cos} \phi {\rm sin}\phi 
\eea
As before the partial width for $D[\bar D] \to K\bar K \pi^+\pi^-$ is given by 
$\Gamma_{1,2} [\bar \Gamma_{1,2}]$; $\Gamma_1 \neq \bar \Gamma_1$ or 
$\Gamma_2 \neq \bar \Gamma_2$ represents direct \cp~violation in the partial width. 
$\Gamma_3 \& \bar \Gamma_3$ constitute \ot~odd correlations. By themselves they do not necessarily 
indicate \cp~violation, since they can be induced by strong final state interactions. However 
\beq 
\Gamma_3 \neq \bar \Gamma_3 \; \; \Longrightarrow \cp~{\rm violation!}
\eeq 
It is quite possible or even likely that a difference in $\Gamma_3$ vs. $\bar \Gamma_3$ 
is significantly larger than in $\Gamma_1$ vs. $\bar \Gamma_1$ or 
$\Gamma_2$ vs. $\bar \Gamma_2$. Furthermore one can expect that differences in detection 
efficiencies can be handled by comparing $\Gamma_3$ with $\Gamma_{1,2}$ and 
$\bar \Gamma_3$ with $\bar \Gamma_{1,2}$. A pioneering search for such an effect has been 
undertaken by FOCUS \cite{PEDRINI}.

\subsubsection{\cp~Violation in the Lepton Sector}
\label{CPVLEPT}

I find the conjecture that baryogenesis is a {\em secondary} phenomenon driven by {\em primary}  leptogenesis a most intriguing and attractive one also for philosophical reasons 
\footnote{For it would 
complete what is usually called the Copernican Revolution 
\cite{ARAB}: first our Earth was removed from the center of the Universe, then in due course our Sun, our Milky Way and local cluster; few scientists believe life exists only on our Earth. Realizing that the stuff we are mostly made out of -- protons and neutrons -- 
are just a cosmic `afterthought' fits this pattern, which culminates in the dawning realization that even {\em our} Universe is just one among innumerable others, albeit a most unusual one.}. Yet then it becomes mandatory to search for \cp~violation in the lepton sector in a most dedicated fashion. 

In Sect. \ref{EDMS} I have sketched the importance of measuring {\em electric dipole moments}  as accurately as possible. The electron's EDM is a most sensitive probe of 
\cp~violation in leptodynamics. Comparing the present experimental and CKM upper bounds, 
respectively  
\beq 
d_e^{exp} \leq 1.5 \cdot 10 ^{-27} \; \; {\rm e \; cm} \; \; \; vs. \; \; \; 
d_e^{CKM} \leq 10 ^{-36} \; \; {\rm e \; cm}
\eeq
we see there is a wide window of several orders of magnitude, where New Physics could surface in an unambiguous way. This observation is reinforced by the realization that New Physics scenarios can 
naturally generate $d_e > 10^{-28}$e cm, while of only secondary significance in $\epsilon_K$, 
$\epsilon^{\prime}$ and sin$2\phi_i$. 

The importance that at least part of the HEP community attributes to finding \cp~violation in leptodynamics is best demonstrated by the efforts contemplated for observing \cp~asymmetries 
in {\em neutrino oscillations}. Clearly hadronization will be the least of the concerns, yet one has to 
disentangle genuine \cp~violation from matter enhancements, since the neutrino oscillations can be studied only in a matter, not an antimatter environment. Our colleagues involved in such endeavours 
will rue their previous complaints about hadronization and remember the wisdom of an ancient 
Greek saying: 

\begin{center}
"When the gods want to really harm you, they fulfill your wishes."   
\end{center} 

\subsubsection{The Decays of $\tau$ Leptons -- the Next `Hero Candidate'}
\label{NEXTHERO}

Like charm hadrons the $\tau $ lepton is often viewed as  a system with a great past, but hardly a 
future. Again I think this is a very misguided view and I will illustrate it with two examples. 

Searching for $\tau ^{\pm} \to \mu ^{\pm} \mu ^+\mu ^-$ (and its variants) -- 
processes forbidden in the SM -- is particularly intriguing, since it involves only `down-type' leptons 
of the second and third family and is thus the complete analogy of the quark lepton process 
$b \to s \bar s s$ driving $B_s \to \phi K_S$, which has recently attracted such strong attention. 
Following this analogy literally one guestimates ${\rm BR}(\tau \to 3 \mu) \sim 10^{-8}$ to be 
compared with the present bound from BELLE  
\beq 
{\rm BR}(\tau \to 3 \mu) \leq 2\cdot 10^{-7} \; . 
\eeq
It would be very interesting to know what the 
$\tau$ production rate at the hadronic colliders is and whether they could be competitive or even superior with the $B$ factories in such a search.  

In my judgment $\tau$ decays -- together with electric dipole moments for leptons and possibly $\nu$ oscillations referred to above -- provide the best stage to search for manifestations of 
\cp~breaking leptodynamics. 

The most promising channels for exhibiting \cp~asymmetries are $\tau \to \nu K \pi$, since due to 
the heaviness of the lepton and quark flavours they are most sensitive to nonminimal Higgs dynamics,  
and they can show asymmetries also in the final state distributions rather than integrated rates 
\cite{KUHN}.  

There is also a {\em unique}  opportunity in $e^+e^- \to \tau ^+ \tau ^-$: since the $\tau$ pair is produced with its spins aligned, the decay of one $\tau$ can `tag' the spin of the other $\tau$. I.e., 
one can probe {\em spin-dependent} \cp~asymmetries with {\em unpolarized} beams. This provides 
higher sensitivity and more control over systematic uncertainties. 

I feel these features are not sufficiently 
appreciated even by proponents of Super-B factories. It has been recently pointed \cite{BSTAU}  
out that based on known physics one can actually predict a 
\cp~asymmetry: 
\beq 
\frac{\Gamma(\tau^+\to K_S \pi^+ \overline \nu)-\Gamma(\tau^-\to K_S \pi^- \nu)}
{\Gamma(\tau^+\to K_S \pi^+ \overline \nu)+\Gamma(\tau^-\to K_S \pi^- \nu)}= 
(3.27 \pm 0.12)\times 10^{-3}
\label{CPKS}
\eeq
due to $K_S$'s preference for antimatter.

\subsection{Future Studies of $B_{u,d}$ Decays}
\label{BFUTURE}

The successes of CKM theory to describe flavour dynamics do {\em not} tell us at all that 
New Physics does not affect $B$ decays; the message is that {\em typically} we can{\em not} count on 
a {\em numerically} massive impact there. Shifting an asymetry by, say, ten percentage points -- for example from 40 \%  to 50 \% -- might already be on the large side. Thus our aim has to be to aim for 
uncertainties that do not exceed a few percent. 

The discussion given in Lect. II shows that an integrated luminosity of 1 $ab^{-1}$ at the $B$ factories will fall short of such a goal for $B_d \to \pi \pi$, $B^{\pm} \to D^{neut}K^{\pm}$ and in particular also 
for the modes driven by $b \to s q \bar q$.  Even ten times that statistics  would not suffice in view of the 
`big picture', i.e. when one includes other rare transitions. Of course we are in the very fortunate 
situation that one of the LHC experiments, namely LHCb, is dedicated to undertaking precise 
measurements of the weak decays of beauty hadrons. Thus we can expect a stream of high quality data to be forthcoming over the next several years. I will briefly address different classes of rare decays with 
different motivations and requirements.  

\subsubsection{Radiative $B$ Decays}
\label{RADDEC}
\subsubsubsection{$B \to \gamma X$}
As already mentioned in Lect. II 
\begin{itemize}
\item 
the branching ratio for $B \to \gamma X_s$ has been measured with good accuracy and in agreement with the SM prediction; 
\item
the photon energy spectrum has been determined 
down to $E_{\gamma} = 1.9$ GeV or even 1.8 GeV; its moments have provided important information 
on the heavy quark parameters, in particular the $b$ quark mass $m_b$. 
\end{itemize}
There is another more subtle observable, for which the SM makes a rather accurate prediction, namely 
the photon polarization: the SM electroweak Penguin operator produces mostly {\em left-}handed photons. New Physics on the other hand can generate right-handed photons as well. They would hardly be noticed in the total rate: since left- and right-handed photons cannot interfere, the rate would be 
{\em quadratic} in the New Physics amplitude. The gluonic counterpart to such a New Physics 
$b\to s \gamma_R$ could however contribute {\em linearly} in amplitude to the \cp~asymmetry in 
$B_d \to \phi K_S$ and other $b\to s q \bar q$ modes and thus become significant there. 

Rather  than measure the photon polarization, which seems hardly feasible, one can infer it from 
measuring angular correlations in 
$B \to \gamma K^{**} \to \gamma (K\pi\pi)$ modes \cite{PIRJOL}.

It has been suggested to distinguish $B \to \gamma X_d$ against $B \to \gamma X_s$ to extract 
$V(td)/V(ts)|$ or to probe for New Physics using a value for $V(td)/V(ts)|$ extracted from 
$\Delta M_{B_d}/\Delta M_{B_s}$ or the overall CKM fit. This does not seem to be a hopeless 
undertaking -- at a Super-B factory.

\subsubsubsection{$B \to l^+l^- X$}

We are just at the beginning of studying $B \to l^+l^-X$, and it has to be 
pursued in a dedicated and comprehensive manner for the following reasons: 
\begin{itemize}
\item 
With the final state being more complex than for $B \to \gamma X$, it is described by a larger number 
of observables: rates, spectra of the lepton pair masses and the lepton energies, their forward-backward asymmetries and \cp~asymmetries. 
\item 
These observables provide independent information, since there is a larger number of effective transition operators than for $B \to \gamma X$. By the same token there is a much wider window 
to find New Physics and even diagnose its salient features. 
\item 
It will take the statistics of a Super-B factory to mine this wealth of information on New Physics. 
\item 
Essential insights can be gained also by analyzing the exclusive channel $B\to l^+l^-K^*$ at hadronic 
colliders like the LHC, in particular the position of the zero in the lepton forward-backward asymmetry. 
For the latter appears to be fairly insensitive to hadronization effects in this exclusive mode 
\cite{HILLER1}. It will be important to analyze quantitatively down to which level of accuracy this feature 
persists. 
\end{itemize}

\subsubsection{Semileptonic Decays Involving $\tau$ Leptons}
\label{SLBTAU}

There are some relatively rare $B$ decays that could conceivably reveal New Physics, although they 
proceed already on the tree level. One well known example is $B^+ \to \tau \nu$ that is sensitive 
to charged Higgs fields. This applies also to semileptonic $B$ decays. As will be described in 
Sect.\ref{ADDING}, 
the Heavy Quark Expansion (HQE) has provided a sturdy and accurate description of 
$B \to l \nu X_c$ that allowed to extract $|V(cb)|$ with less than 2\% uncertainty. With it and other heavy quark parameters determined with considerable accuracy one can predict 
$\Gamma (B\to \tau \nu X_c)$  within the SM and compare it with the data. 
A discrepancy can be attributed to New Physics, presumably in the form of a {\em charged} 
Higgs field. 
Measuring also its hadronic mass moments can serve as a valuable cross check. Such studies will probably require the statistics 
of a Super-B factory. 

This is true also for studying the exclusive channel $B \to \tau \nu D$. As pointed out in Ref.\cite{MIKI}, 
one could find that the ratio $\Gamma (B \to \tau \nu D)/\Gamma (B \to \mu \nu D)$ deviates from its 
SM value due to the exchange of a charged Higgs boson with a mass of even several hundred GeV. 
This is the case in particular for `large tg$\beta$ scenarios' of two-Higgs-doublet models.  
There is a complication, though. Contrary to the suggestion in the literature the hadronic form factors 
do {\em not} drop out from this ratio. One should keep in mind that (i) the contribution from the second form factor $f_-$, which is proportional to the square of the lepton mass, cannot be ignored for 
$B \to \tau \nu D$ and (ii) the form factors  are not taken at the same momentum transfer in 
the two modes. 

These complications can be overcome by Uraltsev's BPS approximation \cite{BPS}.  
Relying on it one can extract $|V(cb)|$ from $B \to e/\mu \nu D$ and compare it with the 
`true' value obtained from $\Gamma _{SL}(B)$. {\em If} this comparison is successful and our 
theoretical control over $B \to l \nu D$ thus validated, one can apply the BPS approximation 
to $B \to \tau \nu D$. Since, as mentioned above, the second form factor $f_-$ can be measured there, 
one has another cross check.

\subsection{$B_s$ Decays -- an Independent Chapter in Nature's Book}
\label{BSDEC}

When the program for the $B$ factories was planned, it was thought that studying $B_s$ transitions 
will be required to construct the CKM triangle, namely to determine one of its sides and the angle 
$\phi_3$. As discussed above a powerful method has been developed to extract 
$\phi_3$ from $B^{\pm} \to D^{neut}K^{\pm}$, the effort has started to obtain 
$|V(td)/V(ts)|$ from $\Gamma (B \to \gamma \rho/\omega)/\Gamma (B \to \gamma K^*)$ 
and $B_s - \bar B_s$ oscillations have already been resolved at the TEVATRON. 
None of this, however, reduces the importance of a future 
comprehensive program to study $B_s$ decays -- on the contrary! With the basic CKM parameters 
fixed or to be fixed in $B_{u,d}$ decays, $B_s$ transitions can be harnessed as powerful probes 
for New Physics and its features. 

In this context it is essential to think `outside the box' -- pun intended. The point here is that several 
relations that hold in the SM (as implemented through quark box and other loop diagrams) are 
unlikely to extend beyond minimal extensions of the SM. In that sense $B_{u,d}$ and $B_s$ decays 
constitute two different and complementary chapters in Nature's book on fundamental dynamics. 

\subsubsection{CP~Violation in Non-Leptonic $B_s$ Decays}
\label{NLBS}

One class of nonleptonic $B_s$ transitions does not follow the paradigm of large \cp~violation in 
$B$ decays \cite{BS80}: 
$$ 
A_{\cp}(B_s(t) \to [\psi \phi]_{l=0}/\psi \eta ) = {\rm sin}2\phi (B_s) {\rm sin}\Delta M(B_s)t 
$$
\beq
{\rm sin}2\phi (B_s) =
{\rm Im}\left[ \frac{(V^*(tb)V(ts))^2}{|V^*(tb)V(ts)|^2}\frac{(V(cb)V^*(cs))^2}{(V(cb)V^*(cs))^2}  \right]
\simeq 2\lambda ^2 \eta   \sim 0.02  \; . 
\label{BSPSISM}
\eeq
This is easily understood: on the leading CKM level only quarks of the second and third families 
contribute to $B_s$ oscillations and $B_s \to \psi \phi$ or $\psi \eta$; therefore on that level there can be no \cp~violation making the \cp~asymmetry Cabibbo suppressed. {\em Yet New Physics of various ilks can quite conceivably generate sin$2\phi (B_s) \sim $ several $\times$ 10 \%.} 

Analyzing the decay rate evolution in proper time of 
\beq 
B_s(t) \to \phi \phi 
\label{BSPHIPHI} 
\eeq
with its direct as well as indirect \cp~violation is much more than a repetition of the 
$B_d(t) \to \phi K_S$ saga: 
\begin{itemize}
\item 
${\cal M}_{12}(B_s)$ and ${\cal M}_{12}(B_d)$ -- the off-diagonal elements in the mass matrices for 
$B_s$ and $B_d$ mesons, respectively -- provide in principle independent pieces of information 
on $\Delta B=2$ dynamics. 
\item
While the final state $\phi K_S$ is described by a single partial wave, namely $l=1$, there are 
three partial waves in $\phi \phi$, namely $l= 0,1,2$. Disentangling the three partial rates and their 
\cp~asymmetries -- or at least separating $l$ = even and odd contributions -- provides a new 
diagnostics about the underlying dynamics. 
\end{itemize}

\subsubsection{Leptonic, Semileptonic and Radiative Modes}
\label{BSSLRAD}

The decays into a lepton pair and to `wrong-sign' leptons should be studied also for $B_d$ mesons; however here I discuss only $B_s$ decays, where one can expect more dramatic effects.

$\bullet$ The mode $B_s \to \mu^+\mu^-$ is necessarily very rare since it suffers from helicity 
suppression $\propto (m(\mu)/M(B_s))^2$ and `wave function suppression' 
$\propto (f_B/M(B_s))^2$, which reflects the practically zero range of the weak interactions. 
Within the SM one predicts 
\beq 
{\rm BR} (B_s \to \mu^+\mu^-)|_{SM} \sim 3 \cdot 10^{-9} 
\eeq
These tiny factors can be partially compensated in some 
large tg$\beta$ SUSY scenarios, where an  enhancement factor of tg$^6\beta$ arises 
\cite{KOLDAETAL}, which could produce a rate at the experimental bound of $10^{-7}$. 

$\bullet$ Due to the rapid $B_s$ oscillations those mesons have a practically equal probability 
to decay into `wrong' and `right' sign leptons. One can then search for an asymmetry in the 
wrong sign rate for mesons that initially were $B_s$ and $\bar B_s$: 
\beq 
a_{SL}(B_s) \equiv \frac{\Gamma (\bar B_s \to l^+X) - \Gamma (B_s \to l^-X)}
{\Gamma (\bar B_s \to l^+X) + \Gamma (B_s \to l^-X)}
\eeq
This observable is necessarily small; among other things it is proportional to 
$\frac{\Delta \Gamma _{B_s}}{\Delta M_{B_s}} \ll 1$. The theoretical CKM predictions are not very precise, yet certainly tiny \cite{LENZNEW}: 
\beq 
a_{SL}(B_s) \sim 2 \cdot 10^{-5} \; , \; a_{SL}(B_d) \sim 4 \cdot 10^{-4} \; ; 
\eeq
$a_{SL}(B_s)$ suffers a suppression quite specific to CKM dynamics; analogous to $B_s \to \psi \phi$ 
quarks of only the second and third family participate on the leading CKM level, which therefore 
cannot exhibit \cp~violation. Yet again, New Physics can enhance $a_{SL}(B_s)$, this time  
by two orders of magnitude up to the 1\% level. 

$\bullet$ As already emphasized $B_s \to \gamma X$ and $B_s \to l^+l^-X$ should be studied in a 
comprehensive manner.  

\subsection{Instead of a Summary: On the Future HEP Landscape -- a Call to Well-Reasoned 
Action}
\label{SUM5}\footnote{Originally I had intended to name this Section `A call to Arms'. Yet recent 
events have reminded us that when the drums of war sound, reason all to often is left behind.}

The situation of the SM, as it enters the third millenium, can be characterized through 
several statements: 
\begin{enumerate}
\item 
There is a new dimension due to the findings on $B$ decays:  one has established the first 
\cp~asymmetries outside the $K^0 - \bar K^0$ complex in four $B_d$ modes-- as predicted 
qualitativly as well as quantitatively by CKM dynamics: 
\begin{itemize}
\item 
\beq 
B_d (t) \to \psi K_S \; ; 
\eeq
\item 
\beq 
B_d(t) \to \pi^+\pi^- \; ; 
\eeq
\item 
\beq 
B_d \to K^+\pi^- \; ; 
\eeq
\item 
\beq 
B_d(t) \to \eta^{\prime}K_S \; . 
\eeq

\end{itemize} 
Taken together with the other established signals -- $K^0(t) \to 2\pi$ and 
$|\eta_{+-}| \neq |\eta_{00}|$ -- we see that in all these cases except for $B_d \to K^+\pi^-$ 
the intervention of meson-antimeson oscillations was instrumental in \cp~violation becoming 
observable. This is why I write $B_d[K^0] (t) \to f$. For practical reasons this holds even for $|\eta_{+-}| \neq |\eta_{00}|$. 

For the first time strong evidence has emerged for \cp~violation in the decays of a charged state, namely 
in $B^{\pm} \to K^{\pm} \rho^0$. 

The SM's success here can be stated more succinctly as follows: 
\begin{itemize}
\item 
From a tiny signal of $|\eta_{+-}| \simeq 0.0023$ one successfully 
infers \cp~asymetries in $B$ decays two orders 
of magnitude larger, namely sin$2\phi_1 \simeq 0.7$ in $B_d(t) \to \psi K_S$. 
\item 
From the measured values of two \cp~insensitive quantities -- $|V(ub)/V(cb)|$ in semileptonic $B$ 
decays and $|V(td)/V(ts)|$ in $B^0 - \bar B^0$ oscillations -- one deduces the existence of 
\cp~violation in $K_L \to 2\pi$ and $B_d (t) \to \psi K_S$ even in quantitative agreement with the data. 
\end{itemize}

We know now that CKM dynamics provides at least the lion's share in the observed \cp~asymmetries. 
The CKM description thus has become a {\em tested} theory. Rather then searching for 
{\em alternatives} to CKM dynamics we hunt for {\em corrections} to it. 
\item 
None of these novel successes of the SM invalidate the theoretical arguments for it being incomplete. 
There is also clean evidence of mostly heavenly origin for New Physics, namely 
\begin{itemize}
\item 
neutrino oscillations, 
\item 
dark matter, 
\item 
presumably dark energy, 
\item 
probably the baryon number of our Universe and 
\item 
possibly the Strong \cp~Problem. 
\end{itemize} 

\item 
Flavour dynamics has become even more intriguing due to the emergence of neutrino 
oscillations. We do not understand the structure of the CKM matrix in any profound way -- and 
neither the PMNS matrix, its leptonic counterpart. Presumably we do understand why they look different, since only neutrinos can possess Majorana masses, which can give rise to the 
`see-saw' mechanism. 

Sometimes it is thought that the existence of two puzzles makes their resolution harder. I feel 
the opposite way: having a larger set of observables allows us to direct more questions to Nature, 
if we are sufficiently persistent, and learn from her answers. 
\footnote{Allow me a historical analogy: in the 1950's it was once suggested to a French politician 
that the 
French government's lack of enthusiasm for German re-unification showed that the French had not learnt to overcome their dislike of Germany. He replied with aplomb: "On the contrary, Monsieur!  
We truly love Germany and are therefore overjoyed that there are two Germanies we can love. Why would we change that?"}
\item 
The next `Grand Challenge' after studying the dynamics behind the electroweak phase transition is 
to find \cp~violation in the lepton sector -- anywhere. 
\item 
While the quantization of electric charge is an essential ingredient of the SM,  it  does not offer any understanding of it. It would naturally be explained through 
Grand Unification at very high energy scales. I refer to it as the `guaranteed New Physics', 
see Sect.\ref{THARG}. 
\item 
The SM's success in describing flavour transitions is not matched by a deeper understanding of the flavour structure, namely the patterns in the fermion masses and CKM parameters. 
For those do not appear to be of an accidental nature. I have referred to the dynamics generating the flavour structure as the 
`strongly suggested' New Physics ({\bf ssNP}), see Sect.\ref{THARG}. 
\item 
Discovering the {\bf cpNP} that drives the electroweak phase transition 
has been the justification for the LHC program, which will come online soon. Personally I am very partisan to the idea that the {\bf cpNP} will be of the 
SUSY type. Yet SUSY is an organizing principle rather than a class of theories, let alone a theory. 
We are actually quite ignorant about how to implement the one empirical feature of SUSY that has been established beyond any doubt, namely that it is broken. 
\item 
The LHC is likely, I believe, to uncover the {\bf cpNP}, and I have not given up hope that the 
TEVATRON will catch the first glimpses of it. Yet the LHC and a forteriori the TEVATRON  are primarily 
discovery machines. The ILC project is motivated as a more surgical probe to map out the salient features of that {\bf cpNP}. 
\item 
This {\bf cpNP} is unlikely to shed light on the {\bf ssNP} behind the flavour puzzle of the SM, although one should not rule out such a most fortunate development. On the other hand New Physics even at the 
$\sim $ 10 - 100 TeV scale could well affect flavour transitions significantly through virtual effects. A comprehensive 
and dedicated program of heavy flavour studies might actually elucidate salient features of the 
{\bf cpNP} that could not be probed in any other way. Such a program is thus 
complementary to the one pursued at the TEVATRON, the LHC and hopefully at the ILC and -- 
I firmly believe  -- actually necessary rather than a luxury to identify the {\bf cpNP}. 

To put it in more general terms: Heavy flavour studies 
\begin{itemize}
\item 
are of fundamental importance, 
\item 
many of its lessons cannot be obtained any other way and 
\item 
they cannot become obsolete. 

\end{itemize} 
I.e., no matter what studies of high $p_{\perp}$ physics at the LHC and ILC will or 
will not show -- comprehensive and detailed studies of flavour dynamics will remain crucial 
in our efforts to reveal Nature's Grand Design. 
\item 
Yet a note of caution has to be expressed as well. Crucial manifestations of New Physics in flavour dynamics are likely to be subtle. Thus we have to succeed in acquiring data as well as interpreting them  
with {\em high precision}. Obviously this represents a stiff challenge -- however one that I believe we can meet, if we prepare ourselves properly as I have exemplified in Sect.\ref{HQTH}.

\end{enumerate}
One of three possible scenarios will emerge in the next several years. 
\begin{enumerate} 
\item 
{\em The optimal scenario}: New Physics has been observed in "high $p_{\perp}$ physics", i.e. through the production of new quanta at the TEVATRON and/or LHC. Then it is {\em imperative} to study the impact of such New Physics on flavour dynamics; even if it should turn out to have none, this is an important piece of information, no matter how frustrating it would be to my 
experimental colleagues. Knowing the typical mass scale of that New Physics from collider data will be of great help to estimate its impact on heavy flavour transitions.  

\item 
{\em The intriguing scenario}: Deviations from the SM have been established in heavy flavour decays -- like the $B \to \phi K_S$ \cp~asymmetry or an excess in $\Gamma (K\to \pi \nu \bar \nu )$ -- without a clear signal for New Physics in high $p_{\perp}$ physics. A variant of this scenario has already emerged through the observations of neutrino 
oscillations. 

\item 
{\em The frustrating scenario}: No deviation from SM predictions have been identified. 

\end{enumerate}
I am optimistic it will be the `optimal' scenario, quite possibly with some elements of the 'intriguing' one. Of course one cannot rule out the `frustrating' scenario; yet we should not treat it as a case for defeatism: a possible failure to identify New Physics in future experiments at the hadronic colliders (or the $B$ factories) does not -- in my judgment -- invalidate the persuasiveness of the theoretical arguments and experimental evidence pointing to the incompleteness of the SM. 
It `merely' means we have to increase the sensitivity of our probes. 
{\bf I firmly believe a 
Super-flavour factory with a luminosity of order $10^{36}$ cm$^{-2}$ s$^{-1}$ or more for the 
study of beauty, charm and $\tau$ decays has to be an integral part of our future efforts towards deciphering Nature's basic code.}  
For a handful of even perfectly measured transitions will not be sufficient for the task at hand -- a 
{\em comprehensive} body of {\em accurate} data will be essential. {\bf Likewise we need a new round 
of experiments that can measure the rates for $K \to \pi \nu \bar \nu$ {\em accurately} 
with sample sizes $\sim {\cal O}(10^3)$ and mount another serious effort to probe the 
muon transverse polarization in $K_{\mu3}$ decays.} 

I will finish with a poem I have learnt from T.D. Lee a number of years ago. It was written by A.A. Milne, 
who is best known as the author of Winnie-the-Pooh in 1926:

\begin{center} 
{\em Wind on the Hill} 
\end{center} 

{\em No one can tell me} 

\noindent 
{\em Nobody knows} 
 
\noindent 
{\em Where the wind comes from,} 

\noindent 
{\em Where the wind goes.}  

\vspace{5mm} 

{\em But if I stopped holding}  

\noindent 
{\em The string of my kite,}   
 
\noindent 
{\em It would blow with the wind}   

\noindent 
{\em For a day and a night.}  

\vspace{5mm} 

{\em And then when I found it,}   

\noindent 
{\em Wherever it blew,}   
 
\noindent 
{\em I should know that the wind}    

\noindent 
{\em Had been going there, too.}  

\vspace{5mm} 

{\em So then I could tell them}   

\noindent 
{\em Where the wind goes ...}    
 
\noindent 
{\em But where the wind comes from}    

\noindent 
{\em Nobody knows. }

\vspace{0.5cm}
One message from the poem is clear: we have to let our `kite' respond 
to the wind, i.e. we have to perform experiments. Yet the second message `... Nobody knows.' is overly 
agnostic: Indeed experiments by themselves will not provide us with all these answers. It means 
one will still need `us', the theorists, to figure out `where the wind comes from'.

In any case, we are at the beginning of an exciting adventure -- and we are most privileged to participate. 

{\bf Acknowledgments:} I truly enjoyed the beautiful setting and warm atmosphere of the school, the kind help from Danielle Metral and Egil Lillestol and the seemingly effortless, yet firm guidance from the 
`Boss' Tord Ekelof. I benefitted from the discussions with the students about both physics and 
World Cup soccer,  
and I am grateful to them for their kindness in including me in their soccer teams. This work was supported by the NSF under the grant number PHY-0355098.


\end{document}